\documentclass[12pt,twoside,openright]{book}
\usepackage{pstricks,graphicx,amssymb,amsmath,enumerate}
\usepackage{listings}  
\usepackage{supertabular}
\usepackage{pstricks,pst-node}
\usepackage{lscape}
\usepackage[british,german]{babel}
\newcommand{\myref}[1]{(\ref{#1})}
\newcommand{\lr}[1]{\left(#1\right)}
\newcommand{\qgrafText}{\emph{Qgraf}\;}
\newcommand{\qgraf}{\emph{Qgraf}}
\newcommand{\up}[1]{\mathbf{{#1}^+}}
\newcommand{\down}[1]{\mathbf{{#1}^-}}
\begin{document}
\pagestyle{empty}

\selectlanguage{german}
\pagestyle{empty}
\begin{titlepage}
\begin{center}
{\Huge{
\textsc{
Methods for\\
the Reduction of\\
Three-Loop QCD Form Factors\\[0.9cm]
}}}
\vspace{-1.cm}
\center{\rule{\textwidth}{0.1mm}}\\[0.9cm]
\vspace{0.5cm}{
{\bf \Large Dissertation}\\[0.9cm]
}{\large{
zur}\\
{
Erlangung der naturwissenschaftlichen Doktorw\"urde} 
\\
{
(Dr.~sc.~nat.)}\\
{
vorgelegt der}\\
{
Mathematisch-naturwissenschaftlichen Fakult\"at}\\
{
der}\\
{
Universit\"at Z\"urich} \\
 {
von}
\\[14mm]
{\Large \bf Beat T\"odtli}
\\[14mm]
 {
von}\\
 {
Altst\"atten SG}\\[10mm]
{\bf Promotionskomitee}\\[0.2cm]
\begin{tabular}{c}
  Prof. Dr. Thomas Gehrmann (Vorsitz)\\
 Prof. Dr. Daniel Wyler\\
 Prof. Dr. Ben Moore\\
Prof. Dr. Ulrich Straumann
\end{tabular}\\[1cm]
Z\"urich, 2008 } 
\end{center}
\end{titlepage}
\selectlanguage{british}
	

\null\vspace{\stretch{1}}
\begin{flushright}
 Dem z\"urcher Steuerzahler
\end{flushright}
\null\vspace{\stretch{2}}

\pagestyle{headings}
\tableofcontents
\cleardoublepage



\chapter{Introduction}
'Modern physics', as it stands one century after Einstein's \emph{annus mirabilis}, has had the privilege of giving scientific footing and empiric evidence to some insights that philosophers had long ago. 
Immanuel Kant already stated in his ``Critique of Pure Reason''~\cite{kantRV} that space and time are inherent concepts of our mind, not properties of the objects we perceive. 
These ideas always seemed too abstract to be taken seriously by non-pro\-fessional philosophers. But nowadays students in special relativity and quantum mechanics, optical engineers, molecular biologists all at some point in their studies face the severe difficulties that a na\"ive realistic world view (possibly even with a notion of universal time) has in the light of the past 100 years of science. We become painfully aware how much our thinking depends on concepts such as space, time and causality-- prerequisites for our mind to work. They are given to us ``a priori'', and we can do nothing but to accept that they don't seem to be adequate in the context of modern physics. 

Nowadays, we're getting used to lasers, the global positioning system and elementary particles such as electrons. But while we have gained a hundred years of experience in dealing with quantum mechanical and relativistic laws and ``objects'', the basic insights of modern physics remain astounding. 

The three-loop quark form factor that this thesis deals with is a relativistic quantum mechanical amplitude. It describes a quantum mechanical probability amplitude (it is a complex number) and it is a real physical object in the same sense as electrons are (quantum mechanical) waves (not classical ``spheres''). It is tempting to think of the quark form factor as a part of the classical probability of an electron scattering off a proton, rather than of a quantum mechanical amplitude. Classical intuition is comfortable, but it must be kept in mind that it is just that: comfort for a mind that is not adapted to the laws of nature it explores.

Is it real? To the critical reader and also to non-experts it must be admitted that modern experimental particle physics has become extremely complex, and therefore measurements are more indirect. To measure the quark form factor, modern particle physics experiments such as LEP, HERA, and very soon the eagerly awaited Large Hadron Collider (LHC) study various processes such as jet production, deep-inelastic scattering, Drell-Yan photo-production etc. A measurement of a physical quantity is a very complex task involving (just to give a rough and incomplete list) fund raising and constructing the facility, testing and calibrating the experiment, performing Monte-Carlo simulations of the collisions in the detector and finally comparing experimentally measured vs. theoretically expected outcomes of the experiment. Nevertheless, to particle physicists, the quark form factor must be real simply because it is a gauge invariant quantity in quantum chromodynamics (QCD), and because the QCD-improved parton model has been tested so successfully in the perturbative regime. But such an experts answer is clearly neither exhaustive nor the only possible one. And since almost the entire rest of this thesis will strive to be professional and answer all questions about the form factor by putting a single number to it, I'd like to take this moment and pass the question on to someone with an entirely different viewpoint. Maybe, if physicists claim the form factor to be real, it should be real also for non-physicists? Maybe it isn't just a number? ``If the quark form factor is real, how does it look like?''-- Please turn the page and let an artist give her answer.

\begin{figure}
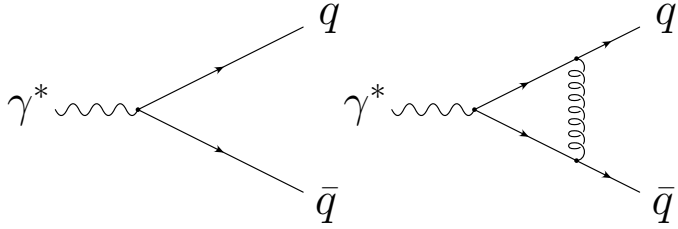

  \begin{center}
	\vspace{1cm}\resizebox{9cm}{!}{
		\includegraphics[width=11cm]{pictures/vertexcorrectiontree.epsi}
		\includegraphics[width=11cm]{pictures/vertexcorrection1loop.epsi}
}
  \end{center}
\caption{The tree-level and one-loop diagrams contributing to the quark form factor.}\label{fig:treeandoneloopdiags}
\end{figure}

\begin{figure}
  \begin{center}
	\vspace{1cm}
	\resizebox{9cm}{!}{
		\includegraphics[width=11cm]{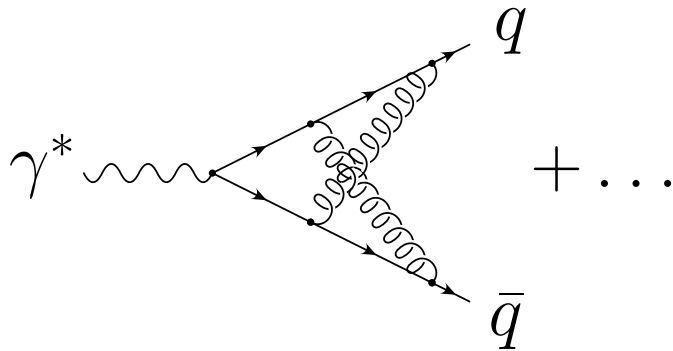}
}
\caption{Diagrams contributing to the two-loop quark form factor. The ellipsis denotes 13 further diagrams.}\label{fig:twoloopdiags}
  \end{center}
\end{figure}

Since it is a QCD quantity, the immediate context of the quark form factor is not so much new physics but rather a precise understanding of ``known'' physics needed to discern small new physics deviations from known physics uncertainties. In perturbation theory the quark form factor is expanded in small coupling constants, the largest of which being $\alpha_s$. The contributions to various orders in this expansion (called ``tree-level'', ``one-loop'', ``two-loop'') can be represented as Feynman graphs. To physicists, Feynman graphs are useful because the converse is also possible: Each Feynman graph corresponds to a quantifiable contribution to  the form factor. The tree-level and one-loop~\cite{Schwinger:1949ra} contributions shown in fig.~\ref{fig:treeandoneloopdiags} are text-book calculations, and the two-loop result in fig.~\ref{fig:twoloopdiags} has been obtained more than 20 years ago~\cite{Gonsalves:1983nq}. 
The three-loop contribution is expected to be a correction of a few percent. Observing it at the LHC might thus be difficult, but possible. For the International Linear Collider (ILC), it clearly is an interesting object, helping to bring the theoretical prediction of several observables to percent-level precision.

In this thesis, only part of the calculation is attempted. Chapter~\ref{chap:Formfactor} will use standard tools such as \qgraf~\cite{Nogueira} and FORM~\cite{Vermaseren:2000nd} to generate the amplitude as a linear combination of three-loop integrals. In chapter~\ref{chap:standardization} we will then have to analyse these integrals closely. When first obtained, they are in a quite unstructured form and we have to devise a (partially) automated way to bring these integrals into a unique standard form. This is necessary because in chapter~\ref{chap:Reduction} we will solve a system of linear equations in those integrals in order to (dramatically) reduce the number of integrals that actually have to be evaluated. These systems of equations would be unnecessarily large without standardisation of the integrals. In view of the fact that so far we have not managed to obtain the final result for the reduction of the three-loop form factor, chapter~\ref{chap:Reduction} will show why this task is so hard, despite the fact that a CPU cluster was at our disposal (http://www-theorie.physik.unizh.ch/~stadel/zbox/start\#zbox2). We will then introduce two immediate improvements (symbolic reduction of reducible topologies and loop-by-loop integration), and also report on improvements on the reduction algorithm we use. 

\begin{landscape}
\thispagestyle{empty}
\begin{pspicture}(18,14)
\end{pspicture}
\begin{center}
Is the three-loop QCD form factor real? If so, how does it look like? An artists impression, courtesy E. Sch\"urmann. 
\end{center}
\end{landscape}



\chapter{The Quark Form Factor}\label{chap:Formfactor}
In this chapter we will introduce the three-loop QCD quark form factor. This is the object we ultimately wish to compute. We will also motivate why its computation should be worth the effort and where it could improve the theory prediction of physical observables. 

\section{Introducing the Quark Form Factor}\label{sec:quarkformfactor}
In perturbative QCD, the quantum mechanical amplitude for the process $\gamma^*\rightarrow q\bar{q}$, ${\cal M}\lr{\gamma^*\rightarrow q\bar{q}}$ is expanded in the strong coupling constant $\alpha_s$. It is then represented as a sum of Feynman graphs:
	\begin{center}
	\includegraphics[width=11cm]{pictures/vertexCorrectionBlobFormfactor.epsi}
	\end{center}
The shaded blob represents the sum over all possible higher order corrections that are allowed by the Feynman rules of QCD. The matrix element ${\cal M}^\mu\left(\gamma^*\rightarrow q\bar{q}\right)$ has an incoming off-shell photon and an outgoing quark-antiquark pair. 

In a theory like QCD that we take to be $CP$-even, the only vectors that can carry the Lorentz index of the external photon are the Dirac matrices $\gamma^\mu$ and the (outgoing) momenta $p_1^\mu$ and $p_2^\mu$ of the external quark and antiquark, respectively. So we may write~\cite{PeskinSchroeder}
\begin{equation}\label{eq:FormfactorAnsatz}
 {\cal M}^{\mu,ss'}\left(\gamma^*\rightarrow q\bar{q}\right)=\lr{-iee_q}\bar{u}^s\!\left(p_1\right)\left[ A \gamma^\mu+B q^\mu+C \left(p_1^\mu-p_2^\mu\right)\right]v^{s'}\!\left(p_2\right)\;,
\end{equation}
where $q=p_1+p_2$ and $e_q$ is $+2/3$ for up- and $-1/3$ for down-type quarks. $\bar{u}$ and $v$ are the spinors for the quark and the antiquark, respectively, and they depend on the particle's spin and momentum. If we considered the process $Z^*\rightarrow q\bar{q}$, couplings proportional to $\gamma^\mu\gamma_5$, $q^\mu\gamma_5$ and $i\sigma^{\mu\nu}q_\nu$ would also be present.

Using the Ward-Identity $q_\mu{\cal M}^\mu=0$ and the Dirac equations of motion, we find that $B=0$. 
Finally, in massless theories with vector-like gauge bosons (like QED and QCD), helicity is conserved, which leads to $C=0$. Taken together, the massless vertex-corrections ${\cal M}^\mu\left(\gamma^*\rightarrow q\bar{q}\right)$ is given by a single \emph{form factor} ${\cal F}\equiv A$. Using Lorentz covariance of ${\cal M}^\mu$ and $\gamma^\mu$ in~\myref{eq:FormfactorAnsatz}, it must be a Lorentz scalar and thus it may only depend on kinematical invariants. Since the only non-vanishing kinematical invariant is $q^2$, we have
\begin{equation}\label{eq:FormfactorMatrixelement}
{\cal M}^{\mu,ss'}_q\left(\gamma^*\rightarrow q\bar{q}\right)= \left(-i e e_q\right){\cal F}\left(q^2\right)\bar{u}_q^{s}\!\left(p_1\right) \gamma^\mu v_q^{s'}\!\left(p_2\right)\;.
\end{equation}
The form factor may now be extracted from~\myref{eq:FormfactorMatrixelement} via
\begin{equation}\label{eq:Formfactorprojection}
 {\cal F}\left(q^2\right)=\sum_{q=1}^{n_f n_c} \frac{1}{2\left(2-d\right) \lr{-i e e_q}q^2}\sum_{s,s'}\bar{v}_q^{s'}\!\left(p_2\right)\gamma_\mu u_q^s\!\left(p_1\right){\cal M}^{\mu,ss'}_q\left(\gamma^*\rightarrow q\bar{q}\right)\;.
\end{equation}
Eq.~\myref{eq:Formfactorprojection} computes the quantity we want, ${\cal F}\lr{q^2}$, out of ${\cal M}^{\mu,ss'}_q\left(\gamma^*\rightarrow q\bar{q}\right)$, which is a sum of three-loop Feynman graphs. In sec.~\ref{sec:FeynmanGraphGeneration} we will describe their generation and how to evaluate the sum over spins in~\myref{eq:Formfactorprojection}. The rest of chapter~\ref{chap:standardization} will be concerned with finding a compact standard representation for the three-loop integrals that we obtain after this summation over spins. This is preparatory work necessary for the reduction to master integrals that will be described in chapter~\ref{chap:Reduction}.


\section{Physics Motivation for the Quark Form Factor}
Having defined the quark form factor ${\cal F}\lr{q^2}$, we next have a look at its physical relevance. It constitutes part of the calculation for various different physical processes, on some of which we shall comment below. Apart from this its infrared behaviour is of special interest since it may be used to determine various (process independent) quantities in the context of resummation. 

The three-loop contribution, being formally ${\cal O}\lr{\alpha \alpha_s^3}$, is expected to contribute a correction of a few percent and thus in many cases it will probably increase the precision of the theory prediction to the percent level and possibly beyond. The corresponding accuracy will be reached in experiments both at the LHC and the ILC. 

\subsection{The Form Factor in $e^+e^-$ Collisions}
\begin{figure}[ht]
 \begin{center}
	\resizebox{5cm}{!}{\includegraphics{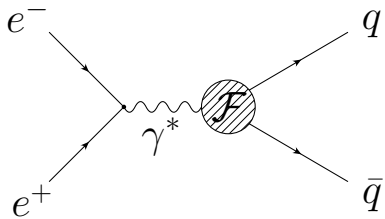}}
 \end{center}
\caption{The form factor enters $e^+e^-\rightarrow 2\mathrm{jets}$.}
\label{diag:epluseminusFormfactor}
\end{figure}

The three-loop QCD form factor directly enters the matrix element for $e^+e^-\rightarrow q\bar{q}$. It may be written down using eq.~\myref{eq:FormfactorMatrixelement}, where the form factor acts as an effective  $\gamma$-$q$-$\bar{q}$-vertex and thus includes the higher order QCD corrections. This effective vertex is denoted by a blob in fig.~\ref{diag:epluseminusFormfactor}. Clearly, if three-loop QCD corrections are included in this way, QED corrections cannot be neglected, and they may not be of the form of fig.~\ref{diag:epluseminusFormfactor}. For example, an additional photon may be exchanged between the external electron and quark line. 

Now the cross section for $e^+e^-\rightarrow q\bar{q}$ is not infrared finite at higher orders in the perturbative expansion, and neither is the form factor. Thus so far the QCD form factor will only summarise an important contribution to physical processes. Other contributions come from radiative corrections with additional external gluons and photons. Physical observables are e.g. the cross sections for $e^+e^-\rightarrow n\mathrm{~jets}$ and all their differential distributions (event shapes, angular and $p_T$-distributions etc.). At present they are known up to (at most) NNLO, and there the contribution from the two-loop vertex correction has certainly been one of the easier parts of the calculation. The $\mathrm{N}^3\mathrm{LO}$ contribution, to which the three-loop form factor would belong, is clearly not yet within reach of a theoretical computation. We may still view the three-loop contribution as a first step in this direction. Knowing the theory value of the form factor to three loops should eventually lead to an improved determination of the strong coupling constant $\alpha_s$ and its running at the ILC. New physics can then be envisaged once the standard model parameters are known precisely enough. For example, the present error on $\alpha_s$ represents the dominant uncertainty on the prediction of the scale for grand unification of the strong and electro-weak interactions~\cite{ILC_RDR2007}.

\subsection{The Form Factor in Deep-Inelastic scattering}
\begin{figure}[ht]
 \begin{center}
	\resizebox{5cm}{!}{\includegraphics{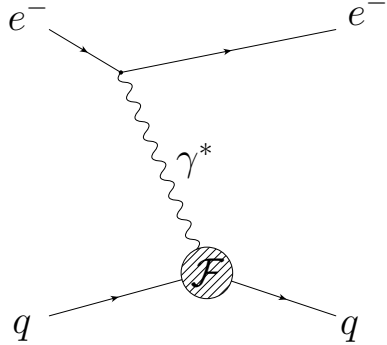}}
 \end{center}
\caption{In deep-inelastic scattering, the exchanged photon probes the target nucleon structure via the form factor.}
\label{diag:DISFormFactor}
\end{figure}

Deep-inelastic scattering (e.g. at HERA, Hamburg) prominently involves the exchange of a hard photon between the incident lepton and the target hadron. At parton level, the photon is incident to a quark line and so the quark form factor enters directly here (see fig.~\ref{diag:DISFormFactor} for an illustration). 
Deep-inelastic scattering experiments usually measure structure functions which
 then allow to determine parton distribution functions. Together with Drell-Yan lepton pair production deep-inelastic scattering processes are among the best suited ones for probing the short-distance structure of hadrons. The uncertainty from higher order perturbative corrections often dominates and so higher order calculations like this three-loop correction are needed. 

\subsection{The Form Factor in Drell-Yan Lepton Pair Production.}
\begin{figure}[ht]
 \begin{center}
	\resizebox{5cm}{!}{\includegraphics{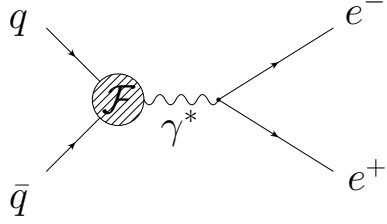}}
 \end{center}
\caption{The Form Factor enters the Drell-Yan cross-section when a virtual photon is exchanged.}
\label{diag:DrellYanFormFactor}
\end{figure}

Data on Drell-Yan lepton pair production, $p\bar{p}\rightarrow \ell^+\ell^-$ (or $pp\rightarrow \ell^+\ell^-$) will reach unprecedented precision at the LHC. It will clearly be possible to see corrections of the order of one percent or even in the per-mille range for certain observables such as the W- or Z-boson mass and $\sin\theta_W$. Fig.~\ref{diag:DrellYanFormFactor} shows how the $q$-$\bar{q}$-$\gamma$ form factor enters the Drell-Yan cross section. To make quantitative predictions also for differential distributions in the Drell-Yan process, the parton distributions for both incident (anti-)quarks are needed:
\begin{equation}
\begin{split}
\frac{d\sigma}{dQ^2dy}\sim\sum_{a,b}\int_{x_A}^1&d\xi_A\int_{x_B}^1d\xi_B f_{a/A}\lr{\xi_A,\mu}\frac{d\hat{\sigma}}{dQ^2dy}f_{b/B}\lr{\xi_B,\mu}
\end{split}
\end{equation}
Here, $f_{a/A}$ denotes the parton distribution function and can be interpreted as the probability to find a parton of type $a$ in the hadron of type $A$ carrying a fraction $\xi_A$ to $\xi_A+d\xi_A$ of the hadron's momentum. They are universal in the sense that they are independent of the hard scattering process. They are best extracted from deep-inelastic scattering experiments. Due to the uncertainties in the parton distribution functions, $\alpha_s$-measurements will not reach the precision of LEP or ILC experiments. Rather, as stated before, the form factor will be an ingredient of ``known physics'' that will be needed when searching for ``new physics''.
\subsection{The Form Factor in the Context of Resummation}
Form factors also play an important role in a different way: Using factorisation and gauge invariance it can be shown that infrared singularities of the form $\ln^n \lr{q^2/\mu^2}$ can be resummed. They exponentiate according to~\cite{Magnea:2000ss}
\begin{equation}
\begin{split}
{\cal F}\lr{\frac{q^2}{\mu^2},\alpha_s\lr{\mu^2},\epsilon}=
\exp\biggl\{&\frac{1}{2}\int_0^{-q^2}\frac{d\xi^2}{\xi^2}\biggl[\phantom{+}K\lr{\epsilon,\alpha_s\lr{\mu^2}}\\
\biggl.&\phantom{\frac{1}{2}\int_0^{-q^2}\frac{d\xi^2}{\xi^2}\biggl[}+G\lr{-1,\bar{\alpha_s}\lr{\frac{\xi^2}{\mu^2},\alpha_s\lr{\mu^2},\epsilon},\epsilon}\biggr]\\&+\frac{1}{2}\int_{\xi^2}^{\mu^2}	\frac{d\lambda^2}{\lambda^2}\gamma_K\lr{\bar{\alpha}\lr{\frac{\lambda^2}{\mu^2},\alpha_s\lr{\mu^2},\epsilon}}\biggr\}\;.
\end{split}
\end{equation}
By matching the general exponentiated expression for the form factor onto a fixed order calculation, resummation coefficients in the exponent may be determined. These in turn allow --after reexpansion of the exponential-- to determine subleading logarithms in the coefficient functions for processes such as deep-inelastic scattering or Drell-Yan lepton pair production. 

\section[Current Status of the Form Factor]{Current Status of QCD and Electroweak Corrections to Form Factors}

\subsection{QED Form Factors}
Vertex corrections are prominent already in typical first year quantum field theory courses when the ${\cal O}\lr{\alpha}$ correction to the anomalous magnetic moment of the electron~\cite{Schwinger:1948iu} is studied. The higher order contributions to the QED vertex can be used to determine the fine structure constant $\alpha$. These calculations have been pushed to ${\cal O}\lr{\alpha^4}$ and turned this observable famous~\cite{Kinoshita:1981vs,Gabrielse:2006gg} for yielding one of the most precise predictions in physics. The muon anomalous magnetic moment is particularly interesting for finding new physics effects~\cite{Czarnecki:2001pv}. Calculations for massive two-loop vertex corrections have also been performed already~\cite{Bonciani:2003te}.

\subsection{QCD Form Factors}
The one-loop QCD vertex correction was derived in 1949 by Schwinger~\cite{Schwinger:1949ra}. The massless two-loop contribution became available by the work of Gonsalves~\cite{Gonsalves:1983nq} in 1983, and also by Kramer and Lampe~\cite{Kramer:1986sg} and van Neerven et al.\cite{Matsuura:1988sm}. Massive external quarks were also considered at two loops. As was already mentioned in sec.~\ref{sec:quarkformfactor}, there are four (CP even) form factors in this case, two vector and two axial vector form factors. They were calculated in~\cite{Bernreuther:2004ih,Bernreuther:2004th,Bernreuther:2005rw}.

Quite recently, in two by now famous papers Moch, Vermaseren and Vogt computed the three-loop QCD splitting functions~\cite{Moch:2004pa,Vogt:2004mw}, from which they derived the divergent part (the $1/\epsilon$ poles in dimensional regularisation) of the three-loop quark form factor~\cite{Moch:2005id}.

\subsection{Electroweak Form Factors}
Due to the additional scales of the W- and Z-boson masses, electroweak corrections to the form factor are much more difficult. One-loop results for observables in $e^+e^-\rightarrow f\bar{f}$ (where $f$ is either a lepton or a quark) have already been available for LEP~\cite{BardinPassarinoBuch}. Full two-loop results do not seem to be in reach yet, though progress is made~\cite{Aglietti:2004tq}. It was then realized that leading and sub-leading ``Sudakov``-type logarithms, $\mathrm{ln}q^2/m_{W,Z}^2$ contribute corrections of up to 10\%~\cite{Ciafaloni:1998xg,Beccaria:1999fk} at $q^2=1\mathrm{~TeV}$. Thus aiming at predictions for the ILC the community has focused on resumming (leading and next-to-leading) logarithms of $q^2$ over $M_{W,Z}$ for various processes (see~\cite{Pozzorini:2001rs} and references therein).

\chapter{Generation and Standardisation}\label{chap:standardization}
We now concentrate on the three-loop QCD contribution to the quark form factor. As a first step, section~\ref{sec:FeynmanGraphGeneration} will explain how to obtain the form factor as a linear combination of three-loop integrals. Once obtained, these Feynman integrals will still be quite unstructured. Sections \ref{sec:compactIntegrals}, \ref{sec:eliminatingVanishingTopologies} and \ref{sec:exploitingSymmetries} then describe how the following goals are met: 
\begin{itemize}
 \item To find a compact representation for all three-loop vertex integrals,
 \item To identify those topologies that vanish trivially, and
 \item To exploit all (obvious) symmetries to make this representation as unique as possible.
\end{itemize}
In this way, we will arrive at a unique representation of any three-loop integral that we will have to compute. The actual computation is attempted in chapter~\ref{chap:Reduction}.

\section{Generating the Three-Loop Quark Form Factor}\label{sec:FeynmanGraphGeneration}
 We would like to obtain the form factor using eq.~\myref{eq:Formfactorprojection} by providing the matrix element ${\cal M}^{\mu,ss'}_q\lr{\gamma^*\rightarrow q\bar{q}}$ explicitly. 

It is well known that Feynman diagrams exhibit factorial growth in external legs and also in loops. We will soon see that for the three-loop form factor, one faces --depending on how diagrams are counted-- several hundred diagrams. Thus it is clearly necessary to automatise the generation and evaluation of these diagrams. A number of tools have been written to this end, e.g. \emph{FeynArts}~\cite{Hahn:2000kx} and \qgraf~\cite{Nogueira}; in appendix~\ref{app:Qgraf} we give further details on how \qgrafText and FORM~\cite{Vermaseren:2000nd} were used for the generation of the quark form factor. 

\qgrafText essentially requires the input of a model file, a style file, and the file \texttt{qgraf.dat}. The model file describes which types of vertices and propagators appear in the diagrams to be generated. \emph{Qgraf}'s output is user-defined according to the style file. When \qgrafText is run, it reads the process definition given in \texttt{qgraf.dat} and then generates information on each possible diagram. The style file then defines how this information is written to the output, a plain text file. The style file may be tailored such that the output can be read into FORM. 

When \qgrafText is run it generates the three-loop diagrams for $\gamma^*\rightarrow q\bar{q}$. The output of \qgrafText contained 244 diagrams.

From here, the Feynman amplitude may be constructed in FORM. The projection onto the form factor, eq.~\myref{eq:Formfactorprojection}, involves summing over spins and colour factors. They lead to traces of Dirac and colour matrices, respectively. FORM has built-in capabilities for evaluating Dirac traces, and colour traces are easily evaluated using
\begin{eqnarray}
 f^{abc}&=&\frac{-i}{T_R}\mathrm{Tr}\lr{T^a\left[T^b,T^c\right]}\;,\\
 T^a_{i_1i_2}T^a_{i_3i_4}&=&T_R\lr{\delta_{i_1i_2}\delta_{i_3i_4}-\frac{1}{N_C}\delta_{i_1i_3}\delta_{i_2i_4}}\;,
\end{eqnarray}
with $T_R=1/2$. The Dirac traces produce scalar products involving loop and external momenta. The propagators involve the loop momenta in a way dictated by \qgraf. For example, the contribution of the diagram discussed in more detail in appendix~\ref{app:Qgraf}, fig.~\ref{diag:Label1}, will involve integrals of the form
\begin{equation}\label{eq:Label1SampleIntegral}
 \int \Biggl(\prod_{i=1}^3 d^dk_i\Biggr)\frac{k_1\cdot k_2,\lr{k_3\cdot q}^2,\mathrm{etc.}}{k_1^2k_2^2k_3^2\lr{\lr{k_1+p_1}^2}^2\lr{k_1+q}^2\lr{k_1-k_2-k_3+p_1}^2}\;.
\end{equation}
In order for this output to be useful, we will next exploit the freedom to shift the loop momenta, in such a way as to bring the propagators into a unique form. The next chapter will be devoted to this task. 

\section{Representing Integrals in a Compact Way}\label{sec:compactIntegrals}
In this section our goal is to set up a compact notation that is capable of representing any possible three-loop integral that we might encounter in our reduction. As a first step, we will have a look at the possible scalar products that appear in the numerators and denominators of our integrals. 

The three-loop vertex-integrals have denominators consisting of propagators, and numerators that are scalar products stemming e.g. from the Dirac traces of the quark line. Numerators and denominators consist of scalar products involving the loop momenta $k_1,k_2,k_3$ and external momenta $p_1$ and $p_{12}=p_1+p_2$. Out of these, 13 different nonvanishing scalar products can be built:\\
\begin{equation}
 \begin{tabular}{lll}\label{eq:scalarProducts}
$k_1\cdot k_1\qquad$& $k_2\cdot k_2\qquad$& $k_3\cdot k_3\qquad$\\
$k_1\cdot k_2\qquad$& $k_1\cdot k_3\qquad$& $k_2\cdot k_3\qquad$\\ 
$k_1\cdot p_1\qquad$& $k_2\cdot p_1\qquad$& $k_3\cdot p_1\qquad$\\
$k_1\cdot p_{12}\qquad$& $k_2\cdot p_{12}\qquad$& $k_3\cdot p_{12}\qquad$\\
$p_{12}\cdot p_{12}\;.$&&
\end{tabular}
\end{equation}

These scalar products span a (real) vector space. The set of possible inverse propagators that might occur in our three-loop integrals are elements in it. So alternatively, we may choose as our basis any set of inverse propagators, e.g.:
\begin{equation}
\begin{tabular}{lll}\label{eq:PropStandardSet}
$\phantom{(k_1}k_1^2\phantom{k_1)^2}\qquad$&$\;\phantom{(k_1}k_2^2\phantom{k_1)^2}\qquad$&$\phantom{(k_2}\,k_3^2\phantom{k_1)^2}$\\
$\lr{k_1-k_2}^2$&
$\lr{k_1-k_3}^2$&
$\lr{k_2-k_3}^2$\\
$\lr{k_1-p_1}^2$&
$\lr{k_1-p_{12}}^2$&\\
$\lr{k_2-p_1}^2$&
$\lr{k_2-p_{12}}^2$&\\
$\lr{k_3-p_1}^2$&
$\lr{k_3-p_{12}}^2$
\end{tabular}
\end{equation}
together with $p_{12}^2$.
The basis transformation from~\myref{eq:scalarProducts} to~\myref{eq:PropStandardSet} reads
\begin{equation}\label{eq:scalarProductToinvProp}
 \begin{split}
k_1\cdot k_1 =& k_1^2\\
k_2\cdot k_2 =& k_2^2\\
k_3\cdot k_3 =& k_3^2\\
k_1\cdot k_2 =& \left[k_1^2 - \lr{k_1-k_2}^2+k_2^2\right]\!/\,2\\
k_1\cdot k_3 =& \left[k_1^2- \lr{k_1-k_3}^2 + k_3^2\right]\!/\,2\\
k_2\cdot k_3 =& \left[k_2^2- \lr{k_2-k_3}^2 + k_3^2\right]\!/\,2\\
k_1\cdot p_1 =& \left[k_1^2- \lr{k_1-p_1}^2\right]\!/\,2\\
k_1\cdot p_2 =& \left[\lr{ \lr{k_1-p_1}^2}- \lr{k_1-p_{12}}^2 + p_{12}^2\right]\!/\,2\\
k_2\cdot p_1 =& \left[k_2^2- \lr{k_2-p_1}^2\right]\!/\,2\\
k_2\cdot p_2 =& \left[\lr{k_2-p_1}^2- \lr{k_2-p_{12}}^2  + p_{12}^2\right]\!/\,2\\
k_3\cdot p_1 =& \left[k_3^2-\lr{k_3-p_1}^2 \right]\!/\,2\\
k_3\cdot p_2 =& \left[\lr{k_3-p_1}^2 -\lr{k_3-p_{12}}^2  + p_{12}^2\right]\!/\,2\;.\\
p_1\cdot p_2 =& p_{12}^2/2\;.
\end{split}
\end{equation}
\begin{figure}
 \begin{center}
	\resizebox{7cm}{!}{\includegraphics{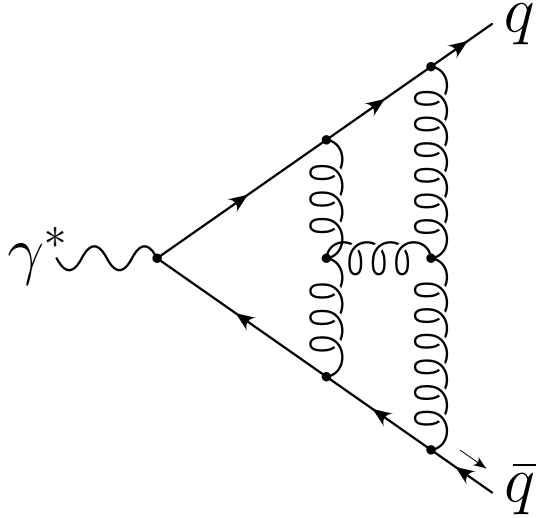}}
 \end{center}
\caption{A sample planar diagram, called $A_{9,1}$. It is the only $t=9$ planar master integral that appears in the reduction of the three-loop quark form factor.}
\label{diag:1790}
\end{figure}


Now a Feynman integral is determined if the propagators and scalar products are given as a linear combination of~\myref{eq:scalarProducts}. Alternatively, we may use \myref{eq:PropStandardSet} together with $p_{12}^2$. In general, each propagator of an arbitrary graph will be given as a linear combination of the propagators in~(\ref{eq:PropStandardSet}) and $p_{12}^2$.

Still aiming towards a compact representation of the three-loop vertex integrals, we will now argue the following: For integrals occurring in planar diagrams, it is possible to shift loop momenta such that all occurring propagators are members of the basis set~(\ref{eq:PropStandardSet}), rather than just a linear combination of them. 
For example, the scalar integral (the one without any scalar products in the numerator) for the diagram in fig.~\ref{diag:1790} can be represented as 
\begin{multline}\label{eq:exampleStandardProps}
\int d^dk_1d^dk_2 d^dk_3\frac{1}{k_1^2k_2^2\lr{k_1-k_2}^2\lr{k_1-k_3}^2\lr{k_2-k_3}^2}\\
\times\frac{1}{\lr{k_1-p_1}^2\lr{k_2-q}^2\lr{k_3-p_1}^2\lr{k_3-q}^2}\\
={\cal T}(-1,-1,0,-1,-1,-1,-1,0,0,-1,-1,-1)\;.
\end{multline}
So the integral is now given by a compact expression indicating the 12 exponents of the (inverse) propagators \myref{eq:PropStandardSet}. If the `propagator' $1/p_{12}^2$ should ever appear, we will simply factor it out of the integral and write it down explicitly. Its exponent could even be inferred by dimensional analysis, since it constitutes the only scale left after the loop integration is performed.

The notation introduced in~\myref{eq:exampleStandardProps} includes the possibility of having the same propagator appearing several times, as is the case in the diagram of fig.~\ref{diag:Label1}, and in many integrals generated in integration-by-parts identities (see below, chap.~\ref{chap:Reduction}). Furthermore, if the numerator of the integral contains scalar products, they may be rewritten as inverse propagators using~\myref{eq:scalarProductToinvProp}.

Thus this notation is capable of expressing any three-loop integral (including scalar products) if its propagators are already shifted such that they are members of a standard set. As a more general example for the same diagram fig.~\ref{diag:1790}, we have
\begin{multline}
  \int d^dk_1d^dk_2 d^dk_3\frac{k_2\cdot p_2}{\lr{k_1^2}^3 k_2^2\lr{k_1-k_2}^2\lr{\lr{k_1-k_3}^2}^2\lr{k_2-k_3}^2}\\ 
                          \times \frac{1}{\lr{k_1-p_1}^2\lr{k_2-q}^2\lr{k_3-p_1}^2\lr{k_3-q}^2}\\
\begin{aligned}
=\frac{1}{2}\Bigl(&-{\cal T}_1(-3,-1,0,-2,-1,-1,-1,0,0,0,-1,-1)\\
&+{\cal T}_1(-3,-1,0,-2,-1,-1,-1,0,1,-1,-1,-1)\\
&+p_{12}^2{\cal T}_1(-3,-1,0,-2,-1,-1,-1,0,0,-1,-1,-1)\Bigr)\;, 
\end{aligned}
\end{multline}
where the subscript $1$ in ${\cal T}_1$ shall specify that the propagators are taken with respect to the propagators in~\myref{eq:PropStandardSet}. We will refer to a (standard) set of propagators such as~\myref{eq:PropStandardSet} as an \emph{auxiliary topology}. A \emph{topology} is specified by indicating a subset of distinct propagators of the auxiliary topology.

\begin{figure}
  \begin{center}
	\resizebox{6cm}{!}{\includegraphics{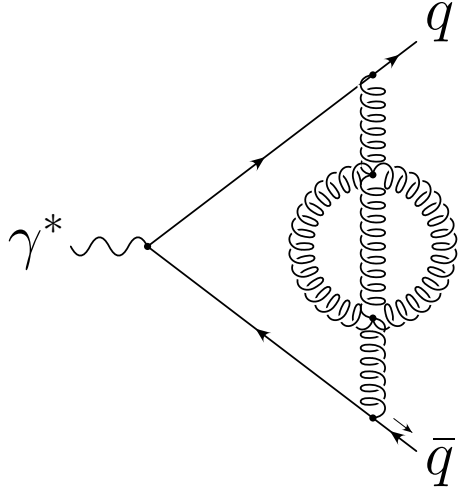}}
  \end{center}
\caption{A diagram whose topology inherently consists of two propagators with the same momentum flow.}\label{diag:Label1}
\end{figure}

\subsection{Auxiliary Topologies}
\begin{figure}
 \begin{center}
	\resizebox{7cm}{!}{\includegraphics{pictures/vertexCorrection3LoopNonPlanar2.epsi}}
 \end{center}
\caption{A sample non-planar diagram belonging to the auxiliary topology number 2.}
\label{diag:nonplanar2}
\end{figure}

One important virtue of the notation introduced in the previous section lies in the fact that usually many different Feynman graphs can be represented in one auxiliary topology. For example it turns out that all planar\footnote{The notions ``planar'' or ``nonplanar'' are properties defined in graph theory. We use them without much formality. Planar diagrams are diagrams that can be drawn on the plane in such a way that propagators only intersect at vertices. Nonplanar graphs are graphs where this is not possible.} diagrams share the auxiliary topology already mentioned earlier:
\begin{equation}
\begin{tabular}{lll}\label{eq:PropStandardSetRestated}
$\phantom{(k_1}k_1^2\phantom{k_1)^2}\qquad$&$\;\phantom{(k_1}k_2^2\phantom{k_1)^2}\qquad$&$\phantom{(k_2}\,k_3^2\phantom{k_1)^2}$\\
$\lr{k_1-k_2}^2$&
$\lr{k_1-k_3}^2$&
$\lr{k_2-k_3}^2$\\
$\lr{k_1-p_1}^2$&
$\lr{k_1-p_{12}}^2$&\\
$\lr{k_2-p_1}^2$&
$\lr{k_2-p_{12}}^2$&\\
$\lr{k_3-p_1}^2$&
$\lr{k_3-p_{12}}^2$\;.
\end{tabular}
\end{equation}

In general however, a single auxiliary topology will not be enough. In particular, non-planar diagrams do not fit into the propagator set~\myref{eq:PropStandardSetRestated}. As an example, the diagram in fig.~\ref{diag:nonplanar2} has the propagators

\begin{equation}\label{eq:PropStandardSetNP2_1}
\begin{split}
& k_2^2,k_3^2,\left(k_1-k_2\right)^2,\left(k_1-k_3\right)^2,\left(k_1-k_2-k_3\right)^2\\
& \left(k_1-p_1\right)^2,\left(k_1-q\right)^2,\left(k_2-p_1\right)^2,\left(k_3-q\right)^2\;.
\end{split}
\end{equation}

Clearly, the inverse propagator $\lr{k_1-k_2-k_3}^2$ in \myref{eq:PropStandardSetNP2_1} is not one of~\myref{eq:PropStandardSetRestated}. So it is necessary to define a new auxiliary topology that contains this propagator and again has 12 linearly independent propagators. We choose
\begin{equation}
\begin{tabular}{lll}\label{eq:PropStandardSetNP2}
$\phantom{(k_1}k_1^2\phantom{k_1)^2}\qquad$&$\;\phantom{(k_1}k_2^2\phantom{k_1)^2}\qquad$&$\phantom{(k_2}\,k_3^2\phantom{k_1)^2}$\\
$\lr{k_1-k_2}^2$&
$\lr{k_1-k_3}^2$&
$\lr{k_1-k_2-k_3}^2$\\
$\lr{k_1-p_1}^2$&
$\lr{k_1-p_{12}}^2$&\\
$\lr{k_2-p_1}^2$&
$\lr{k_2-p_{12}}^2$&\\
$\lr{k_3-p_1}^2$&
$\lr{k_3-p_{12}}^2$\;.
\end{tabular}
\end{equation}
\begin{figure}
 \begin{center}
	\resizebox{7cm}{!}{\includegraphics{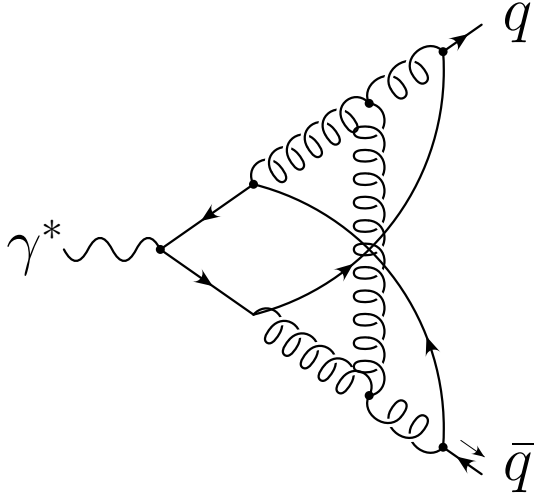}}
 \end{center}
\caption{A sample non-planar diagram belonging to the auxiliary topology number 3.}
\label{diag:nonplanar1}
\end{figure}
\begin{table}
\begin{center}
\begin{tabular}{rlll}
ID& Topology 1 & Topology 2 & Topology 3 \\
1&$k_1$		& $k_1$		&$k_1$\\
2&$k_2$		&$k_2$		&$k_2$\\
4&$k_3$		&$k_3$		&$k_3$\\
8&$k_1-k_2$		&$k_1-k_2$	&$k_1-k_2$\\
16&$k_1-k_3$		&$k_1-k_3$	&$k_1-k_3$\\
32&$k_2-k_3$		&$k_2-k_3$	&$k_1-k_2-k_3$\\
64&$k_1-p_1$		&$k_1-k_3-p_2$	&$k_1-p_1$\\
128&$k_1-p_{12}$	&$k_1-p_{12}$	&$k_1-p_{12}$\\
256&$k_2-p_{1}$	&$k_2-p_{1}$	&$k_2-p_{1}$\\
512&$k_2-p_{12}$	&$k_1-k_2-p_{2}$	&$k_2-p_{12}$\\
1024&$k_3-p_{1}$	&$k_3-p_{1}$	&$k_3-p_{1}$\\
2048&$k_3-p_{12}$	&$k_3-p_{12}$	&$k_3-p_{12}$\\
\end{tabular}
\caption{The auxiliary topologies needed to represent all integrals appearing in the three-loop form factor. Topology 1 corresponds to planar diagrams. The numbers on the left will be convenient to determine the topology ID (see sec.~\ref{sec:exploitingSymmetries} below).
}\label{tab:auxtopos}
\end{center}
\end{table}
Finally, a third auxiliary topology is needed to represent the propagators of diagrams like the one in fig.~\ref{diag:nonplanar1}. Two new propagators are needed: $1/\lr{k_1-k_3+p_1}^2$ and $1/\lr{k_1-k_2+p_2}^2$. We may choose the auxiliary topology
\begin{equation}
\begin{tabular}{lll}\label{eq:PropStandardSetNP3}
$\phantom{(k_1}k_1^2\phantom{k_1)^2}\qquad$&$\;\phantom{(k_1}k_2^2\phantom{k_1)^2}\qquad$&$\phantom{(k_2}\,k_3^2\phantom{k_1)^2}$\\
$\lr{k_1-k_2}^2$&
$\lr{k_1-k_3}^2$&
$\lr{k_2-k_3}^2$\\
$\lr{k_1-k_3-p_2}^2$&
$\lr{k_1-p_{12}}^2$&\\
$\lr{k_2-p_1}^2$&
$\lr{k_1-k_2-p_2}^2$&\\
$\lr{k_3-p_1}^2$&
$\lr{k_3-p_{12}}^2$\;.
\end{tabular}
\end{equation}
We find that all integrals appearing in the three-loop quark form factor can be represented in one of the auxiliary topologies~\myref{eq:PropStandardSetRestated}, \myref{eq:PropStandardSetNP2} and~\myref{eq:PropStandardSetNP3}. Tab.~\ref{tab:auxtopos} summarises the auxiliary topologies that are needed to represent all integrals appearing in the three-loop quark form factor.

In what follows, we will concentrate on the case of planar diagrams. An omitted index $1$ on integrals ${\cal T}\equiv{\cal T}_1$ will be understood. For several reasons, the reduction of planar topologies is the most important and most difficult one: Firstly, 202 out of the 244 diagrams will turn out to be planar below. Furthermore, the reduction algorithm discussed in chapter~\ref{chap:Reduction} will show that integrals of a given topology can be reduced to integrals of so-called \emph{subtopologies}. They have one type of propagator less than the parent topology. Since for the quark form factor there are no non-planar topologies with less than seven propagators, the non-planar topologies will reduce to planar ones at the level of six-propagator topologies. In addition quite a few 8- or 7-propagator subtopologies of non-planar topologies are planar. 

\subsection{Casting Propagators into a Standard Form}
The propagators of the \qgrafText output differ from those in tab.~\ref{tab:auxtopos} by shifts in the loop momenta. For example, the \qgrafText output for the diagram in fig.~\ref{diag:Label1} contains integrals such as the ones given in~\myref{eq:Label1SampleIntegral}. By inspection, one finds that after applying the shift
\begin{eqnarray*}
 k_1&\rightarrow& -k_1',\\ k_2&\rightarrow& k_2'-k_1',\\ k_3&\rightarrow&-k_3'+p_1\;,
\end{eqnarray*}
all propagator momenta are members of the auxiliary topology 1 in tab.~\ref{tab:auxtopos}, up to a sign.
Thus in order to bring the integrals of the \qgrafText output into the compact form defined in the last section, the task is to find this shift and apply it to the \qgrafText output; to rewrite the scalar products as inverse propagators and cast all integrals into the form of one of the auxiliary topologies in tab.~\ref{tab:auxtopos}.

How to find these shifts in the loop momenta? Formally we are given $t$ propagators that are expressed as linear combinations of $\vec{k}=\left\{k_1,k_2,k_3,p_1, p_2\right\}$; i.e. the set of propagator momenta can be written as $A.\vec{k}$ where A is a $t\times 5$-matrix. Similarly the target set of propagator momenta can be written as $B.\vec{k}'$. We are looking for a change of basis $\vec{k}'=S.\vec{k}$ such that $A.\vec{k}=B.\vec{k}'$ \emph{up to a permutation $P$, and up to a sign} in each component. I.e. given $A$ and $B$, find a permutation $P$ and a change of basis $S$ such that $A.\vec{k}=P.B.S.\vec{k}$ up to a sign in each component. A brute force algorithm to find $S$ and $P$ can be set up like this:
\begin{enumerate}
 \item Choose any $P$.
 \item Set up the system $A.\vec{k}=P.B.S.\vec{k}$.
\item Pick the first equation out of the system that contains $k_1$ and extract the loop momentum shift, $k_1=S_{1i}\vec{k}'_i$.
 \item Apply this shift to the system, $\left.A.\vec{k}\right|_{k_1\rightarrow k_1'}=\left.P.B.S.\vec{k}\right|_{k_1\rightarrow S_{1i}k_i'}$ and check whether it might still have a solution. I.e. there should be no equation that does not depend on unprimed loop momenta and that is not a member of $B.\vec{k}'$. \label{item:check}
\item If a solution is still possible, repeat step 3 and 4 with loop momentum $k_2$ (instead of $k_1$), then $k_3$. If not, choose the next $P$ and go to step 2.
\item Upon arriving here, $k_i\rightarrow S_{1j}\vec{k}'_j$ is a solution.
\end{enumerate}
The problem is the permutation $P$- because na\"ively trying all $12!\sim 5\cdot 10^9$ possibilities is out of question. But the check in step~\ref{item:check} prevents a lot of permutations of being fully tested. The average computation time for a given diagram and with $B$ according to the set~\myref{eq:PropStandardSetRestated} was $\sim$ 15 seconds on a 1.7GHz Pentium processor. Diagrams where no solution could be found were all non-planar. 

For planar graphs a different (and faster) approach is worth pointing out: There, a unique shortest path between the external legs exists. Then, the following algorithm can be used: 
\begin{enumerate}
 \item Find the shortest paths between any two of the three external vertices.
 \item Out of these three paths, choose the longest one (or any one of the longest ones).
 \item Prescribe the following loop momenta on the first two legs: 
	\begin{itemize}
	 \item For the $\gamma$-$q$ leg: $k_1$, $k_2$;
	 \item For the $\gamma$-$\bar{q}$ leg: $k_1-p_{12}$, $k_2-p_{12}$;
	 \item For the $q$-$\bar{q}$ leg: $k_1-p_1$, $k_2-p_1$.
	\end{itemize}
 \item Shift the third loop momentum to fit a chosen propagator into any of the 5 $k_3$-dependent propagators. Try all possibilities if necessary.
\end{enumerate}
In the above procedure, steps 1 to 3 fix two of the three loop momenta. This is always possible because there is no nonvanishing three-loop vertex graph whose longest shortest path between external vertices contains just one propagator\footnote{This statement is true for any two- or higher-loop (one-particle-irreducible) vertex graph.}. For the third loop momentum, at most five shifts have to be considered (since in the planar auxiliary topology each loop momentum occurs in five propagators). The only danger here might be snail (or tadpole\footnote{A definition of tadpole and snail topologies is given in the \qgrafText manual, see~\cite{Nogueira}. Loosely speaking, a tadpole contains a propagator that carries zero momentum. Fig.~\ref{diag:vanishing} contains a snail that is not a tadpole.}) diagrams; but \qgrafText has an option to switch off the generation of those diagrams. 

\section{Eliminating Vanishing Topologies}\label{sec:eliminatingVanishingTopologies}
In this section we will identify those loop integrals that evaluate to zero. The reduction using integration-by-parts identities will prove difficult and computationally expensive, so topologies that are zero should not be treated in this way, nor should they be kept in reduction equations of other topologies.

In dimensional regularisation, integrals of the form 
\begin{equation*}
\int\! d^dk\, \frac{1}{k^2} 
\end{equation*}
vanish. They do not depend on any scale and therefore must be zero, simply because applying a rescaling of the loop momenta $k\rightarrow \lambda k$ does not change the value of the integral, and we thus have 
\begin{equation}
\lr{1-\lambda^{d-2}}\int\! d^dk\, \frac{1}{k^2} =0\;.
\end{equation}
 The same holds for integrals of the form 
\begin{equation*}
 \int \!d^dk_1 \, \frac{1}{\lr{\lr{k_1-p_1}^2}^{n_1}\lr{k_1^2}^{n_2}}\;,
\end{equation*}
since by dimensional analysis (the result must have mass dimension $d-2\lr{n_1+n_2}$, and $p_1^2$ is the only Lorentz invariant quantity out of which a mass scale can be formed) they must be proportional to
\begin{equation}
 \lr{p_1^2}^{d/2-\lr{n_1+n_2}}\;, 
\end{equation}
an expression that vanishes for $\operatorname{Re} d>2\lr{n_1+n_2}$. Thus, via analytic continuation, they vanish in the entire complex $d$-plane. Additional scalar products in the numerator do not change this, so we may identify entire topologies (diagrams with a given set of propagators) as vanishing. The generalisation to three-loop integrals brings about a complication, namely that the three-loop integral might factorise into a 1-loop and a 2-loop integral (or three 1-loop integrals). In that case, each factor must be checked separately. 

\begin{figure}
 \begin{center}
\resizebox{8cm}{!}{\includegraphics{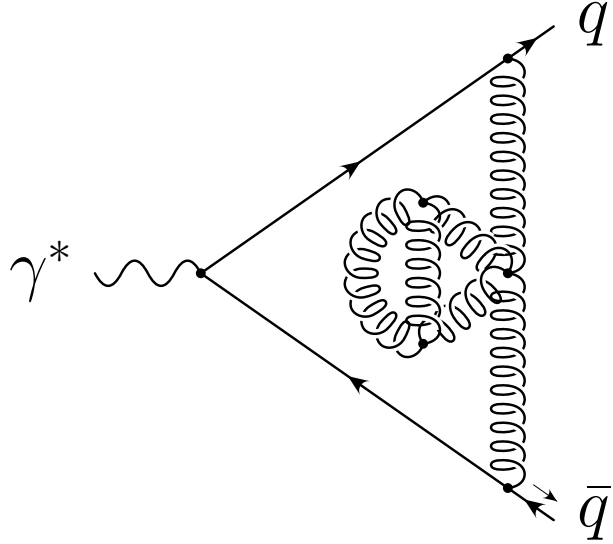}}
 \end{center}
\caption{A vanishing planar diagram. It contains the propagators $k_1-k_2$, $k_1-k_3$,  $k_2-k_3$ and e.g. $k_1-q$. Looking at the propagators, it is easily seen to be vanishing only after a shift $k_2\rightarrow k_2+k_1$, $k_3\rightarrow k_3+k_1$ is applied.}
\label{diag:vanishing}
\end{figure}

Building upon the results obtained in the previous section has one main advantage: It is easy to factorise the integral, because in our given basis, the number of propagators that contain more than one loop momentum is minimal. The integral thus factorises if out of the propagator momenta $k_1-k_2$, $k_1-k_3$, $k_2-k_3$ at most one is present in the given topology. Furthermore, since in our basis no propagator depends on $p_2$ alone, at least one propagator has to be of the form $k_i-q$ and thus introduce the only nonvanishing scale $q^2$ into the integral. If an integral only contains factors $k_i-q$ and $k_j-p_1$, a shift in the loop momenta $k_i\rightarrow k_i+p_1$ shows that the integral only depends on $p_2$ and thus vanishes. To summarise, the following algorithm detected most vanishing planar topologies:
\begin{enumerate}
 \item Bring the integral into a standard form, with propagator momenta out of the set of the auxiliary topology.
 \item Factorise the 3-loop integral into lower-loop integrals as far as possible.
 \item For a topology to be non-vanishing, require each factor to contain at least one propagator $1/\left(k_i-q\right)^2$ and one propagator $1/k_j^2$. 
\end{enumerate}
One type of topology that vanishes but was not detected as such using the above algorithm is shown in fig~\ref{diag:vanishing}. This particular diagram is harmless since \qgrafText has a switch to turn off snail or tadpole diagrams. On the other hand, this example shows that shifts in loop momenta play an important role in the above algorithm, since after a shift in the loop momenta, this topology
\begin{multline}
 \left\{k_1,k_1-p_1,k_1-q,k_1-k_2,k_1-k_3,k_2-k_3\right\}\\ \xrightarrow{
\begin{matrix}
k_2\rightarrow -k_2+k_1,\\
k_3\rightarrow -k_3+k_1
\end{matrix}
}  \left\{k_1,k_1-p_1,k_1-q,k_2,k_3,-\lr{k_2-k_3}\right\}
\end{multline}
will be identified as vanishing since the factor $\int d^dk_2 d^d k_3\frac{1}{\lr{k_3}^2\lr{k_2-k_3}^2}$ has no scale. 
This illustrates again that the propagators that involve two different loop momenta, i.e. $k_1-k_2$, $k_1-k_3$, $k_2-k_3$ should be eliminated if possible such that the factorisation of the integral in the above algorithm is maximal. Which shifts should be tried is not clear (we will discuss this point in the next section). Here, we state that even without any further shifts, all vanishing topologies except the one mentioned above were detected. As we will describe in sec.~\ref{sec:identifySubTopos} one may double-check using integration-by-parts identities. The two methods yield identical results for diagrams that do not contain snails.

\section{Exploiting Symmetries between Topologies}\label{sec:exploitingSymmetries}
If two Feynman integrals are equal, there is usually an easy way to understand this in terms of a symmetry of the integrand. These symmetries could be:
\begin{itemize}
 \item a permutation of the labels of the loop momenta $k_1\leftrightarrow k_2\leftrightarrow k_3$,
 \item the interchange  of $p_1\leftrightarrow p_2$,
 \item a shift of the loop momenta.
\end{itemize}
It should be clear that the first two operations actually transform entire topologies into other topologies (the explicit exponents of the propagators do not matter for this transformation). Shifting loop momenta is more delicate since it is not clear which shifts of loop momenta might be useful, and because not all shifts transform entire topologies in the same way. For this reason we discuss shift and permutation symmetries separately in the next two sections.

\subsection{Symmetries that are Permutations of Momenta}\label{sec:permutationSymmetries}
Permutations of loop momenta correspond to a mere relabelling of the integration variables, i.e. the loop momenta. Interchanging $p_1$ and $p_2$ cannot alter the value of (massless) vertex integrals since $q^2=2\:p_1\cdot p_2$ is the only scale left after the integration has been carried out. In the planar auxiliary topology, the permutation $p_1\leftrightarrow p_2$ can be performed by applying the shift $k_i\rightarrow -k_i +q$ for all $i=1,2,3$ simultaneously, followed by the permutation $p_1\leftrightarrow p_2$. It implies
\begin{equation}
\begin{split}
 p_1\leftrightarrow p_2:&\\
 &{\cal T}\bigl(n_1,n_2,n_3,n_4,n_5,n_6,n_7,n_8,n_9,n_{10},n_{11},n_{12}\bigr)= \\
& \hspace{3cm}{\cal T}\bigl(n_8,n_{10},n_{12},n_4,n_5,n_6,n_7,n_1,n_9,n_2,n_{11},n_3\bigr)\;.
\end{split}
\end{equation}
Similarly, we have 
\begin{equation}
\begin{split}
k_1\leftrightarrow k_2:&\\
 &{\cal T}\bigl(n_1,n_2,n_3,n_4,n_5,n_6,n_7,n_8,n_9,n_{10},n_{11},n_{12}\bigr)= \\
& \hspace{3cm}{\cal T}\bigl(n_2,n_1,n_3,n_4,n_6,n_5,n_9,n_{10},n_7,n_8,n_{11},n_{12}\bigr)\;,\\
k_1\leftrightarrow k_3:&\\
&{\cal T}\bigl(n_1,n_2,n_3,n_4,n_5,n_6,n_7,n_8,n_9,n_{10},n_{11},n_{12}\bigr)= \\
& \hspace{3cm}{\cal T}\bigl(n_3,n_2,n_1,n_6,n_5,n_4,n_{11},n_{12},n_9,n_{10},n_{7},n_{8}\bigr)\;,\\
k_2\leftrightarrow k_3:&\\
&{\cal T}\bigl(n_1,n_2,n_3,n_4,n_5,n_6,n_7,n_8,n_9,n_{10},n_{11},n_{12}\bigr)= \\
& \hspace{3cm}{\cal T}\bigl(n_1,n_3,n_2,n_5,n_4,n_6,n_7,n_8,n_{11},n_{12},n_9,n_{10}\bigr)\;.
\end{split}
\end{equation}
Clearly, such relations should be determined and topologies related by them should be identified, reducing the number of topologies that have to be evaluated.

If two topologies have such a relation, we need a prescription which of the two topologies should be mapped onto the other. An ordering of topologies is provided by their \emph{topology} $\mathrm{ID}$ \emph{number}, given as the number $\in\left\{n\in \mathbf{N},n\leq 2^{12}-1\right\}$ whose binary representation simply indicates which of the 12 propagators out of the set~\myref{eq:PropStandardSet} are present in this topology. All integrals in this topology then receive the same $\mathrm{ID}$ number. E.g. for the topology~\myref{eq:exampleStandardProps} (see fig.~\ref{diag:1790}) we write 
\begin{multline}
 {\cal T}(-1,-1,0,-1,-1,-1,-1,0,0,-1,-1,-1)\\\rightarrow\mathrm{ID}=1 + 2 + 8 + 16 + 32 + 64 + 512 + 1024 + 2048=3707=1 1 1 0 0 1 1 1 1 0 1 1b\;.
\end{multline}
We will add a corresponding subscript to the planar topology ${\cal T}={\cal T}_{\mathrm{ID}}$.
Implementing permutations of loop momenta is now straightforward. In this example, permuting $k_1\leftrightarrow k_3$ yields 
\begin{multline}
  {\cal T}_{3707}(-1,-1,0,-1,-1,-1,-1,0,0,-1,-1,-1)\\
\xrightarrow{
k_1\leftrightarrow k_3
} {\cal T}_{1790}(0,-1,-1,-1,-1,-1,-1,-1,0,-1,-1,0)
\end{multline}
with an $\mathrm{ID}=1790 \le 3707$. The permutation $p_1\leftrightarrow p_2$ corresponds to the shift $k_i\rightarrow -k_i +q,p_1\leftrightarrow p_2$ for all $i=1,2,3$ simultaneously, which in this case does not lower the $\mathrm{ID}$ number.

\subsection{Symmetries that are not Permutations}
There are also more complicated permutations that one might consider trying.
For example the topology with $\mathrm{ID}=187$ has the propagator momenta
\begin{equation*}
 k_1, \quad k_2,\quad  k_1-k_2,\quad  k_1-k_3,\quad  k_2-k_3, \quad k_1-q\;.
\end{equation*}
Applying the shift
\begin{equation}\label{eq:nontrivShift}
 \begin{split}
 k_1&\rightarrow k_1'\\ 
 k_2&\rightarrow -k_2'+k_1\\ 
 k_3&\rightarrow -k_3'+k_1\;,
\end{split}
\end{equation}
one ends up with topology 175 that has momenta 
\begin{equation*} 
k_1,\quad  k_2,\quad  k_3,\quad  k_1-k_2, \quad k_2-k_3, \quad k_1-q\;.
\end{equation*}
But under the shift~\myref{eq:nontrivShift} additional scalar products such as $k_2-p_1$, $k_3-p_1$ transform into a linear combination of inverse standard propagators, and thus integrals in the topology 187 are expressed as several terms in topology 175. Since this might enlarge the size of integration-by-parts identities (see sec.~\ref{sec:IBP_LI}), it is not clear whether they are of any use there. We did not use these identities at all, although they might be of moderate use to reduce the number of subtopologies that appear as integrals in the form factor.

We have not answered the general question of what group of (linear) loop momentum shifts would map subsets of the auxiliary topology 1 (the propagators of a topology) into subsets of the same set~\myref{eq:PropStandardSet}. If the sets of propagators are different, it seems that only in rare cases there exists a shift that is not a permutation of loop or external momenta. One of the few examples was given in eq.~\myref{eq:nontrivShift}, another in fig.~\ref{diag:vanishing}.
\begin{figure}
 \begin{center}
	\resizebox{8cm}{!}{\includegraphics{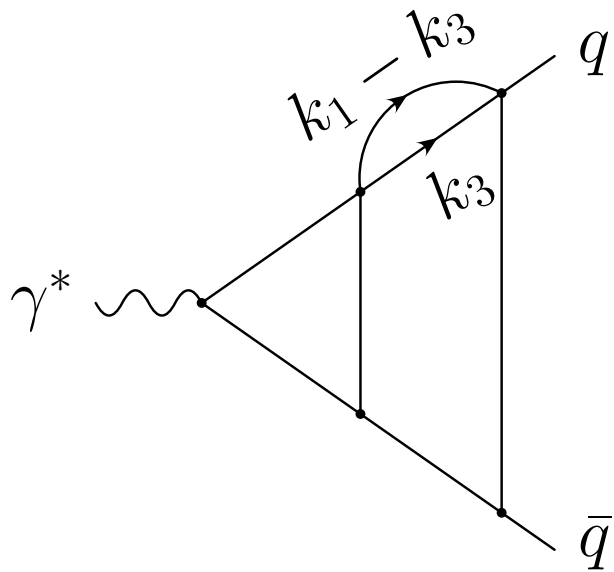}}
\caption{The topology with ID=734 has a bubble insertion and thus an additional symmetry: The two propagators with momenta $k_3$ and $k_1-k_3$ can be interchanged with a shift $k_3\rightarrow -k_3+k_1$.}\label{diag:734}
\end{center}
\end{figure}
The far more frequent case is that a shift permutes the propagators of a given topology. This happens for example whenever the integral contains a two-point function (also called a \emph{bubble insertion}) as in fig~\ref{diag:734}. There, the shift $k_3\rightarrow -k_3+k_1$ interchanges the two indicated propagators and all other propagators are left unchanged (but not the scalar products). Thus we obtain what is called a \emph{symmetry identity}
\begin{equation}\label{eq:symmetryIdentityExample}
{\cal T}_{734}-\left.{\cal T}_{734}\right|_{k_3\rightarrow-k_3+k_1}=0
\end{equation}
The significance of symmetry identities will be discussed in sec.~\ref{sec:estimateScaleOfComputation}. Basically, if a symmetry is present, it is possible to gain a lot of relations between integrals with little computational effort. Since these symmetry identities are found by inspection of the diagram anyhow, it was important that the symmetry considerations in sec.~\ref{sec:permutationSymmetries} together with the algorithm in sec.~\ref{sec:eliminatingVanishingTopologies} were enough to bring down the number of different topologies to about 200. Their symmetry properties were then implemented by hand. Drawing those diagrams, no relations except the ones mentioned here were apparent.

\pagestyle{empty}
\cleardoublepage
\pagestyle{headings}
\chapter{Methods for the Reduction}\label{chap:Reduction}
In the previous chapters we have generated the three-loop QCD quark form factor ${\cal F}\lr{q^2}$ as a linear combination of three-loop topologies ${\cal T}$. Obvious symmetries have been used and all vanishing topologies were removed. At that stage, our representation of the form factor consisted of 2951 integrals. Our next main task will be to use integration-by-parts identities in an automated way to relate these integrals to a small number of master integrals. The computation of some of these master integrals is very involved and not part of the present work.  By now most of the master integrals have been computed \cite{Gehrmann:2006wg,Heinrich:2007at}, such that the value of the form factor (it is just a number) will be in reach once the reduction of the form factor to master integrals is achieved. 

\section{Integration-By-Parts and Lorentz Invariance Identities}\label{sec:IBP_LI}
Integration-by-parts identities~\cite{Chetyrkin:1981qh} (IBPs) are relations that hold generally in dimensional regularisation. They express the fact that dimensionally regulated integrals are Poincar\'e covariant. Consider a general integral $I$ depending on loop momenta $k_1\ldots k_n$: For an infinitesimal translation $k_i\rightarrow k_i+ v_j$, we have
\begin{equation}
 {I}\lr{k_1\ldots,k_i,\ldots,k_n}={ I}\lr{k_1\ldots,k_i+ v_j,\ldots,k_n}\;.
\end{equation}
Expanding in $ v_j$, we get
\begin{equation}
  v_j^\mu\:\frac{\partial}{\partial k_i^\mu}\:{ I}\lr{k_1\ldots,k_i,\ldots,k_n}=0\;.
\end{equation}
Integrating by parts the IBP identities in our case read 
\begin{equation}\label{eq:IBPs}
 \int \frac{d^d k_1}{\left(2\pi\right)^d}\frac{d^d k_2}{\left(2\pi\right)^d}\frac{d^d k_3}{\left(2\pi\right)^d} \:\frac{\partial}{\partial k_i^\mu}\:v_j^\mu\: {\cal I}=0
\end{equation}
with $v_j\in\left\{k_1,k_2,k_3,p_1,q\right\}$, and where ${\cal I}$ denotes the integrand of a three-loop (massless vertex) integral ${I}$. 

Acting with the derivative in \myref{eq:IBPs} on $v^\mu{\cal I}$ generates a number of terms, with one exponent of a propagator incremented by one and with $v^\mu$ contracted with a four-momentum. The scalar products can be rewritten using~\myref{eq:scalarProductToinvProp}, yielding a linear combination of functions ${\cal T}$. In this way IBP identities relate ${\cal T}$'s with different (integer) arguments. Thus IBP identities form a system of equations that are linear in the subtopologies ${\cal T}$. 

We call ${\cal I}$ a \emph{seed integral} since it generates the IBP identities. Below, we will often call the integrals that appear in the IBP identities of a seed integral its \emph{subtopologies}. Note that $\frac{d}{d k^\mu} k^\mu=d$ in dimensional regularisation, so that the coefficients in the system of IBP identities contain the symbol $d=4-2\epsilon$.

It has soon been realized that the number of IBP identities grows faster than the number of unknowns. This apparently overconstrains the system. In fact the system is underconstrained, so that its solution relates the subtopologies it contains to so called \emph{master integrals}. These are integrals with a minimal sum of exponents of propagators $r$ and sum of exponents of scalar products $s$ that cannot be reduced to integrals with a smaller number of propagators $t$ using IBP identities. In our case, we expect all master integrals to have $r=t$, $s=0$. 
In section~\ref{sec:LaportaAlgorithm} we describe an algorithm~\cite{Laporta:2001dd} to use IBP identities in an automated way to express more complicated integrals as a linear combination of simpler ones, and eventually in terms of master integrals.

Just as IBP identities express the translation invariance of dimensionally regulated integrals, we may also exploit Lorentz invariance in the same manner~\cite{Gehrmann:1999as}: An infinitesimal Lorentz transformation on an external momentum reads $p_i^\mu\rightarrow p_i^\mu+\delta p_i^\mu=p_i^\mu+\delta\epsilon_\nu^\mu\, p_i^\nu$ with $\delta\epsilon_\mu^\nu=-\delta\epsilon_\nu^\mu$. Again, expanding the identity 
\begin{equation}
 { I}\lr{p_1,\ldots,p_n}={I}\lr{p_1+\delta p_1,\ldots,p_n+\delta p_n}
\end{equation}
we obtain
\begin{equation}
 \delta\epsilon_\nu^\mu\left(p_1^\nu\frac{\partial}{\partial p_1^\mu}+\ldots+ p_n^\nu\frac{\partial}{\partial p_n^\mu}\right){\cal I}\lr{p_1,\ldots,p_n}=0
\end{equation}
or, using the antisymmetry of $\delta\epsilon^\mu_\nu$,
\begin{equation}
\left(p_1^\nu\frac{\partial}{\partial p_1^\mu}-p_1^\mu\frac{\partial}{\partial p_1^\nu}+\ldots+ p_n^\nu\frac{\partial}{\partial p_n^\mu}-p_n^\mu\frac{\partial}{\partial p_n^\nu}\right){\cal I}\lr{p_1,\ldots,p_n}=0\;.
\end{equation}
For three-point functions, we have $n=2$, i.e.
\begin{equation}
 \lr{p_1^\nu\frac{\partial}{\partial p_1^\mu}-p_1^\mu\frac{\partial}{\partial p_1^\nu}+p_2^\nu\frac{\partial}{\partial p_2^\mu}-p_2^\mu\frac{\partial}{\partial p_2^\nu}}{\cal I}\lr{p_1, p_2}=0\;.
\end{equation}
By contracting these equations with antisymmetric combinations of $p_1,\ldots,p_n$ we get the \emph{Lorentz invariance} (LI) identities. For $n=2$, the only such combination is $p_{1,\mu}-p_{2,\mu}$ and we get
\begin{equation}
 \lr{p_1^\mu\frac{\partial}{\partial p^\mu_{1}}-p_2^\mu\frac{\partial}{\partial p^\mu_{2}}}{\cal I}\lr{p_1, p_2}=0\;,
\end{equation}
or with $q=p_1+p_2$,
\begin{equation}
 \lr{p_1^\mu\frac{\partial}{\partial p_1^\mu}+2 p_1^\mu\frac{\partial}{\partial q^\mu}-q^\mu\frac{\partial}{\partial q^\mu}}{\cal I}\lr{p_1,q}=0\;.
\end{equation}

LI identities have recently been proven to be linearly dependent on IBP identities~\cite{Lee:2008tj}. While it is possible to obtain these relations from IBP identities, this is computationally expensive because a larger system of IBP identities must be solved. For that reason we used LI identities at all stages of our reduction.

The question of which integrals have to be used to generate IBP and LI identities will be discussed in sec.~\ref{sec:estimateScaleOfComputation}.

\section{The Laporta-Algorithm}\label{sec:LaportaAlgorithm}
In~\cite{Laporta:2001dd} an algorithm (by now simply called the \emph{Laporta algorithm}) was proposed that makes use of IBP (and LI) identities to reduce integrals to master integrals in an automated way. The algorithm solves a system of IBP identities using a specific ordering of the integrals according to their complexity. To quantify this, we now define the three numbers mentioned in the previous section that characterise an integral ${\cal T}\lr{n_1,\ldots,n_{12}}$ ($n_i\in \mathbf{Z}$):
\begin{itemize}
 \item[t:] the number of propagators: $t=\sum_{i=1}^{12} \theta \lr{-n_i-1}$
 \item[r:] the sum of the exponents of the propagators: $r=\sum_{i=1}^{12} \left|n_i\right| \theta \lr{-n_i}$
 \item[s:] the sum of the exponents of the scalar products: $s=\sum_{i=1}^{12} \left|n_i\right| \theta \lr{n_i}$
\end{itemize}
(where $\theta\lr{x}=1$ for $x\ge0$ and $0$ otherwise; remember that propagators have $n_i<0$). The Laporta algorithm instructs us to order integrals first according to their $t$, then their $r$, and finally their $s$. Integrals with the same $t$, $r$, $s$ will be further ordered according to the number of terms in their coefficient, and then according to the total number of terms in their equation. 

The system of IBPs may then be solved using Gauss' algorithm (further detail on the algorithm are given in sec.~\ref{sec:Solve}). All the identities extracted from the system of IBP identities should map more complex integrals (according to the ordering just given) to simpler ones. If enough identities are solved, the solutions should express all desired integrals as linear combinations of master integrals only.

\section{Identifying Subtopologies}\label{sec:identifySubTopos}
It is a good idea to look ahead and try to estimate the scale of the calculation we are aiming at. The form factor has already been introduced and expressed as a linear combination of three-loop integrals. We will next try to estimate the size of the system of IBP identities that has to be solved in order to reduce the integrals in the form factor to master integrals. The master integrals themselves will also be identified in the process. 

To start we should identify the topologies that might arise when we reduce our form factor. Here, we do not yet care about which integrals appear in the reduction. We will have to perform a reduction for the integrals in these topologies by solving IBP identities. Of course, equivalent topologies (e.g. those that differ by a permutation of loop momenta) only have to be reduced once since we know the equivalence classes of topologies (see sec.~\ref{sec:exploitingSymmetries} and the appendix~\ref{app:TopoPermutations}). 

The $t-1$ subtopologies for a topology with $t$ propagators arise in the following way: In an IBP identity a scalar product is formed by the contraction of $v^\mu$ with a vector in the numerator of $\frac{\partial}{\partial k^\mu} {\cal I}$. When it is expressed as a linear combination of inverse propagators, an inverse propagator may cancel one of the existing propagators. Thus subtopologies arise by removing one propagator from the given topology\footnote{In symmetry identities, up to $s$ propagators may be cancelled.}. Which of the propagators are cancelled can only be determined by writing down the IBP identities. However, we need to write down the IBP identities for the topology with $r=t$ and $s=0$ only: Any possible cancellation of a scalar product against a propagator will show up here, and therefore all $t-1$ subtopology ID numbers can be determined in this way.

Two interesting properties of the subtopologies can easily be derived from the IBP identities of the $r=t$, $s=0$ seed integral of a topology: Firstly if the topology is vanishing, this will almost always be apparent after the system of the 15 IBP identities from the $r=t$, $s=0$ seed integral is solved. Secondly, if the topology is reducible, the reduction equation for the $r=t$, $s=0$ integral can often (but not always) be extracted from the same solution of 15 IBP identities. Thus we arrive at a natural procedure to get a first insight into the subtopologies that have to be reduced:
\begin{enumerate}
 \item Create a list $L_{ID}$ of the ID numbers of all topologies present in the form factor.
 \item Set $t_0$ equal to the maximum number of propagators present in the topologies in $L_{ID}$; in our case, $t_0=9$.
 \item For every topology in $L_{ID}$ with $t=t_0$, generate the IBP identities for its $r=t$, $s=0$ seed integral and 
	\begin{itemize}
	 \item store all subtopology ID numbers found in the identities and add them to $L_{ID}$;
	 \item solve the system, and determine from the solution whether the topology is vanishing or else whether it is reducible or (possibly) irreducible.
 	\end{itemize}
 \item Set $t_0=t_0-1$ and repeat the previous step.
\end{enumerate}
This creates a list of all (sub-)subtopologies that will occur during the reduction. When generating the IBP identities, the information on the equivalence of certain topologies derived in chapter~\ref{chap:Formfactor} should be used. 

One subtlety should still be pointed out: Several topologies in fact are reducible although they are not found to be so using the above algorithm. One example is the topology with ID=734 (fig.~\ref{diag:734}). Their reduction equations are found only after the inclusion of additional IBP identities, e.g. up to $r=t+1$, $s=1$. Since these systems are already quite big, it makes sense to check reducibility first using the $r=t$, $s=0$ seed integral only, and checking the potentially irreducible graphs in a second run using e.g. $t\leq r\leq t+1$, $s\leq 1$ seed integrals. 

\begin{figure}
	\resizebox{15cm}{!}{\includegraphics[width=11cm]{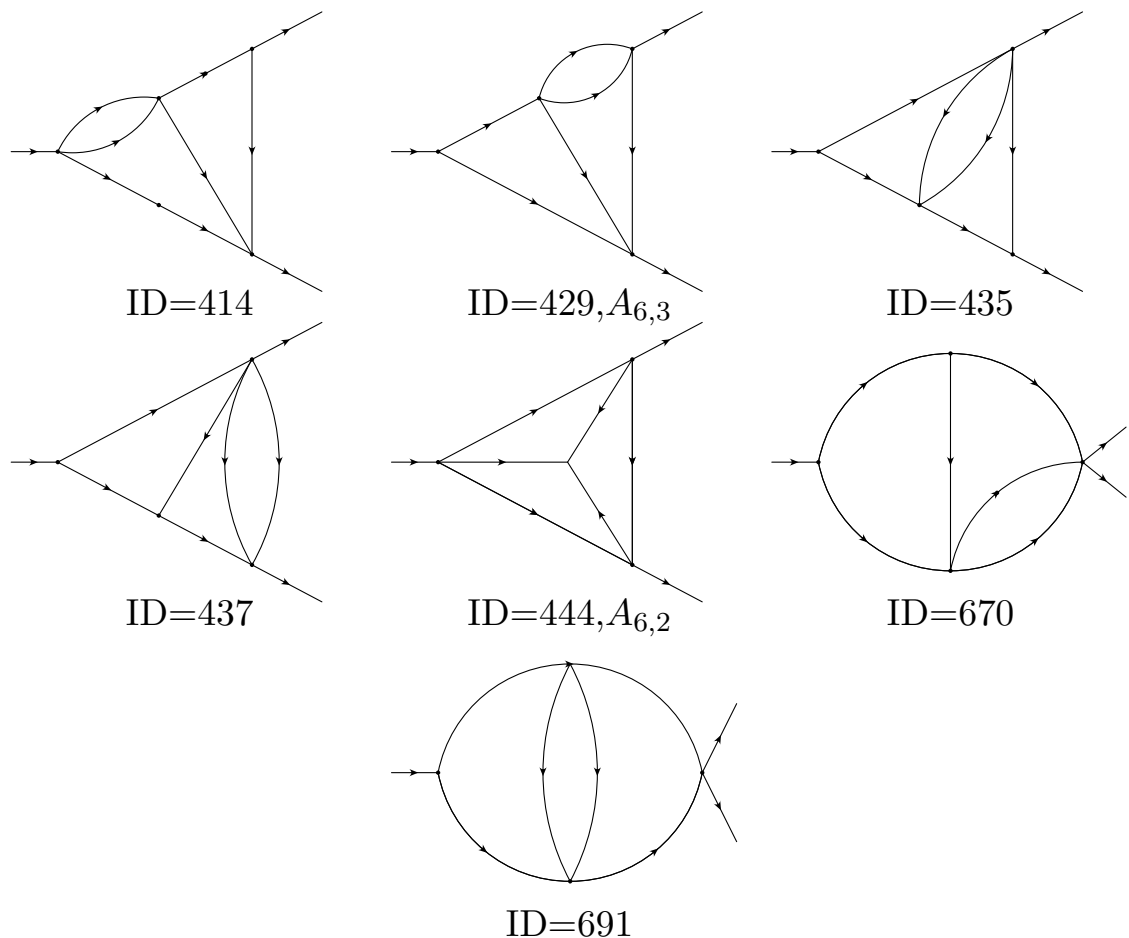}}
\caption{The ``hard'' 6-propagator topologies appearing in the reduction, i.e. those that are not integrable loop-by-loop or for which no symbolic reduction equations were found. The topologies are denoted using their binary ID number as well as (if present) the name used in~\cite{Gehrmann:2006wg}. They split into two groups: There are the irreducible topologies (the middle column, consisting of the topologies 429, 444 and 691) that are not integrable loop-by-loop. The second group (first and third column) consist of reducible topologies for which no symbolic reduction equations were found. All these graphs have to be reduced using explicit numeric IBP-identities. Note that the first four and also the last two topologies are bubble-insertions into the same 2-loop topology. }\label{fig:t6IrreducibleTopos}
\end{figure}

\begin{figure}
	\resizebox{15cm}{!}{\includegraphics[width=11cm]{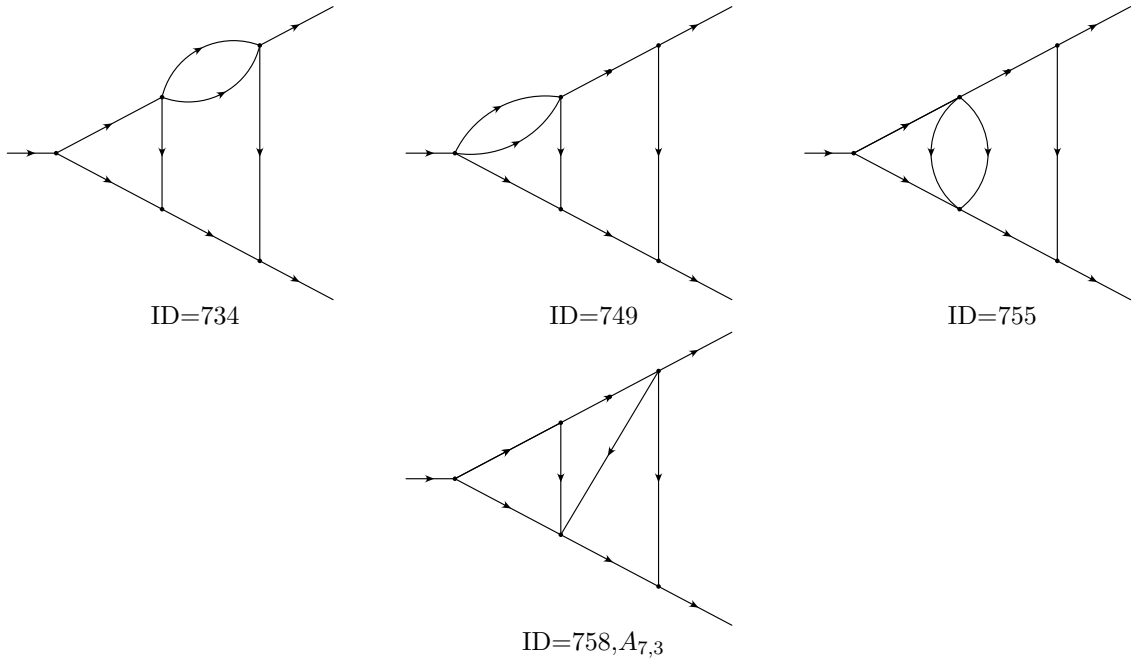}}
  \caption{The ``hard`` 7-propagator topologies appearing in the reduction. The topologies are denoted using their binary ID number as well as (if present) the name used in~\cite{Gehrmann:2006wg}. They should be reduced using IBP-identities where the exponents of propagators are written out explicitly. Note that the topologies in the first row are bubble-insertions into the same 2-loop topology. They are all reducible.}\label{fig:t7IrreducibleTopos}
\end{figure}

Table~\ref{tab:NrOfTopos} gives the number of nonvanishing subtopologies and irreducible subtopologies found in the application of the above algorithm and after double-checking the irreducibility of the topologies. Obvious symmetries (such as permutations of the loop momenta, see sec.~\ref{sec:exploitingSymmetries}) have already been taken into account. Out of the topologies, some can be integrated loop-by-loop or reduced using symbolic IBP identities and do not pose a problem (see chap.~\ref{chap:5PropIntegrals} and sec.~\ref{sec:symbolicIBPs}). All others (called ``hard'' here) are shown in fig.~\ref{fig:t6IrreducibleTopos} for $t=6$ and in fig.~\ref{fig:t7IrreducibleTopos} for $t=7$. They have to be reduced using explicit numeric IBP identities\footnote{Note though that the first four topologies in fig.~\ref{fig:t6IrreducibleTopos} ($t=6$) and the first three topologies in fig.~\ref{fig:t7IrreducibleTopos} ($t=7$) contain a two-point function inserted into the same two-loop topology. One option that we have not yet fully explored is to reduce the two-loop topology with one exponent of a propagator kept generic. This introduces the symbolic exponent as a second symbol (apart from the space-time dimension $d$), but reduces the number of equations per seed integral from 16 to 7, and also the number of seed integrals per $r$-$s$-block is reduced considerably.}. The irreducible master integrals for $t=4,5$ are given later in fig.~\ref{fig:t5irredtopos}.

\begin{table}
\begin{center}
\begin{tabular}{|c|c|c|c|}
\hline
t & \# of IDs & \# of irred. IDs & \# of ``hard'' IDs\\ 
\hline
9 &  3 & 1 &1\\ 
8 & 25 & 0&0\\
7 & 46 & 1&4\\
\hline
6 & 33 & 3&7\\
5 & 10 & 4&0\\
4 &  1 & 1&0\\
\hline
\end{tabular}
\end{center}
\caption{The number of topologies and irreducible topologies found as subtopologies in the three-loop quark form factor. The third column shows the number of topologies that have to be reduced numerically. Symmetries are already taken into account.}\label{tab:NrOfTopos}
\end{table}

The result that there is only one topology left at $t=4$ is expected: Generally, for any diagram the relation 
\begin{equation}
 L-1=t-V
\end{equation}
holds, where $L$ is the number of loops and $V$ the number of vertices. Diagrams with only one vertex are snail diagrams and vanish (in massless, dimensionally regulated computations). Thus $t\geq4$ follows. Moreover, for $t=5, L=3$ it follows that $V=3$. But $V=3$ means that all integrals contain at most one three-point function. They can be integrated loop-by-loop, as will be discussed in chap.~\ref{chap:5PropIntegrals}. For $t\geq 6$ the integrals become complicated, and we have to resort to the Laporta algorithm to reduce integrals to master integrals. 


\section{A First Estimate of the Scale of the Computation}\label{sec:estimateScaleOfComputation}
We have now gained an overview of which topologies have to be reduced using the Laporta algorithm. We next ask which $t$-$r$-$s$-classes of integrals might appear in our reduction.

Given a seed integral ${\cal I}_{t,r,s}$ for the massless quark form factor, the derivative acting on a denominator will increment $r$ by one. At the same time, the vector $v^\mu$ from the IBP identity will be contracted with a vector that was obtained from the differentiation via the chain rule. This  yields an additional scalar product in the numerator. The results are integrals of the type\footnote{In case the scalar product is $p_1.p_2$ we do not count this as a scalar product that increments $s$ since it factors out of the integral and does not increase the complexity of the integral.} ${\cal I}_{t,r+1,s+1}$ or ${\cal I}_{t,r+1,s}$
. 
When expressed as a linear combination of inverse propagators, the scalar product may cancel with one propagator, yielding either ${\cal I}_{t,r,s}$ or ${\cal I}_{t-1,r,s}$ integrals. Finally, integrals of the type ${\cal I}_{t,r,s-1}$ may be generated when the derivative acts on a scalar product in the numerator. It may remove one loop momentum from the scalar product and substitute it by an external momentum. To summarise, 
\begin{equation}\label{eq:subtopotrs}
t,r,s\xrightarrow{\mathrm{generate~IBPs}}\left\{
\begin{array}{l}
{t,r+1,s+1}\\
{t,r+1,s}\\
{t,r,s}\\
{t,r,s-1}\\
{t-1,r,s}\\
\end{array}\right.\;.
\end{equation}
The topologies ${\cal I}_{t-1,r,s}$ are the simplest in this list and will appear in the final reduction equations. The important observation here is that if a seed integral ${\cal I}_{t,r,s}$ is used to generate IBP identities, this introduces subtopologies ($t-1$ integrals) with \emph{the same $r$ and $s$}.

Concerning possible subtopologies, symmetry identities are quite different from IBP identities. They mostly relate integrals that have the same $t$, $r$ and $s$ as the seed integral. This is a significant advantage: To obtain relations for integrals with a given $r$ and $s$, IBP identities are generated using mostly the seed integrals with $r-1$, $s-1$, whereas the symmetry identities are generated using the seed integrals with the same $r$ and $s$. Clearly, there are many more seed integrals of the type $r$, $s$ than of the type $r-1$, $s-1$ or $r-1$, $s$. Symmetry identities thus provide a significant amount of information, even though there is mostly a single symmetry identity present, compared to the 15 IBP identities. Indeed we will see shortly that there is roughly an order of magnitude more $r+1$, $s+1$ than $r$, $s$ seed integrals, such that if present, a symmetry identity provides a number of relations between $r$-$s$ integrals comparable to the one of IBP identities.

 After the generation and basic manipulation of the three-loop quark form factor, we find by inspection that the most difficult integrals to be computed have 
\begin{equation}\label{eq:highesttrs}
\begin{array}{l}
 t=9\\
 r=9\\
 s=4\;.
\end{array}\qquad
\end{equation}
The corresponding topologies finally reduce to $t\geq6$ topologies that are irreducible, or to $t=5$ topologies that can be integrated loop-by-loop. 

Next, we compute the number of seed integrals of a given topology as a function of $t$, $r$ and $s$. This is simple combinatorics, with the result~\cite{Gehrmann:1999as}
\begin{equation}
 N\left({\cal I}_{t,r,s}\right)=\left(\!\!
\begin{array}{c}
 r-1\\
 r-t
\end{array}\!\!\right)\left(\!\!
\begin{array}{c}
 12-t+s\\
 s
\end{array}\!\!\right)\;.
\end{equation}
Since we have 15 IBP identities and one LI identity, the corresponding number of equations for seed integrals with a given $t$, $r$ and $s$ is just $N_{t,r,s}=16 N\left({\cal I}_{t,r,s}\right)$. The explicit numbers relevant for us are given in table~\ref{tab:NrOfEqsXX}. Since we will evaluate the integrals explicitly for $t=5$, table~\ref{tab:NrOfSeedEqsT5} gives the number of seed integrals in a given $r$-$s$-block, $N_{5,r,s}$.

\begin{table}
\begin{center}
\begin{tabular}{|c|c|c|c|c|c|}
\hline
\multicolumn{6}{|c|}{$t=9$} \\ 
\hline
&s=0&s=1&s=2&s=3&s=4\\
\hline
r=9&16 & 64 & 160 &320&560\\
\hline
\multicolumn{6}{}{}\\
\hline
\multicolumn{6}{|c|}{$t=8$} \\ 
\hline
&s=0&s=1&s=2&s=3&s=4\\
\hline
r=8&16 & 80 & 240 & 560 & 1120 \\
r=9&128 & 640 & 1920 & 4480 & 8960\\
\hline
\multicolumn{6}{}{}\\
\hline
\multicolumn{6}{|c|}{$t=7$} \\ 
\hline
&s=0&s=1&s=2&s=3&s=4\\
\hline
r=7&16 & 96 & 336 & 896 & 2016\\
r=8&112 & 672 & 2352 & 6272 & 14112\\
r=9&448 & 2688 & 9408 & 25088 & 56448\\
\hline
\multicolumn{6}{}{}\\
\hline
\multicolumn{6}{|c|}{$t=6$} \\ 
\hline
&s=0&s=1&s=2&s=3&s=4\\
\hline
r=6&16 & 112 & 448 & 1344 & 3360 \\
r=7&96 & 672 & 2688 & 8064 & 20160\\
r=8&336& 2352 & 9408 & 28224 & 70560\\
r=9&896 & 6272 & 25088 & 75264 & 188160\\
\hline
\end{tabular}
\end{center}
\caption{The number of equations generated from a given $r$-$s$-block of seed integrals for a system with 12 independent scalar products. It grows quickly with $r$ and $s$ and with decreasing $t$.}\label{tab:NrOfEqsXX}
\end{table}

\begin{table}
\begin{center}
 \begin{tabular}{|c|c|c|c|c|c|}
 \hline
\multicolumn{6}{|c|}{$t=5$} \\ 
\hline
&s=0&s=1&s=2&s=3&s=4\\
\hline
r=5&1 & 8 & 36 & 120 & 330 \\
r=6&5 & 40 & 180 & 600 & 1650\\
r=7&15 & 120 & 540 & 1800 & 4950 \\
r=8&35 & 280 & 1260 & 4200 & 11550 \\
r=9&70 & 560 & 2520 & 8400 & 23100\\
\hline
\end{tabular}
\caption{The number of seed integrals in a given $r$-$s$-block for a system with 12 independent scalar products.}\label{tab:NrOfSeedEqsT5}
\end{center}
\end{table}

The numbers are not very pleasing. The code \emph{Solve} that will be discussed in sec.~\ref{sec:Solve} will turn out to be able to solve systems with up to $\sim3000$ equations. In sections~\ref{sec:SolveImprovements} and \ref{sec:SolveExtensions} we will improve this code in various ways. This will allow us to solve systems with close to $10000$ equations. But clearly, we cannot afford to reduce the topologies with $t$, $r$, $s$ as in eq.~\myref{eq:highesttrs} using seed integrals up to $r=9$, $s=4$ for $t=7,8,9$. We would have to expect subtopologies with $t=6$, $r=9$, $s=4$. To reduce those topologies, we expect it to be necessary to solve a system of equations with an order of magnitude more equations than could be handled using \emph{Solve}: Tab.~\ref{tab:NrOfEqsXX} predicts 188160 equations to be solved in the $r=9$, $s=4$ block alone. The fact that thesis does not present the full reduction of the three-loop quark form factor shows that ultimately we have not been able to fully address this challenging problem. We will discuss various aspects of it including some attempts to solve such large systems in the remainder of this thesis.

On the other hand, for 9- (and 8-)propagator topologies the number of equations is quite moderate. We may go ahead and reduce these systems without much difficulty, given that we have a CPU cluster at our disposal.

\subsection{The Hierarchy of Integration-by-Parts Identities}\label{sec:hierarchy}
\begin{figure}
  \begin{center}
	\resizebox{10cm}{!}{\includegraphics[width=11cm]{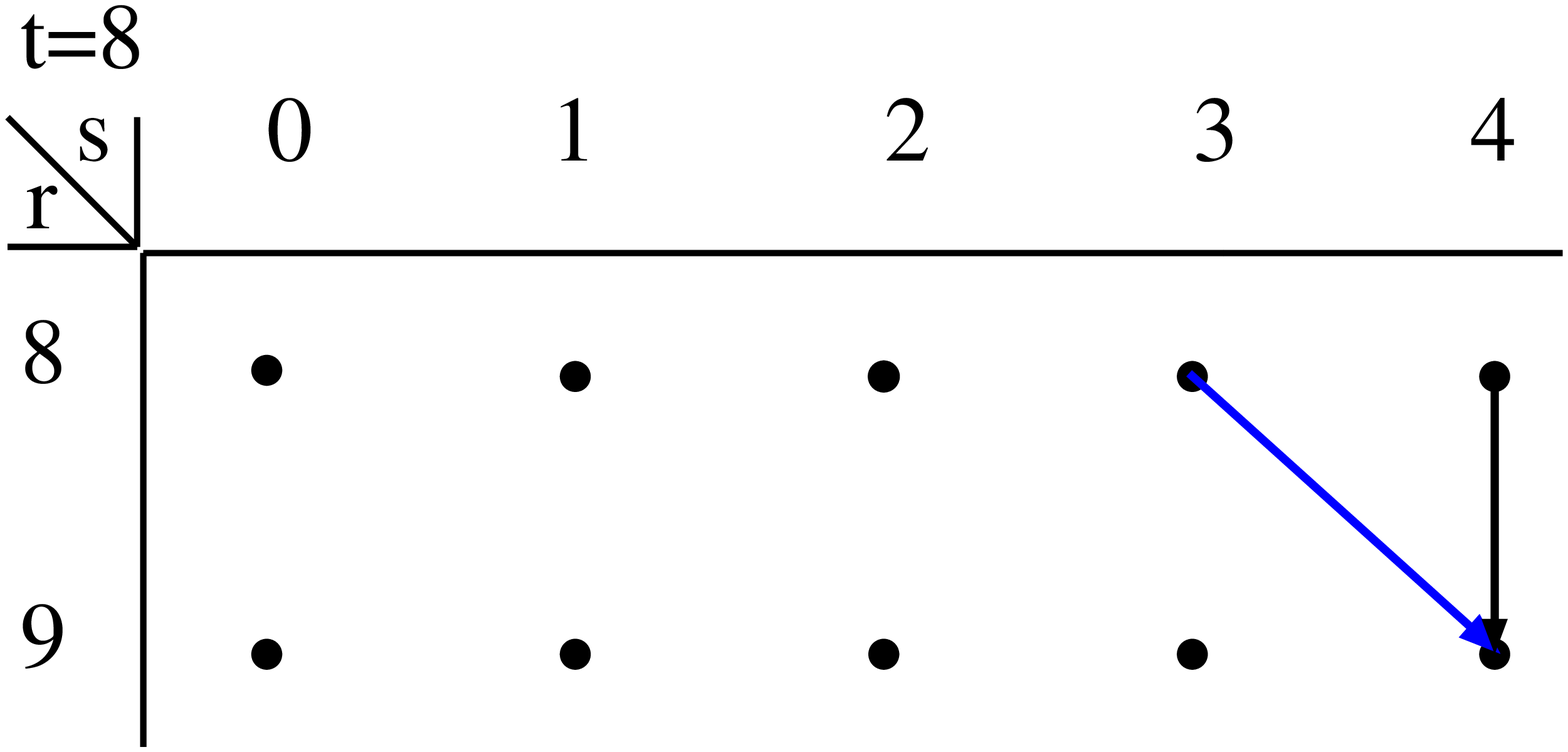}}
  \end{center}
\caption{Some integration-by-parts identities are more useful than others: To reduce integrals with $t=8$, $r=9$, $s=4$, we may use seed integrals with e.g. $r=8$, $s=3$ (dashed blue arrow) or $r=8$, $s=4$ (solid black arrow). But $r=8$, $s=3$ seed integrals lead to IBP identities without any $t=7$, $r=8$, $s=4$ subtopologies, whereas IBP identities from $r=8$, $s=4$ seed integrals do contain these more complicated subtopologies.}\label{fig:rsplot}
\end{figure}

The most immediate objection to the above argument that the system of IBP identities will grow unmanageably large is that to reduce an integral ${\cal I}_{t,r,s}$, there might be no need to generate and solve IBP identities from seed integrals with the same values for $t$, $r$, $s$. As is apparent from~\myref{eq:subtopotrs}, a seed integral ${\cal I}_{t,r-1,s-1}$ or ${\cal I}_{t,r-1,s}$ also generates identities involving ${\cal I}_{t,r,s}$, and their $t-1$ subtopologies are of the form ${\cal I}_{t-1,r-1,s-1}$ and ${\cal I}_{t-1,r-1,s}$, respectively. This is illustrated again in fig~\ref{fig:rsplot}. For the reduction of the integrals in \myref{eq:highesttrs}, $t=9$, $r=9$, $s=4$ seed integrals will be necessary. This will lead to subtopologies with $t=8$, $r=9$, $s=4$. For their reduction, one may use the following seed integrals:\\
\begin{center}
\begin{tabular}{|l|l|l|}
\hline
seed integral & r=9,s=4 generated in&$t=7$ subtopologies\\
\hline
 $t=9$, $r=9$, $s=4$& $t-1$, $r$, $s$&none\\
 $t=8$, $r=8$, $s=3$& $t$, $r+1$, $s+1$&$r=8$, $s=3$\\
 $t=8$, $r=8$, $s=4$& $t$, $r+1$, $s$&$r=8$, $s=4$\\
 $t=8$, $r=9$, $s=4$& $t$, $r$, $s$&$r=9$, $s=4$\\
\hline
\end{tabular}
\end{center}
The first line corresponds to so-called \emph{left-over} equations. These solutions of the system of IBP identities  contain only integrals of type $t-1$, $r$, $s$. They might relate different topologies and reduce the number of $t=8$, $r=9$, $s=4$ topologies needed to be reduced by other identities. The second line corresponds the dashed blue arrow in fig.~\ref{fig:rsplot}. These identities are more useful than the ones in the third line (corresponding to the solid black arrow in fig.~\ref{fig:rsplot}) since they do not contain $r=8$, $s=4$ 7-propagator subtopologies. Finally, identities using $t=8$, $r=9$, $s=4$ seed integrals should be avoided in any case since they lead to $t=7$, $r=9$, $s=4$ subtopologies.

To summarise, in our context it is crucial to note that some IBP identities are more useful than others, since they contain less complicated subtopologies. If several identities reduce the same integral, one must use the identity that was generated from seed integrals with the lowest possible $r$ and $s$. In practice, this means that different collections of the same set of IBP identities must be solved and substituted in. For example one should use the solution for the system with $t=8$, $r=8$, $s\leq 3$ before using the solution for the system with $t=8$, $r=8$, $s\leq 4$. A routine dedicated to finding in such an incremental way the solution of a system of IBP identities will be presented in sec.~\ref{sec:perlmerge}.

\section{Using Symbolic Integration-by-Parts Identities}\label{sec:symbolicIBPs}
As a result of the above investigations, we find that all $r=9$ and all $s=4$ integrals could be reduced to $t=8$, $r=8$, $s=4$ or $t=8$, $r=9$, $s=3$ integrals. However, using IBP identities from seed integrals with $t=8$, $r=8$ and $s\leq 4$ in the way described in the previous section, it was not possible to reduce all 8-propagator integrals. 

As a next step, one could include $t=8$, $r=9$ seed integrals into the system of IBPs, producing $t=7$, $r=9$ subtopologies. This consequence cannot be avoided even after the inclusion of symmetry identities. However, as we will show now, the fact that all 8-propagator topologies are reducible (see tab.~\ref{tab:NrOfTopos}) leads to an elegant way of reducing them using just 15 symbolic IBP identities and one LI identity (per topology). The subtopologies will even tend to be simpler when using symbolic IBP identities as compared to using ``numeric'' IBP identities with higher $r$ and $s$. 

By \emph{symbolic} IBP identities, we mean IBP identities that are written down for general exponents $n_1,\ldots,n_{12}$. Symbolic integration-by-parts identities have been used already in~\cite{Chetyrkin:1981qh}. It is perhaps easiest to illustrate their use in the case of the one-loop vertex-correction, fig.~\ref{fig:treeandoneloopdiags}. The three-propagator integrals that appear in its evaluation are of the form
\begin{equation}\label{eq:1loopthreepoint}
 {\cal I}_{\mathrm{1-loop}}\lr{n_1,n_2,n_3}=\int d^d k \frac{1}{\lr{k^2}^{n_1}\lr{\lr{k-p_1}^2}^{n_2}\lr{\lr{k-q}^2}^{n_3}}\;.
\end{equation}
Again with $q=p_1+p_2$. When writing down the IBP identities for this case, 
\begin{eqnarray}
 \frac{\partial}{\partial k^\mu} k^\mu {\cal I}_{\mathrm{1-loop}}\lr{n_1,n_2,n_3}&=&0\;,\label{eq:symbIBPs1}\\
 \frac{\partial}{\partial k^\mu} p_1^\mu {\cal I}_{\mathrm{1-loop}}\lr{n_1,n_2,n_3}&=&0\;,\label{eq:symbIBPs2}\\
 \frac{\partial}{\partial k^\mu} q^\mu {\cal I}_{\mathrm{1-loop}}\lr{n_1,n_2,n_3}&=&0\;,\label{eq:symbIBPs3}
\end{eqnarray}
it is useful to introduce the following operators:
\begin{eqnarray}
 \up{1}{\cal I}_{\mathrm{1-loop}}\lr{n_1,n_2,n_3}={\cal I}_{\mathrm{1-loop}}\lr{n_1+1,n_2,n_3}\;,\\
 \up{2}{\cal I}_{\mathrm{1-loop}}\lr{n_1,n_2,n_3}={\cal I}_{\mathrm{1-loop}}\lr{n_1,n_2+1,n_3}\;,\\
 \up{3}{\cal I}_{\mathrm{1-loop}}\lr{n_1,n_2,n_3}={\cal I}_{\mathrm{1-loop}}\lr{n_1,n_2,n_3+1}\;,
\end{eqnarray}
and similarly for $\mathbf{1}^-$, $\mathbf{2}^-$, $\mathbf{3}^-$. Using this notation the three IBPs \myref{eq:symbIBPs1}, \myref{eq:symbIBPs2}, and \myref{eq:symbIBPs3} read
\begin{eqnarray}
\begin{aligned}
\bigl(n_2\down{1}\up{2}+n_3\down{1}\up{3}&- n_3 q^2\up{3}\\
&-d+2n_1+n_2+n_3\bigr){\cal I}_{\mathrm{1-loop}}\lr{n_1,n_2,n_3}\!\!&=&\:\:0\;,
\end{aligned}\label{eq:1loopsymbIBPs1}\\
\begin{aligned}
 \bigl(-n_2\down{1}\up{2}-n_3 \down{1}\up{3}&+n_3 q^2 \up{3}\\
+n_3\down{2}\up{3}&+n_1 \up{1}\down{2}-n_1+n_2\bigr) {\cal I}_{\mathrm{1-loop}}\lr{n_1,n_2,n_3}\!\!&=&\:\:0\;,
\end{aligned}\label{eq:1loopsymbIBPs2}\\
\begin{aligned}
\bigl(n_1\up{1}\down{2}-&n_1\up{1}\down{3}+n_1 q^2 \up{1}\\ 
-&n_2 \up{2}\down{3}+n_3\down{2}\up{3}+n_2-n_3\bigr) {\cal I}_{\mathrm{1-loop}}\lr{n_1,n_2,n_3}\!\!&=&\:\:0\;.
\end{aligned}\label{eq:1loopsymbIBPs3}
\end{eqnarray}
The most complicated integrals in \myref{eq:1loopsymbIBPs1} and \myref{eq:1loopsymbIBPs2} are contained in those terms that do not involve a lowering operator $\down{i}$, i.e. the terms $\pm q^2 n_3\up{3}$. We eliminate these terms by adding \myref{eq:1loopsymbIBPs1} and \myref{eq:1loopsymbIBPs2}. We get 
\begin{equation}\label{eq:symbequation}
 \left(n_3\down{2}\up{3}+n_1\up{1}\down{2}-d+n_1+2 n_2+n_3\right){\cal I}_{\mathrm{1-loop}}\lr{n_1,n_2,n_3}=0\;.
\end{equation}
For general $n_i$, all terms seem equally complicated, and the Laporta algorithm would thus instruct us to solve for the first or second term, since their coefficients are small. But this would introduce inverse coefficients $1/n_1$ or $1/n_3$ which would actually become singular when the reduction of the topology should take place. Indeed, solving for the second term in~\myref{eq:symbequation} and relabelling the $n_i$'s, one gets
\begin{equation}\label{eq:symbDoesNotWork}
\begin{split}
 {\cal I}_{\mathrm{1-loop}}\lr{n_1,n_2,n_3}=&\frac{d-n_1-2 n_2-n_3-1}{n_1-1}\:{\cal I}_{\mathrm{1-loop}}\lr{n_1-1,n_2+1,n_3}\\
&- n_3\,{\cal I}_{\mathrm{1-loop}}\lr{n_1-1,n_2,n_3+1}
\end{split}
\end{equation}
which cannot be evaluated for $n_1=1$.

Instead, eq.~\myref{eq:symbequation} should simply be solved for ${\cal I}_{\mathrm{1-loop}}\lr{n_1,n_2,n_3}$. We may write this as an operator equation
\begin{equation}
1=\frac{1}{d-n_1-2 n_2-n_3}\lr{n_1 \up{1}\down{2}+n_3\down{2}\up{3}}\;,
\end{equation}
or more explicitly as
\begin{equation}
\begin{split}
   {\cal I}_{\mathrm{1-loop}}\lr{n_1,n_2,n_3}=\frac{1}{d-n_1-2 n_2-n_3}\Bigl(&n_1\,{\cal I}_{\mathrm{1-loop}}\lr{n_1+1,n_2-1,n_3}\\&+n_3\,{\cal I}_{\mathrm{1-loop}}\lr{n_1,n_2-1,n_3+1}\Bigr)\;.
\end{split}
\end{equation}
Here, the fact that the inverted coefficient contains the symbol $d$ prevents the denominator from becoming singular. For irreducible topologies, the inverted coefficients do not contain the symbol $d$ and thus the equation is not applicable to the master integral. 

Even for reducible topologies, it is by no means obvious how symbolic identities may be used in general. In the above example it was possible to combine two equations to eliminate a term that does not have an operator $\down{i}$, $n_3 q^2\up{3}$. This is not always possible, and so some reducible topologies have to be reduced using numeric IBP identities.

To generalise this to three-loop (reducible) integrals, IBP identities are solved just as in the Laporta algorithm\footnote{For the evaluation of $r$ and $s$, set all $n_i$ equal to unity. Drop the ordering according to $t$.} until topologies with the same $r$ and $s$ as the seed integral are reached. At that point, integrals with coefficients containing a symbol $d$ should be favoured. The corresponding integral usually has no lowering or raising operators acting on ${\cal I}$.

The symbolic identity found in this way may then be applied repeatedly until one propagator is cancelled completely, and the topology thus changes. This substitution has to be detected and a further application of the symbolic identity has to be prevented. The next section will report in more detail on the reformatting necessary for symbolic IBP identities.

\subsection{Symbolic Integration-by-Parts Identities for Irreducible Topologies}
So far the experience gained with symbolic IBP identities for reducible topologies has not let to any significant insight in how symbolic IBP identities could be used for irreducible topologies. Symbolic IBP identities for irreducible topologies always seem to have coefficients that would become singular when applied to certain integrals, especially to the master integral. When they are applied recursively to a general integral, some exponents might be lowered until ultimately they are not applicable any more since that would again lead to singular coefficients just as in \myref{eq:symbDoesNotWork}.

A second caveat lies in scalar products (i.e. propagators with non-positive exponents) whose exponents are decremented upon application of the symbolic identity. If no scalar product is present in the integral onto which the identity is applied, an additional propagator is generated, which is clearly not desirable. So the application of the identity to those cases should also be prevented, \emph{except} if the coefficient of those integrals becomes zero. In this case the identity may still be applied. 

Still, for sufficiently large exponents $n_i$ some of these identities do contain valuable information in that they are indeed able to reduce $r$ and $s$ for certain integrals. Generating symbolic IBP identities is computationally cheap, and applying them is worth trying at least in those cases where otherwise large $r$-$s$ blocks of IBP identities would have to be evaluated. But most of the identities actually are not useful at all, in that they relate a given integral of a certain complexity to many integrals of the same complexity. Thus symbolic identities should only be used if the identity, when applied, in fact reduces $r$ or $s$ of the integral it was applied to. 

From the symbolic identities contained in the solution of the system of (symbolic) IBP identities, the ones must be selected that match the above criteria. All other identities are discarded. The identity then has to be rewritten as FORM code in such a way that it is only applied when appropriate. Altogether our code that analyses the identities is rather complex. It was thus written in the more versatile language of Mathematica~\cite{mathematica}. 

The following sample code illustrates the final form of the identities (we have introduced ellipses to condense the code):
\begin{lstlisting}
id Topo1(-7,1694,[],m1?,n2?,...,m12?)= 
     deltap_(1+n2)*Topo1(-7,1694,[],m1,n2,...,m12) 
  +  delta_(1+n2)*Topo2(-7,1694,[],m1,n2,...,m12);
id Topo1(-7,1694,[],m1?pos0_,...,n8?neg0_,0,n10?neg0_,...)= 
     Topo2(-7,1694,[],m1,...,n8,0,n10,...);
id Topo1(-7,1694,[],m1?,n2?,...,m12?)=
  -  m1/(1+n2)*Topo1(-7,1694,[],-1+m1,1+n2,...,-1+m9,...,m12)
  -  n3/(1+n2)*Topo1(-7,1694,[],m1,1+n2,...,m12)
  +  Topo1(-7,1694,[],m1,1+n2,...,m12)
     ...
  +  m12/(1+n2)*Topo1(-7,1694,[],m1,1+n2,...,-1+m12)
  +  n10/(1+n2)*Topo1(-7,1694,[],m1,1+n2,...,m12)
  +  n8/(1+n2)*Topo1(-7,1694,[],m1,1+n2,...,m12);
id Topo1(-7,1694,[],m1?,0,n3?,...,m12?)= 
     Topo4(-7,0,[],m1,0,n3,...,m12);
\end{lstlisting}
The denominator of this identity would become singular if evaluated at $n_2=-1$. Thus lines \verb|1-3| prevent the identity to be evaluated if $n_2=-1$ by renaming \verb|Topo1| to \verb|Topo2| if $n_2=-1$.

On the right-hand-side of the identity (line 7) the scalar products $m_1$, $m_9$ and $m_{12}$ are lowered, but only when $m_9$ is lowered from 0 to -1 there is a danger of introducing a higher-$t$ topology, since the coefficient of the other integrals contain a factor $m_1$ and $m_{12}$, respectively. They make these terms vanish if $m_1=0$ or $m_{12}=0$. Thus two lines are written (lines \verb|4-5|) that prevent the case where $m_9$ is lowered, but no such code is written to prevent the lowering of $m_1$ or $m_{12}$.

Finally, lines \verb|14-15| rename topologies where one propagator disappeared, as is the case with the exponent of the second propagator ($-n_2$) that is lowered in this identity, $n_2\rightarrow n_2+1$. In this way the whole code block can simply be iterated a few times until it is not applicable any more.

Also, note that in fact this identity has correctly been selected as useful: all the right-hand-side topologies have the exponent of the second propagator ($-1-n_2$) lowered by one with respect to the left-hand-side, a property that the symbolic IBP identities (right after their generation) usually do not have. It only arises in some of the equations that constitute the solution of the system of IBP identities.

The identity as it stands in this form can be safely applied to the form factor expression that should be reduced. With some luck, high-$r$-$s$ subtopologies can be reduced using these identities. But mostly integrals do not find a reduction equation because in our application ($t\leq 9$, $6\leq r\leq 9, s\leq 4$) most exponents of propagators and scalar products are still 1 and 0, respectively, and because symbolic IBP identities for irreducible topologies usually become singular at some $n_i=-1$. Therefore it is quite likely that for most integrals none of the identities matches all criteria listed above.

\section{The Code \emph{Solve}}\label{sec:Solve}

\emph{Solve} is the name of a code by E. Remiddi that implements the Laporta algorithm as outlined in sec.~\ref{sec:LaportaAlgorithm}. Given a system of linear equations, the algorithm of \emph{Solve} can be described as follows:
\begin{enumerate}
\item Calculate the ordering of the terms using FORM~\cite{Vermaseren:2000nd}. This involves calculating $t$, $r$ and $s$ for every integral according to their definition and counting the number of terms in each coefficient. Also, the number of terms in each equation is computed approximately.\label{alg:calculatetrs}
\item Find the integral with the highest $t$, $r$, $s$, with the smallest coefficient in the smallest equation. \label{alg:calctrsNeq}
\item Solve its equation for this integral. The step 
\begin{equation}\label{eq:identityToSubstituteIn}
c_0 {\cal I}_0+ \sum_{i=1}^n c_i {\cal I}_i=0\rightarrow {\cal I}_0=-\frac{1}{c_0}\lr{\sum_{i=1}^n c_i {\cal I}_i}
\end{equation}
involves inverting the coefficient $c_0$. It is crucial to factor the coefficient if it contains symbols like $d$, the dimension of space-time. This is done using Maple~\cite{maple}.\label{alg:invertcoefficient}
\item Substitute the equation into the system. Store it for later use (see step~\ref{alg:triagTodiag}). 
\item Perform partial fractioning to limit the growth of the size of the coefficients. Check that this removes the equation picked in step~\ref{alg:calctrsNeq} from the system. This and the previous step are computationally expensive.
\item Repeat all previous steps until no more equations are left. The system is then in a triagonal form and all identities are extracted.
\item Substitute identities with some $t$, $r$ and $s$ that were extracted at a later cycle into those with the same $t$, $r$ and $s$ that were extracted earlier.\label{alg:triagTodiag}
\item Read out the identities, sorted according to the $t$, $r$ and $s$ of their left-hand sides; resubstitute them into the original system of IBPs to check that they indeed are a solution to it.
\end{enumerate}

\emph{Solve} is written in C. It acts as a coordinator between various programs that are invoked to perform specific tasks; Maple for example is invoked to invert and factor the coefficient $c_0$. Then \emph{Solve} writes FORM code for loading the system of equations, and Maple's output is used to write code to substitute the identity~\myref{eq:identityToSubstituteIn} into all others. For the actual substitution FORM is run on this code and \emph{Solve} then processes FORM's output. \emph{Solve} also keeps a file that stores the extracted identities and in the end rewrites them for the final output.

The code \emph{Solve} was used extensively in the preparation of this thesis. \emph{Solve}'s main virtue is the fact that it invokes FORM to do all algebraic manipulations. FORM is arguably the fastest computer algebra system to work with large symbolic expressions and so for large systems of IBP identities, \emph{Solve} is quite efficient. The part of \emph{Solve}'s code that computes the ordering of the terms in the system of equations is written as a plain FORM include file and thus can easily be adapted by the user.

\section{Improvements on \emph{Solve}}\label{sec:SolveImprovements}
A number of extensions and improvements have been introduced while trying to make \emph{Solve} fit the needs of the form factor reduction. Most of them try to address the fact that the system of linear equations for the three-loop form factor grows large in intermediate steps of the solution process, and thus \emph{Solve} has to handle large expressions as efficiently as possible. These improvements are not specific to the form factor calculation and thus calling them improvements of \emph{Solve} itself seems justified. In the following, we will comment on improvements of several steps in \emph{Solve}'s algorithm outlined above and we also introduce a few new features:
\begin{itemize}
 \item[1.,2.]Computing $t$, $r$ and $s$ for a given topology ${\cal T}$ is actually quite fast. Still, there is no need to recompute them for all topologies in every step as suggested in \emph{Solve}'s algorithm. Rather, we may multiply $c_0$ by a new symbol $\lambda^{-1}$ and update $t$, $r$ and $s$ only for integrals that come with a factor $\lambda$. For the next step, all $\lambda$'s are removed by setting them equal to one.\\ 
Computing the number of terms of a given subexpression in FORM is quite challenging: For efficiency reasons, FORM has no general possibility to operate on several terms at the same time. Counting terms can be implemented using FORM's \texttt{collect} instruction to cast all terms to be counted into a single function's argument. Then it is possible to count the number of terms, e.g. using \texttt{splitarg}. However, FORM limits the size and number of arguments, and for long equations this method of counting terms does not seem viable. Earlier versions of \emph{Solve} thus could not distinguish the size even of moderately large equations. An alternative that turned out to be rather fast is based on the idea that we do not actually need the size of all equations; instead, we only need to identify among the equations with the smallest $t$, $r$, $s$ and with the smallest coefficient the topology that is part of the smallest equation. Therefore we propose the following specific algorithm to perform the task in step~\ref{alg:calctrsNeq} of \emph{Solve}'s algorithm:
\begin{enumerate}[{1}.,{2}.-i)]
 \item Compute $t$, $r$ and $s$ if needed. To count the number of terms in each coefficient, use \texttt{bracket}, \texttt{collect} and \texttt{nterms} to count the number of terms in the coefficient. If collect fails\footnote{FORM's collect statement has a special second argument to deal with this case.}, set the number of terms in the coefficient to a large value. 
 \item To extract the smallest equation, take a copy of the system of IBP identities and set all coefficients of all integrals equal to one. This drastically reduces the size of the system, making it fast to handle.
 \item Apply a \texttt{.sort}. The coefficient of each integral is now equal to the number of terms in the equation. Use \texttt{bracket} and \texttt{collect} to move this coefficient into the argument of the function according to which FORM sorts the terms.
\item Use \texttt{bracket}, and then \texttt{firstbracket\_} to extract the label of the equation with the smallest $t$, $r$, $s$, number of terms in the coefficient and number of terms in the equation.
\item Use dollar variables\footnote{For earlier versions of FORM, one may also use \texttt{multiply}.} to pass the information on the equation just selected to every term in the system of equations. Adapt the ordering information such that the given equation is sorted before all others.
\end{enumerate}
Thus, information that is present locally with every integral (i.e. $t$, $r$, $s$) is computed locally. Information that has to be collected from all over an equation is not computed for every equation, but instead only the term needed for the next substitution step is singled out. The information on the size of the coefficients is actually not local either and can be computed very much in the same way as the number of terms in the equation. But the difference to the method used above does not seem to be big because large coefficients tend not to be factorisable in a simple way. This provides a strong reason not to use them, since partial fractioning is very expensive in that case. 
\item[3.] When substituting in an identity, it is applied to the system containing the equation that gave rise to the identity. It is thus crucial that this equation indeed reads $0=0$ after substitution and is thus removed from the system. \emph{Solve} explicitly checks this after every substitution of an equation. This in turn requires and insures that partial fractioning is done properly after every substitution cycle.\\
Now our partial fractioning routine in FORM is not able to handle completely general denominators.
Factoring $1/c_0$ into small factors that can be handled by \texttt{Parfrac.prc} might not be possible, and the above-mentioned check would fail. Thus a subroutine has been devised written using Mathematica's~\cite{mathematica} powerful language; it has to 
\begin{enumerate}[a)]
 \item try to factor the coefficient;
 \item analyse the factorised coefficient and decide what \emph{Solve} should do with the current identity.
\end{enumerate}
The information on the outcome of the above operations is handed back to \emph{Solve}, together with the inverted coefficient (in case the inversion was successful). \emph{Solve} then writes code according to the following four scenarios:
\begin{enumerate}[a)]
\item The coefficient is a pure number. No partial fractioning is needed in this case.
\item The coefficient is indeed factored into terms that can and must be handled by \texttt{Parfrac.prc}. In this case do invoke the partial fractioning routine.
\item The coefficient is not factored into terms that can be handled by \texttt{Parfrac.prc}. Write code that eliminates the equation from the system. For the final check of the solution of all IBPs, a list of dropped equations must be kept.
\item The coefficient is vanishing. This may happen since partial fractioning is not used at every step, and FORM does not recognise terms like $\frac{d}{d-4}-1 - \frac{4}{d-4}$ as vanishing. No substitution has to be made, but \texttt{Parfrac.prc} is invoked.
\end{enumerate}

\item[4.] Here, only minor changes were needed. The factor $\lambda$ was already introduced earlier. Also, FORM cannot substitute large identities using a single \texttt{id} statement. Thus,
insted of 
\begin{lstlisting}[numbers=none,frame=none,firstnumber=1]{Name}
id T=T1+T2+...+T50+T51+...;
.sort
\end{lstlisting}
constructions like
\begin{lstlisting}[numbers=none,frame=none,firstnumber=1]{Name}
id T=+T1+T2+...T50+sumUp;
.sort
id sumUp=sumUp+T51+...;
.sort
\end{lstlisting}
were needed already to cope with medium sized expressions.

\item[5.] For small systems of IBP identities, their solution finally contains only a rather limited set of factors in their coefficients. Their partial fractioning may be entered by hand. This approach revealed its weaknesses in the application of \emph{Solve} to the three-loop QCD quark form factor. The growing number of different factors appearing in the denominator of the solution of large systems made it difficult to anticipate all possible denominators. Rather, it became necessary to systematically perform partial fractioning for a general polynomial that contains factors of the form $\alpha_0+\alpha_1 d$, $\alpha_1,\alpha_2\in\mathbf{Q}$. Instead of using symbols such as $[d-4]$, $[3*d+2]^{-1}$ etc. and implement partial fractioning as
\begin{lstlisting}[numbers=none,frame=none,firstnumber=1]{Name}
s [d-4],[3*d+2];
...
id [d-4]*[3*d+2]^(-1)=1/3-14/3*[3*d+2]^(-1);
\end{lstlisting}
for all combinations, we switch to FORM's functions and use $\verb|b(d,-4)|$ and $\verb|1/3*B(d,2/3)|=1/3*1/\lr{d+2/3}$ instead. Then, partial fractioning may be implemented as 
\begin{lstlisting}[numbers=none,frame=none,firstnumber=1]{Name}
cf b,B;
...
id b(n1?)*B(n1?,n2?)=1-b(n2)*B(n1,n2);
\end{lstlisting}
As can be seen from this example, it is crucial to enforce an ordering in the arguments of $\verb|B|(..)$: if $\verb|n1|$ is a number and $\verb|n2|=d$, the above substitution should not be made- otherwise FORM runs into an endless loop.

The use of FORM's \texttt{repeat...endrepeat} statement produced stack overflows when applied to large expressions and had to be replaced by
\begin{lstlisting}[numbers=none,frame=none,firstnumber=1]{Name}
#do k=1,1
if ( match(<Pattern>) );
     redefine k "0";
     id <Pattern>=..;
endif;
.sort
#enddo
\end{lstlisting}

\item[7.] In the final check, identities that were dropped have to be accounted for. These equations are of course not solved by the identities collected, so the check will fail unless the dropped equations are removed separately. 
\end{itemize}

Taken together, the above improvements allowed for the solution of systems of IBPs that contained larger equations and more complicated denominators. Important improvements were the reliable and general partial fractioning of the coefficients of the integrals and the accurate and fast counting of the number of terms in large equations.

\section{Extensions of \emph{Solve}}\label{sec:SolveExtensions}
The code \emph{Solve} as described in sections~\ref{sec:Solve} and~\ref{sec:SolveImprovements} is essentially an implementation of Gauss' algorithm to solve systems of linear equations. We have described several improvements on the efficiency, generality and stability. 
Still, further improvements are necessary to handle the size of the systems of IBP identities that were encountered. Most importantly we have not yet made any use of our knowledge about the structure of IBP identities. The following facts are crucial:
\begin{itemize}
 \item Systems of IBP identities are sparse.
 \item There are subsystems that have non-vanishing coefficients (almost) only in specific blocks in the coefficient matrix.
\end{itemize}

The first point follows immediately from the number of IBP identities per topology and $r$-$s$-block (see tab.~\ref{tab:NrOfEqsXX}) and the fact that IBP identities are generated from seed integrals using differentiation, eq.~\myref{eq:IBPs}. Using the chain and product rule, one can estimate that an IBP identity will typically contain ${\cal O}\lr{10}$ terms. Tab.~\ref{tab:NrOfEqsXX} shows clearly that there are usually many more equations than that, and so most of the entries in the matrix of coefficients will be zero. 

The second point follows from the $r$-$s$ structure of IBP identities: We have already classified integrals into $t$-$r$-$s$ blocks, and we have seen in \myref{eq:subtopotrs} that seed integrals from a given $r$-$s$-block mainly relate integrals of three $r$-$s$-blocks: Mostly $r+1,s+1$, but also $r+1,s$ and $r,s$ integrals. 

\begin{figure}
  \begin{center}
	\resizebox{10cm}{!}{\includegraphics[width=11cm]{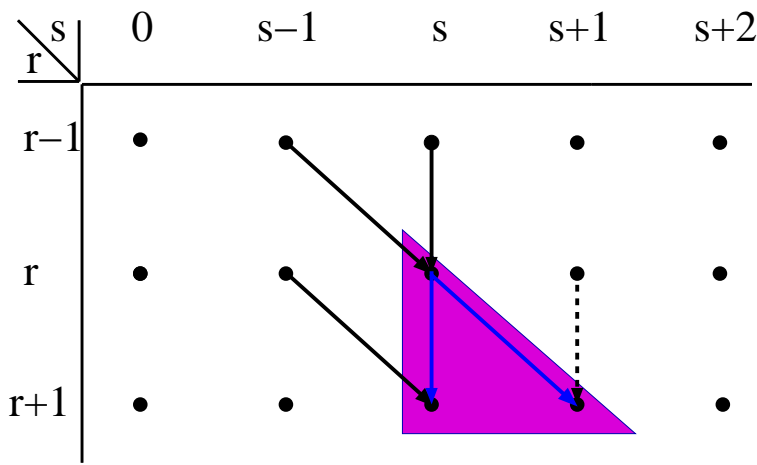}}
\caption{An illustration of how $r$-$s$-blocks of IBP identities interfere with each other. IBP identities generated from the $r$-$s$-block contain mainly integrals in the triangle $\lr{r,s}$, $\lr{r+1,s}$, $\lr{r+1,s+1}$. They must be merged with identities stemming from three blocks with lower $r$ and $s$: $\lr{r,s-1}$, $\lr{r-1,s}$ and $\lr{r-1,s-1}$. Identities from the $r$-$s+1$-block also mix in (dashed line) but since their seed integrals have higher $r$ and $s$, we may argue that IBP identities from $r$-$s$ seed integrals mix into IBP identities from $r$-$s+1$ seed integrals, and not vice versa.}\label{diag:overlap_of_rsblocks}
  \end{center}
\end{figure}

Considering these two facts, it seems natural to divide large systems of IBP identities into subsystems of identities that originate from seed integrals within a given $r$-$s$-block. These identities mix strongly, and once their solutions are found, most of the substitutions of identities will be done. The overlap with the solution of a different $r$-$s$-block of seed integrals will be quite small; it is significant only for the $r$, $s-1$, the $r-1$, $s$ and the $r-1$, $s-1$ block, see fig.~\ref{diag:overlap_of_rsblocks}. 
The next section thus presents routines that deal with the merging of such (sub-)systems of equations.

We recall that dividing up a system of IBP identities into $r$-$s$-blocks is also useful in the light of sec.~\ref{sec:hierarchy}. There we saw that seed integrals in a higher $r$-$s$-block have more complicated $t-1$-subtopologies and that their use should be avoided if possible. Thus it is in fact useful to merge and apply identities with an increasing number of $r$-$s$ blocks of seed integrals, in contrast to simply merging all available $r$-$s$-blocks.

\subsection{\emph{MergeIds} and \emph{RewriteIds}}
The standard task that we would like to describe in this section is the following: Given the solution of a system of IBP identities generated using seed integrals of a certain set of $r$-$s$ blocks, and the solution of the IBP identities of an additional $r$-$s$ block (usually adjacent in $r$-$s$-space); to find the solution of the combined system of IBP identities. 

A simple way to achieve this is to read in the two systems of IBP identities and restart \emph{Solve}. A routine that we call \emph{MergeIds} scans the identities it receives and writes FORM code that builds up the FORM expression containing the system of equations:
\begin{lstlisting}
l zeroes= sumUp;
.sort
id sumUp=sumUp+Label(1)*(Save(lhs1)-lhs1);
id lhs1=rhs1;
.sort
id sumUp=sumUp+Label(2)*(Save(lhs2)-lhs2);
id lhs2=rhs2;
...
id Save(x?)=x;
id sumUp=0;
\end{lstlisting}
(the factor \verb|Label()| distinguishes different equations). This system is then read into \emph{Solve} that solves the system as described in section~\ref{sec:Solve}.

There is one special situation where calling \emph{Solve} is not necessary: Sometimes it is clear that it is necessary to substitute the identities of one set into the other, but not vice versa. This is the case for example when the left-hand sides of identities have different $r$ or $s$. It is then enough to read in the identities of one set, apply the second set of identities and write out the identities of the first set again. \emph{Solve} is not needed in this case. We call this code \emph{RewriteIds}:
\begin{lstlisting}
cf End;
l zeroes= 
  + Label(lhs1)
  + Label(lhs2)
  + ...
  ;
.sort
id Label(lhs?)=Label(lhs)*(lhs+End);
.sort
#include 1stIdentitySet
#include 2ndIdentitySet
#call Parfrac.prc
.sort
b Label;
print+s;
\end{lstlisting}
The contents of the bracket of \verb|Label(lhs)*(..)| is the right-hand-side of the equation, with the second set of identities substituted in. The last term of each right-hand-side is \texttt{End}, making it easy for a simple Perl script to read out the identities. We may note here that the generalisation to more that two sets of identities is straightforward.

We would also like to point out that the two code samples above illustrate two different ways of reading in identities. The second one must be used with care, since if \texttt{lhs1}=\texttt{lhs2}, \texttt{id lhs1=rhs1} will be applied both to \texttt{lhs1} and \texttt{lhs2} and the second identity is lost. Therefore, the above (outdated) version of \emph{RewriteIds} only works on merged identities (where all left-hand sides are different). Of course, this limitation is easily removed:
\begin{lstlisting}
cf End;
l zeroes= sumUp;
.sort
id sumUp=sumUp+Label(lhs1)*(lhs1+End);
id lhs1=rhs1;
.sort
id sumUp=sumUp+Label(lhs2)*(lhs2+End);
id lhs2=rhs2;
...
id sumUp=0;
#call Parfrac.prc
.sort
b Label;
print+s;
\end{lstlisting}

\emph{MergeIds} and \emph{RewriteIds} perform partial tasks of solving systems of equations: \emph{MergeIds} substitutes into each other equations with the same left-hand-side using \emph{Solve}. \emph{RewriteIds} substitutes equations that reduce the integrals on the right-hand-side. Rewriting identities using \emph{RewriteIds} often condenses the identities noticeably which is important for keeping small the size of terms involving subtopologies of subsequent (e.g. higher-$t$) IBP identities.

\subsection{\emph{PerlMerge}}\label{sec:perlmerge}
\emph{MergeIds} and \emph{RewriteIds} are two routines that provide much flexibility in the choice of how to combine subsets of IBP identities. In our context, they are mainly useful for processing large systems of IBP identities that are split into $r$-$s$-blocks, which can be solved more easily. Using this flexibility however requires a lot of manual interaction, and in this section we will present a new routine, called \emph{PerlMerge} that takes over some of this manual work. To illustrate the need for further automatisation, consider the reduction of the $t=7$ topology with $ID=734$: merging all identities from seed integrals with $r=7$, $s\leq2$ with those of $r=7$, $s=3$, we have the following $t$, $r$, $s$-tuples for the left-hand sides of the solutions: For $r=7$, $s\leq2$,
\begin{center}
\begin{tabular}{|lll|}\hline
t&r&s\\
\hline7&8&3\\
\hline7&8&2\\
\hline7&8&1\\
\hline7&8&0\\
\hline7&7&2\\
\hline7&7&1\\
\hline
\end{tabular}\label{tab:Perlmerge1}
\end{center}
and for $r=7$, $s=3$,
\begin{center}
\begin{tabular}{|lll|}\hline
t&r&s\\\hline
7&8&4\\\hline
7&8&3\\\hline
7&7&3\\\hline
7&7&2\\\hline 
\end{tabular}\label{tab:Perlmerge2}
\end{center}
We clearly see that to merge these two sets of equations, there is no need to run \emph{Solve} over the identities with $r=8$, $s=4$ again: There are not two sets of identities that would have to be merged. We may leave the $r=8$, $s=4$ identities aside and start merging the two $r=8$, $s=3$ sets of identities. There it happens that identities with $r< 8$ or $s< 3$ emerge. Those identities would have to be merged with other identities from the two sets. But in any case it is not necessary to merge all identities with $r\leq 8$, $s\leq 3$ at the same time. In this way \emph{Solve} can be sped up significantly, since \emph{Solve} has to process much smaller sets of identities at any given time: Identities whose lhs have a smaller $t$-$r$-$s$ than the ones currently processed do not need to be loaded into \emph{Solve} yet.

The code (written in Perl) accepts a list of identity files as generated by \emph{Solve}. Here is the algorithm: 
\begin{enumerate}
 \item Extract the $t$, $r$ and $s$ of the left-hand sides of all identity files in the list. Sort the identity files accordingly.
 \item Determine the highest $t$-$r$-$s$ triple of all identity files in the list.
 \item If there are two or more files with this $t$-$r$-$s$ triple, merge these identities. Otherwise copy the identity file to the directory storing the results.\label{enum:repeat}
 \item Update the file list with the files containing the solutions of the merging process in the previous step. Update the list of $t$-$r$-$s$ triples.
 \item Cycle through the list of sorted $t$-$r$-$s$ triples and determine the next lower triple. Go to step~\ref{enum:repeat}.
\end{enumerate}
The above algorithm is designed to be used with large systems of equations which is broken into smaller subsystems. They are then merged successively. As already mentioned one possible way of breaking up the systems is into $r$-$s$-blocks. 

One further application concerns the solution of a single (large) $r$-$s$-block e.g. $t=6$, $r=8$, $s=2$ (or $s=3$). It might help breaking it up into smaller, tractable parts again. Their solution can be parallelised. We then use \emph{PerlMerge} to merge these subsystems gradually. When doing so, at some point the solution will start breaking up into parts with different $r$ and $s$ for the left-hand sides. They do not have to be treated at the same time anymore. Thus the systems to be merged with \emph{Perlmerge} will be smaller than the original system. 

We found that for the partitionings we tried the gain was small. For example, dividing the system into the 16 pieces that arise when each IBP or LI operator in~\myref{eq:IBPs} is applied to all seed integrals, the systems typically decrease in size by about 10\%. At some point the merged subsystems of equations again become large, and the gain of $\sim$ 10\% did not help much there.

We should stress again that in the cases where it was possible to obtain the solutions of the single $r$-$s$ blocks of identities, \emph{PerlMerge} and \emph{RewriteIds} did speed up the solutions of those topologies significantly, to the point that e.g. the solution for $t=7$, $r\leq8$, $s\leq2$ (without having the necessary reduction of all $t=6$ integrals at hand) would have taken much longer (if it could be obtained at all) using just a single run of \emph{MergeIds}.

Fig.~\ref{fig:pstricksFlowchart} shows a flow chart summarising how we approached the solution of large systems of IBP identities, combining all the tools discussed in this thesis. 

\psset{unit=1cm}
\psset{framearc=0.07}
\begin{figure}\caption{A flow-chart of how \emph{PerlMerge}, \emph{Solve} and \emph{RewriteIds} work together on large systems of IBP identities. The system is split up into $r$-$s$-blocks $\lr{r_1,s_1},\ldots,\lr{r_n,s_n}$. A given $r$-$s$ block's IBP identities are solved and then merged with lower-$r$-$s$ blocks. This yields identities whose left-hand-sides (lhs) are all different. RewriteIds then recursively substitutes these identities into each others right-hand-side (rhs) to yield the most compact expression possible. Then the next $r$-$s$-block is solved and added to the system.}\label{fig:pstricksFlowchart}
\hspace{3mm}\begin{pspicture}(-5,-10)(5,10)
\rput[B](0,9){\Rnode{rsblock}{\psframebox{$r_i$-$s_i$ block }}}
\rput[t](0,8){\ovalnode{GenSeed}{\shortstack{generate seed integrals\\generate IBP identities}}}
\ncline{->}{rsblock}{GenSeed}
\rput[c](0,5.5){\Rnode{rsIBPs}{\psframebox{IBP identities for block $r_i$, $s_i$}}}
\ncline{->}{GenSeed}{rsIBPs}
\rput[c](6.8,5.5){\ovalnode{Substtmin1}{\shortstack{substitute solutions \\for $t-1$ subtopologies}}}
\ncline{->}{Substtmin1}{rsIBPs}

\rput[c](0,4.0){\psscaleboxto(0,1.5){\ovalnode{Solve}{\emph{Solve}}}}



\ncline{->}{rsIBPs}{Solve}


\rput[c](0,2.5){\rnode{rssolns}{\psframebox{solution of $r_i$-$s_i$-block}}}
\ncline{->}{Solve}{rssolns}

\rput[c](6,2.5){\rnode{rsoldsolns}{\psframebox{\shortstack{solutions of \\previous $r$-$s$-blocks}}}}

\rput[c](3,0){\psscaleboxto(0,1.8){\ovalnode{PerlMerge}{\emph{PerlMerge}}}}


\ncdiag[angleA=90,armA=0,angleB=-2,armB=0]{->}{rsoldsolns}{rsIBPs}
\ncline{->}{rsoldsolns}{PerlMerge}
\ncline{->}{rssolns}{PerlMerge}

\rput[c](3,-2.5){\rnode{MergedriSolns}{\psframebox{\shortstack{solutions of all\\ $\lr{r_1,s_1}\ldots \lr{r_i,s_i}$\\blocks (lhs unique)}}}}
\ncline{->}{PerlMerge}{MergedriSolns}

\rput[c](3,-4.5){\psscaleboxto(0,1.5){\ovalnode{RewriteIds}{\emph{RewriteIds}}}}



\ncline{->}{MergedriSolns}{RewriteIds}

\rput[b](3,-7.2){\rnode{RewrittenriSolns}{\psframebox{\shortstack{solutions of all\\ $\lr{r_1,s_1}\ldots \lr{r_i,s_i}$\\blocks with rhs simplified}}}}
\ncline{->}{RewriteIds}{RewrittenriSolns}

\rput[c](3,-8.5){\ovalnode{nextstep}{\shortstack{move to next $r$-$s$-block:\\$i\rightarrow i+1$}}}
\ncline{->}{RewrittenriSolns}{nextstep}

\ncangle[angleA=0, angleB=0, armB=0.5]{->}{nextstep}{rsoldsolns}

\ncangle[angleA=180, angleB=180, armA=3, armB=3]{->}{nextstep}{rsblock}
\end{pspicture}
\end{figure}

\pagestyle{empty}
\cleardoublepage
\pagestyle{headings}

\chapter{Reduction of the 5-Propagator Topologies}\label{chap:5PropIntegrals}
In the previous chapters we have set up a detailed analysis of how the Laporta algorithm could be employed for the reduction of the three-loop quark form factor to master integrals. We saw that this task is quite involved due to the large number of subtopologies and IBP identities that are present at the level of topologies with 7 or less propagators. Quite a variety of tools were developed in order to tackle different types of topologies: Reducible topologies may be reduced using symbolic IBP identities, some topologies turned out to be integrable loop-by-loop, and for irreducible topologies that are not integrable loop-by-loop we devised routines such as \emph{PerlMerge}, \emph{MergeIds} and \emph{RewriteIds} that all aim to tune up \emph{Solve} in the hope to be able to solve the large systems of IBP identities needed to reduce the form factor to master integrals. 

Here we should stress that for the topologies (at $t=6$ and $t=7$) where explicit numeric IBP identities have to be solved and where the number of IBP identities grows large even for singel $r$-$s$-blocks of seed integrals, it is crucial that (lower-$t$) subtopologies are reduced as far as possible before starting to solve the system of IBP identities. During the solution of a system of IBP identities, many identities are substituted into others. If all these identities contain many lower $t$ subtopologies, the equations quickly become very large. It is thus mandatory to reduce subtopologies to their master integrals first in order to simplify the part of the identity containing lower-$t$ subtopologies.

As we noted in sec.~\ref{sec:identifySubTopos}, at the level of $t=5$ all topologies can be integrated loop-by-loop. At the same time the number of seed integrals is so large that using IBP identities is clearly hopeless. This chapter is thus going to describe how $t=5$ (or $t=4$) topologies can be calculated loop-by-loop. We first discuss how to evaluate a single such integral in sec.~\ref{sec:singleIntegrableIntegral}. Since we need a lot of these integrals, we then also discuss the automatisation of their evaluation in sec.~\ref{sec:parallel5Prop}. To keep the size of the results small, we will rewrite all integrals in terms of their $r=t$, $s=0$ integral. These might not be master integrals. In section~\ref{sec:reductionEquations} we will give their reduction identities to master integrals.

\section[Integration of Bubble-Insertion Vertex Integrals]{Integration of Multi-Loop Bubble-Insertion Vertex Integrals} \label{sec:singleIntegrableIntegral}
\begin{figure}
  \begin{center}
	\resizebox{6cm}{!}{\includegraphics[width=11cm]{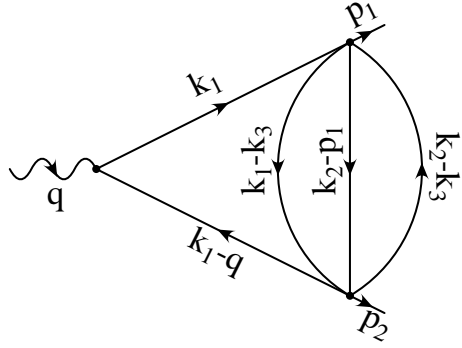}}
\caption{A sample 5-propagator topology that can be integrated loop-by-loop. After integrating out two bubble-insertions (loop momenta $k_2$ and $k_3$), the standard vertex integral (with non-integer exponents) is recovered.}\label{fig:t5integrableplot}
  \end{center}
\end{figure}

 We start with the following basic results for the 1-loop bubble and triangle integral:
\begin{eqnarray}
 I_{\alpha\beta}\lr{q}&=&
\int \frac{d^d k}{\lr{2\pi}^d}\frac{1}{\lr{-k^2-i\,\eta}^\alpha\lr{-\lr{k-\ell}^2-i\,\eta}^\beta}\\
&=&\lr{-\ell^2-i\,\eta}^{\frac{d}{2}-\alpha-\beta} f_{\alpha,\beta}\label{eq:twopoint}\\
I_{\alpha\beta\gamma}\lr{q}&=&
\int \frac{d^d k}{\lr{2\pi}^d}\frac{1}{\lr{-k^2-i\,\eta}^\alpha\lr{-\lr{k-p_1}^2-i\,\eta}^\beta\lr{-\lr{k-q}^2-i\,\eta}^\gamma}\\
&=&\lr{-q^2-i\,\eta}^{\frac{d}{2}-\alpha-\beta-\gamma} g_{\alpha,\beta,\gamma}\label{eq:threepoint}
\end{eqnarray}
where a small imaginary part $i\,\eta$ is understood in every scalar product or propagator. $f_{\alpha,\beta}$ and $g_{\alpha,\beta,\gamma}$ consist of Euler gamma functions:
\begin{eqnarray}
f_{\alpha,\beta}&=&\frac{i}{\lr{4\pi}^{d/2}}\:\frac{ \Gamma
   \left(\frac{d}{2}-\alpha \right) \Gamma
   \left(\frac{d}{2}-\beta \right) \Gamma
   \left(-\frac{d}{2}+\alpha +\beta \right)}{\Gamma \lr{\alpha }
   \Gamma \lr{d-\alpha -\beta } \Gamma \lr{\beta }}\label{eq:f}\\
g_{\alpha,\beta,\gamma}&=&\frac{i}{\lr{4\pi}^{d/2}}\:\frac{ \Gamma
   \left(\frac{d}{2}-\alpha -\beta \right) \Gamma
   \left(\frac{d}{2}-\beta -\gamma \right) \Gamma
   \left(-\frac{d}{2}+\alpha +\beta +\gamma \right)}{\Gamma
   \lr{\alpha } \Gamma \lr{d-\alpha -\beta -\gamma } \Gamma \lr{\gamma }}\label{eq:g}\qquad .
\end{eqnarray}
Eq.~\myref{eq:twopoint} replaces the two propagators of a bubble insertion, $1/k^2$ and $1/\lr{k-\ell}^2$  with a single propagator that has the same external momentum flow $k-(k-\ell)=\ell$. Thus for integrals that consist of bubble insertions into the 1-loop two- or three-point function (such as fig.~\ref{fig:t5integrableplot}), the number of loops may be reduced by replacing repeatedly the left-hand-side of \myref{eq:twopoint} by its right-hand side. The propagators that replace the bubble insertion then have an exponent that depends on the dimensionality $d$, but do not otherwise complicate the integral. 

From this follows a straightforward way to implement the evaluation of multi-loop integrals that consist entirely of two-point bubble insertions into a two- or three-point function:
\begin{enumerate}
 \item Count the number of times $k_1$, $k_2$ and $k_3$ appear in the integrand. Choose as the next loop momentum one that appears in a minimal number of propagators\footnote{The way our loop momenta are assigned to the propagators is crucial for this way of singling out the first two-point function to be integrated. Without this, one would have to consider the topological information of the diagram to find two-point insertions (i.e. two propagators connecting the same two vertices).}. Call it $k'$.
 \item Shift the loop momentum $k'\rightarrow k$ to fit the propagators into the standard form \myref{eq:twopoint} or \myref{eq:threepoint}. Determine $\ell$.
 \item Collect the scalar products in the numerator that depend on $k$, and isolate the tensor loop integral ($\int d^dk k\cdot p_1\rightarrow p_1^\mu\int d^dk  k_{\mu}$, etc.).\label{item:separateTensorIntegral}
 \item Perform the tensor reduction, as will be described below in sec.~\ref{sec:tensorreduction}.  This yields the tensor integral as a scalar integral times a tensor formed out of $\ell$ in eq.~\myref{eq:twopoint} or $p_1$ and $q$ in~\myref{eq:threepoint}. 
\item Evaluate the loop integral. This yields a new propagator $\lr{-\ell^2-i\,\eta}^{\frac{d}{2}-\alpha-\beta}$ times a coefficient $f_{\alpha,\beta}$ or $g_{\alpha,\beta,\gamma}$.
 \item Contract the tensor in $\ell$ with the rest of the amplitude that has been split apart in step~\ref{item:separateTensorIntegral}. Try to cancel scalar products against propagators. 
 \item Go to step 1, proceed with the evaluation of the next loop.
\end{enumerate}

To illustrate the above steps (without tensor reduction), consider as an example the diagram shown in fig.~\ref{fig:t5integrableplot}. We proceed as follows: $k_1$ appears in three different propagators, but $k_2$ and $k_3$ only in two and thus we clearly have two two-point functions to integrate. As a first step, either the integration over $k_2$ or $k_3$ is possible. We choose to integrate over $k_2$. The propagators containing $k_2$, $1/\lr{k_2-p_1}^2$ and $1/\lr{k_2-k_3}^2$ have to be shifted e.g. by $k_2\rightarrow k_2+p_1$ to fall into the form~\myref{eq:twopoint}, with $k=k_2$ and $\ell=k_3-p_1$ . Note that $\ell^2$ is indeed a member of the propagator set~\myref{eq:PropStandardSetRestated}. Thus in the next iteration we are able to integrate over $k_3$. The exponent of one of the propagators is now $d$-dependent. Finally, in a third iteration we encounter the integral~\myref{eq:threepoint}.

For fig.~\ref{fig:t5integrableplot} this procedure yields two $f_{\alpha,\beta}$ coefficients and one $g_{\alpha,\beta,\gamma}$. Of course the dependence on $q^2$ in our massless calculation is fixed by the dimension of the integrals: There will always be a factor $\lr{-q^2-i\,\eta}^{\lr{3d-2r+2s}/2}$.

The functions $f_{\alpha,\beta}$ and $g_{\alpha,\beta,\gamma}$ in~\myref{eq:f} and~\myref{eq:g} are rather lengthy. But since only Gamma functions are involved, these expressions become quite short when divided by $f_{1,1}$ and $g_{1,1,1}$: Using the Pochhammer symbol
\begin{equation}
 \lr{a}_n=\frac{\Gamma\lr{a+n}}{\Gamma\lr{a}}=a\lr{a+1}\ldots\lr{a+n-1}\;,
\end{equation}
 we have for example
\begin{equation}\label{eq:Pochhammer}
 \frac{f_{\alpha,\beta}}{f_{1,1}}=\frac{1}{\Gamma\lr{\alpha}\Gamma\lr{\beta}}\frac{\lr{2-\frac{d}{2}}_{\alpha+\beta-2}\lr{d-\alpha-\beta}_{\alpha+\beta-2}}{\lr{\frac{d}{2}-\alpha}_{\alpha-1}\lr{\frac{d}{2}-\beta}_{\beta-1}}\;.
\end{equation}

 Equivalently, we may divide any integral that can be evaluated loop-by-loop by its scalar $r=t$ integral. Using partial fractioning then reduces the expression to a form similar to those obtained by IBP identities: rational polynomials in $d$ times the scalar $r=t$ integral.

\section{Tensor Reduction}\label{sec:tensorreduction}
For integrals with $s\leq4$ the tensors that might appear in the above procedure are 
\begin{eqnarray}
 &&\int \frac{d^d k}{\lr{2\pi}^d}\frac{\left\{1,k^\mu,k^\mu k^\nu,k^\mu k^\nu k^\rho,k^\mu k^\nu k^\rho k^\sigma\right\}}{\lr{-k^2}^\alpha\lr{-\lr{k-\ell}^2}^\beta}\;,\\
&&\int \frac{d^d k}{\lr{2\pi}^d}\frac{\left\{1,k^\mu,k^\mu k^\nu,k^\mu k^\nu k^\rho,k^\mu k^\nu k^\rho k^\sigma\right\}}{\lr{-k^2}^\alpha\lr{-\lr{k-p_1}^2}^\beta\lr{-\lr{k-q}^2}^\gamma}\;.
\end{eqnarray}
In this section we derive the reduction of these integrals to the scalar integrals~\myref{eq:twopoint} and~\myref{eq:threepoint}. As an explicit example, we take the integral
\begin{equation}
 I^{\mu\nu}_{\alpha,\beta}\lr{\ell}=\int \frac{d^d k}{\lr{2\pi}^d}\frac{k^\mu k^\nu}{-\lr{k^2}^\alpha\lr{-\lr{k-\ell}^2}^\beta}\;,
\end{equation}
but allow for a natural generalisation to higher tensors. We first construct a list containing all possible Lorentz 2-tensors that might arise in this integral:
\begin{equation}
 V^{\mu\nu}=\left(
\begin{array}{c}
\ell^\mu \ell^\nu\\ g^{\mu\nu}
\end{array}\right)\;.
\end{equation}
The tensor integral can now be written as
\begin{equation}\label{eq:tensordecomp}
 I^{\mu\nu}_{\alpha,\beta}=V^{T,\mu\nu}.C\;,
\end{equation}
where $C=\left(\begin{array}{c}C_1\\C_2\end{array}\right)$ is the coefficient vector that is to be found. Multiplying \myref{eq:tensordecomp} from the left with $V_{\mu\nu}$ and inverting, we get
\begin{equation}
 C= \left(V^{\rho\sigma}.V_{\rho\sigma}^T\right)^{-1}.V_{\mu\nu}.I^{\mu\nu}_{\alpha,\beta}\;.
\end{equation}
Substituting this back into \myref{eq:tensordecomp} we arrive at
\begin{equation}\label{eq:tensorreduction}
  I^{\mu\nu}_{\alpha,\beta}=V^{T,\mu\nu}.\left(V^{\rho\sigma}.V_{\rho\sigma}^T\right)^{-1}.V_{\kappa\lambda}.I^{\kappa\lambda}_{\alpha,\beta}\;.
\end{equation}
The result~\myref{eq:tensorreduction} holds not just for general $\alpha$ and $\beta$, but generalises without difficulty to higher rank tensors. However, the matrix $\left(V^{\rho\sigma}.V_{\rho\sigma}^T\right)$ has entries that depend on the dimensionality $d=g_{\mu\nu} g^{\mu\nu}$, and thus its inversion becomes cumbersome for tensors of rank higher than four.

\noindent As an example, we get
\begin{equation}
	I^\mu_{\alpha,\beta}\lr{q}=\frac{1}{2q^2} q^\mu \lr{-I_{\alpha-1,\beta}\lr{q}+ I_{\alpha,\beta-1}\lr{q} + q^2 I_{\alpha,\beta}\lr{q}}\;.
\end{equation}
The result for $I^{\mu\nu}_{\alpha,\beta}$ is 
\begin{equation}
\begin{split}
 I^{\mu\nu}_{\alpha,\beta}\lr{q}=& -\frac{q^\mu
   q^\nu}{q^2}I_{\alpha -1,\beta }\lr{q}\\
&\begin{split}+\frac{d q^\mu q^\nu-q^2g^{\mu\nu}}{4 (d-1) \lr{q^2}^2}\biggl(&\lr{q^2}^2 I_{\alpha ,\beta}\lr{q} +2q^2\left[ I_{\alpha
   -1,\beta }\lr{q}+ I_{\alpha ,\beta -1}\lr{q}\right]
   \\
&+I_{\alpha -2,\beta }\lr{q}-2 I_{\alpha
   -1,\beta -1}\lr{q}+I_{\alpha ,\beta -2}\lr{q}\biggr)\;.
\end{split}
\end{split}
\end{equation}

\section{Parallelising the Evaluation of 5-Propagator Integrals}\label{sec:parallel5Prop}
So far we have discussed the evaluation of a single $t=5$ integral. We will next comment on how to organise the calculation of all $t=5$ integrals needed for the reduction and how to store the results. 

\begin{figure}
  \begin{center}
	\resizebox{12cm}{!}{\includegraphics[width=11cm]{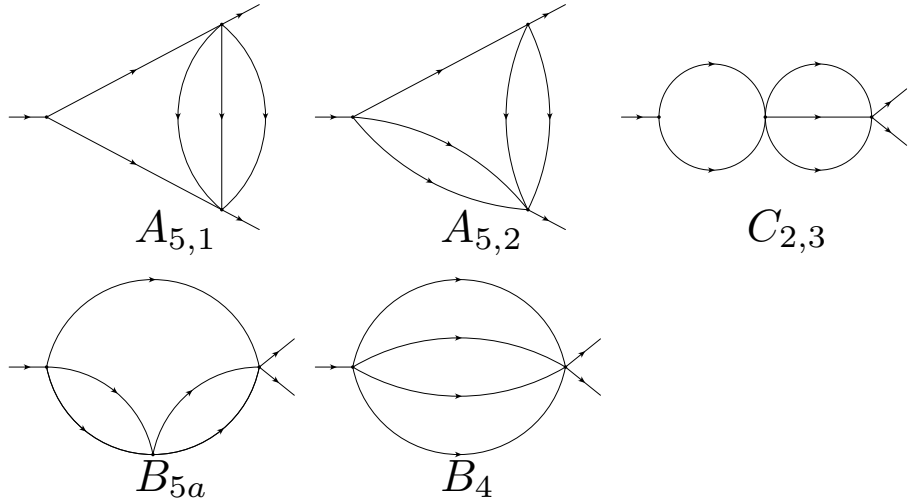}}
\caption{The 5- and 4-propagator irreducible subtopologies. All $t\leq 5$ integrals can be integrated loop-by-loop and ultimately be expressed as a linear combination of these five master integrals. This is the form most suited for a substitution into the $t=6$ system of IBP identities.}\label{fig:t5irredtopos}
  \end{center}
\end{figure}

The 5-propagator integrals appear as subtopologies in the $t=6$ IBP identities. If these would be solved without a prior reduction of the 5-propagator topologies, the solutions would contain unmanageably large expressions involving various 5-propagator integrals. Thus in order to keep those systems small, it is mandatory to reduce the $t=5$ integrals first. Some of the $t=5$ topologies are reducible, in which case they reduce (using symbolic IBP identities) to the only non-vanishing $t=4$ (irreducible) topology. The irreducible $t=5$ and $t=4$ topologies are shown in fig.~\ref{fig:t5irredtopos}. Their integrals must be evaluated as described above.

For the evaluation of integrable subtopologies there are two possible approaches: Either we extract the integrals needed from the $t=6$ system of IBP identities and integrate those topologies only once the $t=6$ system is set up. This results in a possibly rather small set of integrals that has to be computed for a given set of IBP identities; but there is a danger of recomputing the same integral several times if they show up in different sets of IBP identities. A second possibility is to tabulate all integrals in a given $r$-$s$ range (say $t\leq r\leq 9$, $0\leq s\leq 4$) and compute them all, but only once. 

The choice depends on the context. For loop-by-loop integrable $t=7,8$ topologies\footnote{There is no loop-by-loop integrable $t=9$ topology needed for the quark form factor.}, it was affordable not to integrate them when they appear as subtopologies in IBP identities of higher-$t$ topologies. There we would first reduce the integrals and integrate the (less than 1000) loop-by-loop integrable integrals only once they show up in the reduction result for the form factor. An important side remark here is that loop-by-loop integrable topologies may not only be irreducible themselves, but they may also be reducible and thus contain difficult (irreducible and not loop-by-loop integrable) subtopologies. Therefore it is important to detect these topologies in the result of the form factor and integrate them instead of reducing them further. This of course requires determining for every topology whether it is integrable or not (which we did by hand). 

For the 5 and 6 propagator loop-by-loop integrable topologies we chose instead to evaluate all $r\leq 9$, $s\leq 4$ integrals of the irreducible $t=4,5$ topologies needed. This job can be nicely parallelised and the computational effort is acceptable: Table~\ref{tab:NrOfSeedEqsT5} counts a total number of 62370 integrals per topology with $r\leq 9$, $s\leq 4$. Using table~\ref{tab:NrOfTopos}, we are to compute $6\times 62370=374220$ integrals. The computation takes at most a few seconds per integral, so we need at most a few CPU-weeks. The results are then compiled into several FORM tablebases~\cite{Vermaseren:2002rp}. This has the advantage of a short look-up time and the results can easily be reused whenever needed. Also, the tablebases contain the results in a compressed form. The total size of the tablebases was 221 MB.

For the implementation of the algorithm of sec.~\ref{sec:singleIntegrableIntegral}, one important restriction to keep in mind is that the computation should not be implemented in Mathematica~\cite{mathematica} or any other software with expensive licenses, since they do not allow for parallelisation. Here again FORM~\cite{Vermaseren:2000nd} was our product of choice. 

\section[Reduction Equations for $t=5,6$ Integrals]{Reduction Equations for Scalar $t=5,6$ Integrals to Master Integrals}\label{sec:reductionEquations}
In this chapter we have so far discussed the evaluation of (mainly $t=5$ or $t=6$) integrals that are integrable loop-by-loop. After their evaluation in terms of Euler gamma functions, we used eq.~\myref{eq:Pochhammer} to express them in terms of their $r=t$, $s=0$ integral. In this way all gamma functions cancelled, and the coefficient consisted of a polynomial in $d$, eq.~\myref{eq:Pochhammer}, similarly to IBP identities. The $r=t$, $s=0$ integral however might still be reducible. This can easily be found out by solving a few IBP identities. We will next give the reduction equations for the loop-by-loop integrable topologies that appeared as subtopologies of $t=6$ and $t=7$ topologies.
The $t=5$ reducible subtopologies obviously reduce to ${\cal T}_{172}\equiv B_4$:
\begin{eqnarray*}
 {\cal T}_{173}&=&\frac{1}{q^2}\lr{2+\frac{1}{d-3}} {\cal T}_{172}\;,\\
 {\cal T}_{174}={\cal T}_{186}={\cal T}_{668}&=&\frac{1}{q^2}\lr{6+\frac{12}{d-4}-\frac{1}{d-3}} {\cal T}_{172}\;,\\
 {\cal T}_{236}={\cal T}_{434}&=&-\frac{1}{q^2}\lr{4+\frac{6}{d-4}} {\cal T}_{172}\;,\\
{\cal T}_{428}&=&-\frac{1}{q^2}\lr{3+\frac{5}{2\lr{d-3}}+\frac{1}{2\lr{d-3}^2}} {\cal T}_{172}\;.\\ 
\end{eqnarray*}
\lstset{numbers=none}
For the $t=6$ reducible subtopologies, the following master integrals appear on the right-hand-side: ${\cal T}_{158}\equiv B_{5a}$, ${\cal T}_{412}\equiv A_{5,2}$, ${\cal T}_{433}=A_{5,1}$, ${\cal T}_{662}=C_{2,3}$.
\begin{eqnarray*}
  {\cal T}_{222}&=&-\frac{4}{q^2}\lr{1+\frac{1}{d-4}} {\cal T}_{158}\;,\\
  {\cal T}_{237}&=&-\frac{4}{\lr{q^2}^2}\lr{2+\frac{3}{d-4}} {\cal T}_{172}\;,\\
\end{eqnarray*}
\begin{eqnarray*}
  {\cal T}_{238}&=&-\frac{4}{\lr{q^2}^2}\lr{6+\frac{17}{d-4}+\frac{12}{\lr{d-4}^2}} {\cal T}_{172}\;,\\
  {\cal T}_{414}&=&\frac{4}{q^2}\lr{\frac{2}{3}+\frac{1}{d-4}-\frac{1}{9d-30}}\lr{{\cal T}_{412}-{\cal T}_{158}}\;, \\
  {\cal T}_{430}&=&-\frac{1}{\lr{q^2}^2}\lr{18+\frac{46}{d-4}+\frac{24}{\lr{d-4}^2}-\frac{1}{d-3}} {\cal T}_{172}\;,\\
{\cal T}_{435}&=&\frac{1}{\lr{q^2}^2}\lr{-8-\frac{24}{d-4}+\frac{20}{3d-10}} {\cal T}_{172}+\frac{1}{q^2}\lr{4+\frac{8}{d-4}-\frac{4}{3d-10}} {\cal T}_{433}\;,\\
{\cal T}_{437}&=&\frac{4}{q^2}\left[\lr{\frac{2}{3}+\frac{1}{d-4}-\frac{1}{9d-30}} {\cal T}_{412}+\lr{1+\frac{2}{d-4}-\frac{1}{3d-10}} {\cal T}_{433}\right]\;,\\
{\cal T}_{438}&=&-\frac{1}{\lr{q^2}^2}\lr{24+\frac{92}{d-4}+\frac{116}{\lr{d-4}^2}+\frac{48}{\lr{d-4}^3}} {\cal T}_{172}\\
&&+\frac{1}{q^2}\lr{4+\frac{8}{d-4}+\frac{4}{\lr{d-4}^2}} {\cal T}_{412}\;,\\
{\cal T}_{670}&=&\frac{1}{\lr{q^2}^2}\lr{12+\frac{24}{d-4}-\frac{2}{d-3}}{\cal T}_{172}\\
&&+\frac{1}{q^2}\left[\lr{4+\frac{4}{d-4}}{\cal T}_{158}-\lr{3+\frac{4}{d-4}}{\cal T}_{662}\right]\;,\\
{\cal T}_{726}&=&-\frac{1}{q^2}\lr{3+\frac{4}{d-4}}{\cal T}_{662}\;,\\
{\cal T}_{741}&=&-\frac{2}{q^2}\lr{1+\frac{1}{d-4}}{\cal T}_{662}\;,\\
{\cal T}_{1686}&=&-\frac{1}{q^2}\lr{3+\frac{4}{d-4}}{\cal T}_{662}\;.
\end{eqnarray*}

\chapter{Current Status and Outlook}
In chapters~\ref{chap:Formfactor} to~\ref{chap:5PropIntegrals} we have set up tools that we hoped to be sufficient to reduce the three-loop quark form factor to master integrals. We will next describe our experience made so far with the reduction of the three-loop quark form factor. Since we have not yet completed the reduction, sec.~\ref{sec:SolveApplicationToFF} will outline the progress made so far and highlight in which form the difficulties anticipated above were encountered. Sec.~\ref{sec:outlook} will give an outlook on possible future applications of our tools. In sec.~\ref{sec:partialresults} we present a partial result. 

\section{Applying \emph{Solve} to the Three-Loop Form Factor}\label{sec:SolveApplicationToFF}
As already stated in sec.~\ref{sec:estimateScaleOfComputation}, the reduction of $t=9$ and $t=8$ topologies do not pose a problem, though the improvements in \emph{Solve} did help (for example, large, non-factorisable coefficients are already encountered in the reduction of $t=8$ topologies and would have caused earlier versions of \emph{Solve} to interrupt). It was affordable to take a top-down approach, where no $t-1$ subtopologies are reduced before solving the system from $t$-propagator seed integrals. Not reducing the 7-propagator subtopologies has the disadvantage that the intermediate expressions in the reduction of $t=8$ topologies become rather lengthy. But in this way the following information could be gained: At the level of $t=7$ propagator topologies, no $r=9$, $s=4$ topologies were left. In fact, a single $r=9$, $s=3$ topology remained that was already present in the amplitude for the form factor. It turned out to be reducible using symbolic IBP identities, and no $t=6$, $r=9$, $s=3$ subtopologies emerged. In fact, after applying also the symbolic reduction equations (and integrating loop-by-loop where possible) no $t=7$, $r=9$ integrals were left. The worst irreducible topologies that could not be integrated loop-by-loop had ($t=7$) $r=8$, $s=3$. 

One might have hoped that at $t=7$, using seed integrals up to $r=7$, $s=3$ only, all such $t=7$, $r=8$, $s=3$ integrals would reduce. This was not the case, so seed integrals with $t=7$, $r=8$, $s=3$ must be solved.

At this point we tried to start the proper bottom-up approach: $t-1$-subtopologies should be reduced to a form as compact as possible and substituted into the $t$-propagator topologies. We have already seen that 5-propagator topologies can simply be integrated out and then related to master integrals. Ideally the next step would be to reduce all needed $t=6$ topologies (11 in total) up to $r=8$, $s=3$, such that the $t=7$ IBP identities contained the least possible number of $t<7$ integrals. This so far was not possible because of the large number of $t=6$, $r=8$, $s=3$ IBP identities that have to be solved (as was illustrated in tab.~\ref{tab:NrOfEqsXX}). Indeed, even identities from $t=6$, $r=7$, $s=4$ seed integrals sometimes grow too large for an efficient handling using \emph{Solve} as described. 

This is where the reduction stands at the moment. The amplitude still contains ${\cal O}\lr{10}$ $t=7$, $r=8$, $s=3$ integrals, and presumably once they are reduced a few complicated $t=6$, $r=8$, $s=3$ integrals will be present. Also $t=6$, $r=8$, $s=0$ integrals that are not reduced using $t=6$, $r=8$, $s=0$ seed integrals might eventually pose a problem. 

An idea still to be investigated is that equations with a long tail of $t-1$ subtopologies could be shortened by introducing abbreviations. The tools to do so are already set up, and first experiences are positive: With some patience, the reduction of IBP identities from $t=7$, $r=8$, $s=3$ seed integrals might become possible without reducing any $t=6$ subtopologies by simply hiding them in (nested) abbreviations. Also, these systems seem to contain enough equations to overconstrain the whole $t=7$, $r=8$, $s=3$ block of integrals, such that once a solution could be obtained, it would reduce all these integrals instead of just relating them to each other. Using the solutions of lower-$r$-$s$ seed integrals, also the remaining (e.g. $t=7$, $r=8$, $s=2$) integrals might find a reduction equation.

\section{Outlook}\label{sec:outlook}
So far we have dealt exclusively with one out of three auxiliary topologies needed in the reduction of the three-loop quark form factor. Once the reduction of the planar auxiliary topology to master integrals is achieved the reduction of the remaining two auxiliary topologies is clearly the most immediate next step necessary to obtain a complete (gauge invariant) result of the three-loop quark form factor. We have pointed out however that this task will be a smaller one since most of the subtopologies encountered there will be the planar ones discussed here already.

Upon completion of this reduction, several immediate extensions are possible using the methods devised in this thesis. They can be applied without significant changes to any massless three-loop QCD vertex correction. The three-loop QCD form factors of the gluon, Higgs- and $Z$-boson are thus of primary interest. These calculations will profit heavily from the results of the photon form factor. In the case of the gluon form factor new topologies are expected to arise. The Higgs form factor will probably contain an additional scalar product in the numerator of its $t=9$ integrals which might pose a formidable challenge for the reduction. On the other hand the $Z$-boson form factor is expected to be very similar to the photon form factor since the difference only lies in the vertex Feynman rules and does not affect the topologies of the diagrams.

More generally speaking, in the quest for higher order perturbative calculations the niche of the tools discussed in this thesis lies in processes where the systems of IBP identities are too large for other implementations of the Laporta algorithm such as AIR~\cite{Anastasiou:2004vj}. AIR is very user-friendly and well tested but the solution of the system of IBP identities is carried out entirely in MAPLE and not in FORM. For large system this is expected to be a problem. 

Of course, \emph{Solve} and its extensions devised in this thesis do not rely on a particular physical process. If the \qgrafText Feynman rules are implemented, non-QCD (e.g. electroweak, supersymmetric,...) processes can be considered, too. Occasionally we have specialised in vertex graphs, e.g. in the the standardisation of the loop integrals. But the reduction itself only involves the solution of a general system of linear equations. Setting up and solving the system of equations only involves minor adjustments (e.g. of the coefficients of the momenta that define the propagators of the auxiliary topology). The reduction is then automated to a large degree but a lot of control over the details of the reduction can be maintained if necessary.


Summarizing we might say that the Laporta algorithm has brought about great advantages for many applications, but it also shows its limits: The prize for the automatization of the reduction process is the need to solve a large system of linear equations. These systems quickly become large for present-day computers. Still, we are not aware of any other method with which an analytic result for three-loop form factors would be in reach.

\section{Partial Results}\label{sec:partialresults}
The reduction of the 3-loop quark form factor is yet to be completed. However, we may already present one partial result. Collaborators reported that the evaluation of the 9-propagator master integral (with topology ID=1790) is actually much harder than evaluating the integral 
\begin{equation}
 {\cal T}_{1790}\lr{0,-1,-1,-1,-1,-1,-1,-1,1,-1,-1,0}\;,
\end{equation}
i.e. the integral where the integrand has an additional scalar product $\lr{k_2-p_1}^2$ in the numerator. The reduction of this integral is much easier than the reduction of the full amplitude for the form factor: At $t=6$ the most difficult integrals that need to be reduced have $r=8$ and $s=0$. The corresponding reduction equations could be found, and we successfully cross-checked our result with our collaborators who obtained the same result independently using AIR~\cite{Anastasiou:2004vj}. We obtain 

\begin{equation}
\begin{split}
 {\cal T}_{1790}(0,-1,&-1,-1,-1,-1,-1,-1,1,-1,-1,0)=\\
&+\frac{c_{4,1}}{\lr{q^2}^4} A_{4,1}\\
&+\frac{c_{5,1}}{\lr{q^2}^3} A_{5,1}+\frac{c_{5,2}}{\lr{q^2}^3} A_{5,2}+\frac{c_{2,3} }{\lr{q^2}^3}C_{2,3}+\frac{c_{5a}}{\lr{q^2}^3} B_{5a}\\
&+\frac{c_{6,1}}{\lr{q^2}^2} A_{6,1}+\frac{c_{6,2}}{\lr{q^2}^2} A_{6,2}+\frac{c_{6,3}}{\lr{q^2}^2}
   A_{6,3}\\
&+\frac{c_{7,3}}{q^2}  A_{7,3}+q^2 c_{9,1} A_{9,1}
\end{split}
\end{equation}
with
\begin{eqnarray*}
c_{4,1}&=&-\frac{1327872}{25
   \left(d-\frac{22}{5}\right) }+\frac{34600}{(d-4)
   }+\frac{250888}{9 (d-4)^2 }+\frac{67736}{3
   (d-4)^3 }+\frac{89168}{9 (d-4)^4 }+\frac{5056}{3
   (d-4)^5 }\\&&+\frac{13984}{5
   }+\frac{32760}{
   \left(d-\frac{14}{3}\right)}\;,\nonumber\\
c_{5,1}&=&\frac{3400}{(d-4)
   }+\frac{4880}{3 (d-4)^2 }+\frac{1840}{3 (d-4)^3
   }+\frac{320}{3 (d-4)^4
   }+96-\frac{15288}{5 
   \left(d-\frac{22}{5}\right)} \;,  \\
c_{5,2}&=&\frac{32928}{25
   \left(d-\frac{22}{5}\right) }-\frac{304}{9 (d-4)
   }+\frac{32}{(d-4)^2 }-\frac{128}{9 (d-4)^3
   }-\frac{64}{9 (d-4)^4 }-\frac{2944}{45
   }\\
   &&-\frac{14000}{9 
   \left(d-\frac{14}{3}\right)}\;,
   \\
c_{2,3}&=&\frac{1792}{(d-4)
   }+\frac{856}{(d-4)^2 }+\frac{336}{(d-4)^3
   }+\frac{64}{(d-4)^4 }+\frac{216}{5
   }-\frac{40768}{25 
   \left(d-\frac{22}{5}\right)}\;,\\
c_{5a}&=&-\frac{6904}{3 (d-4)
   }-\frac{1456}{(d-4)^2 }-\frac{640}{(d-4)^3
   }-\frac{352}{3 (d-4)^4
   }-192+\frac{8232}{5 
   \left(d-\frac{22}{5}\right)}\;,\\
c_{6,1}&=&-\frac{1176}{25
   \left(d-\frac{22}{5}\right) }+\frac{16}{3 (d-4)
   }+\frac{128}{45 }+\frac{1400}{27 
   \left(d-\frac{14}{3}\right)}\;,   \\
c_{6,2}&=&-\frac{200}{9
   }-\frac{20}{ (d-4)}-\frac{40}{9 
   (d-4)^2}\;,
   \\
c_{6,3}&=&\frac{216}{(d-4)
   }+\frac{88}{(d-4)^2 }+\frac{16}{(d-4)^3
   }+\frac{144}{5 }-\frac{4032}{25 
   \left(d-\frac{22}{5}\right)}\;,
   \\
c_{7,3}&=&\frac{8}{25
   \left(d-\frac{22}{5}\right) }-\frac{6}{5 }\;,
   \\
c_{9,1}&=& -\frac{4
   }{25
   \left(d-\frac{22}{5}\right)}+\frac{1}{d-4}-\frac{9
   }{10}\;.
\end{eqnarray*}

\section{Summary and Conclusions}
In this thesis we have attempted to calculate a part of the three-loop QCD contribution to the process $\gamma^*\rightarrow q\bar{q}$. We generated the 244 Feynman diagrams and projected them onto the form factor. The integrals in this expression had to be analysed and standardised in a semi-automated way, since even after standardisation the expression contained 2951 complicated three-loop integrals. The ultimate goal was to use integration-by-parts identities --linear equations in the unknown integrals-- that in principle relate all integrals to a set of master integrals. Most of these master integrals have already been computed by our collaborators~\cite{Heinrich:2007at}. But the reduction to master integrals proved difficult due to the large number of equations to be solved for the irreducible topologies. A final result is still missing, and so the main chapters in this thesis described the various specialised methods and algorithms that we hoped to push to a sophistication level sufficiently high in order to achieve a result. 

In the meantime --after having finished this thesis-- the quark and gluon form factors have been computed to three loops~\cite{Baikov:2009bg}. Congratulations to the authors!

\pagestyle{empty}
\cleardoublepage

\pagestyle{plain}
\addcontentsline{toc}{chapter}{Acknowledgements}
\chapter*{\center{Acknowledgements}}

\begin{center}
Whose merit is this work? I will not claim to know the answer, nor to understand the question.
Before turning to the more conventional way of showing my gratitude and respect for other people's contributions, my thoughts are with all those whose contributions I am not aware of. With the third hominid in prehistory that burnt his hands when trying to grab fire; with that unknown Egyptian girl from Amarna that could not afford to learn reading; with that Viking sailor that drowned on his return from North America and with Sitting Bull's grand mother; 
with the voters and politicians that granted my salary and 
with the colleague next doors who I just happen not to think of right now, though I should.

Thomas Gehrmann's 
experience and 
advice were indispensable in almost every single part of this thesis. His devotion to
 the needs of 
his students seems unparallelled to me. Who else would set up a special lecture series just because his students wish so?

Special thanks also go to all my collaborators.
It was a great pleasure to tutor Quantum Field Theory with Gionata Luisoni, especially for all the fruitful follow-up puzzles that we addressed together. Andrea Ferroglia and Tobias Motz received a big share of my questions
and 
always took their time for them. Daniel Ma\^itre, Tobias Huber, David Noth, Ayres Freitas, Pedro Schwaller, Nico Greiner, Gabor Somogyi, Cedric Studerus, Christian Kurz, Lucia Hosekova and Hana Tajuddin were no less helpful whenever I needed them.
Thanks to Douglas Potter, Roland Bernet, Peter Englmaier and Joachim Stadel for providing a formidable computing environment. 
I thank Pedro Schwaller and Tobias Motz for their helpful reading of this manuscript, and Tobias Huber for his patience when comparing our results. I thank Esther Meier, SuzAnne Wilde and Regina Schmid for running the secretary.

I also thank Esther Sch\"urmann for her inspiring illustration of the form factor from an artist's point of view. 

Less direct contributions are by no means less important. 
I had countless great evenings with Daniel Ma\^itre, Katja and Pedro Schwaller et al. at our weekly Doppelkopf round. I owe moving moments to the outstanding performances of Franz Welser-M\"ost and all the crew of the opera house of Zurich. I offer my deep gratitude to them-- and to Zurich's taxpayers who pay for the student tickets.

Finally-- how could I possibly have words left for what I owe my family?
\end{center}

\begin{appendix}
\pagestyle{headings}
\addtocontents{toc}{\contentsline {chapter}{Appendices}{}}
\chapter{Qgraf}\label{app:Qgraf}
\qgraf~\cite{Nogueira} is a tool for the fast generation of Feynman diagrams. After reading in the Feynman rules, it essentially provides the information that defines each diagram. This information is then printed according to a user-defined style file. This section describes in some (but not every) detail the files and configurations used for the generation of the three-loop quark form factor. 

Firstly, \qgrafText needs a model file listing propagators and the possible vertices of the particles in the Lagrangian:
\begin{lstlisting}[caption=the file qgraf.dat,numbers=left,frame=lines,firstnumber=1]{Name}
% propagators

[Q,Qbar,-]
[Ghost,Antighost,-]
[Gluon,Gluon,+]
[Photon,Photon,+, external]

% vertices

[Qbar,Photon,Q]
[Qbar,Gluon,Q]
[Gluon,Gluon,Gluon]
[Gluon,Gluon,Gluon,Gluon]
[Antighost,Gluon,Ghost]
\end{lstlisting}
This is a model file for QCD, plus an additional photon that couples to quark-antiquark pairs. The keyword \verb|external| in line 6 specifies that photons appear only as initial or final states in the diagram. \qgrafText generates the diagrams using the Feynman rules in this model file. 

\qgraf's output is user-defined according to the style file. The one we used reads
\begin{lstlisting}[caption=the style file mystyle.sty for \qgraf,numbers=left,frame=lines,firstnumber=1]{Name}
<prologue>
*
* file generated by <program>
*
<command_loop><command_line_loop>* <command_data><back>
<end><end>*

<diagram>(*@\label{line:diagloopstart}@*)

id sumup = sumup + 
 Label(<diagram_index>)*
 (<sign><symmetry_factor>)*
 <in_loop>
 pol<field>(<field_index>,<momentum>)*<end>
 <out_loop>
 pol<field>(<field_index>,<momentum>)*<end>

<propagator_loop> (*@\label{line:proploopstart}@*)
 prop<field>(<field_index>,<dual-field_index>,
<back><momentum>)*
<back><end>(*@\label{line:proploopend}@*)
<vertex_loop>(*@\label{line:vertexloopstart}@*)
 vert<vertex_degree>(<vertex_index>,
<back><ray_loop><field>(<field_index>,
<back><momentum>),<end><back>
<back>)*
<back><end><back>;(*@\label{line:vertexloopend}@*)
.sort(*@\label{line:diagloopend}@*)
<epilogue>

<exit>
\end{lstlisting}

In the diagram section (lines \ref{line:diagloopstart}-\ref{line:diagloopend}), \qgrafText loops over all diagrams, external particles, propagators and vertices and provides information for various quantities that define parts of the diagram. For example, in the propagator loop (lines \ref{line:proploopstart}-\ref{line:proploopend}), the quantities \verb|<field_index>| and \verb|<dual-field_index>| are provided for every propagator. They indicate which two fields are connected by a propagator. In the vertex and ray loop (lines \ref{line:vertexloopstart}-\ref{line:vertexloopend}), quantities like \verb|vertex_index| and again \verb|<field_index>| are available to define the vertices. 

Before being able to run \qgraf, the process must be specified for which the diagrams should be generated. This is done in the file qgraf.dat. The information given there includes the name of the output file (e.g. \verb|gammaq|\-\verb|qbar3Loop.out|), the style and model file to be used, the initial and final states of the process together with their momenta, the numbers of loops, the root name to be used for the loop momentam and various options. 
\begin{lstlisting}[caption=the file qgraf.dat,numbers=left,frame=lines,firstnumber=1]{Name}
output = 'gammaqqbar3Loop.out' ;
style = 'mystyle.sty' ;
model = 'QCD_QED' ;
in = Q[p1], Qbar[p2], Photon[mq];
out = ;
loops = 3 ;
loop_momentum = k ;
options =  onshell, nosnail ;
\end{lstlisting}

\begin{figure}
  \begin{center}
	\resizebox{8cm}{!}{\includegraphics{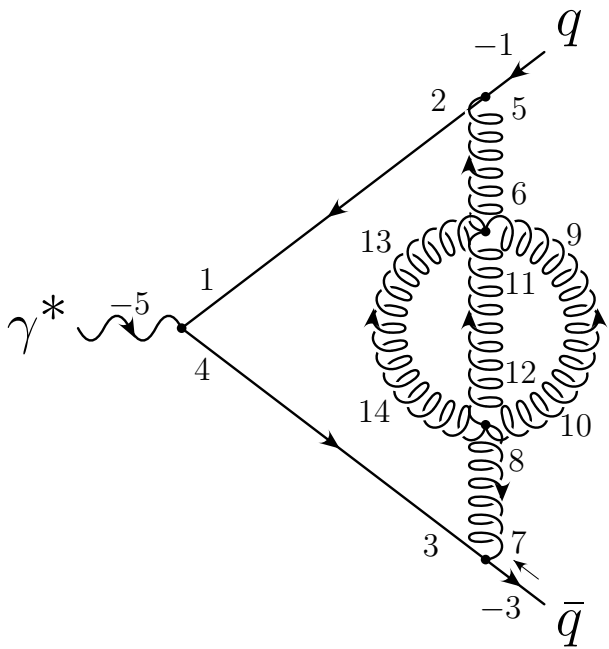}}
  \end{center}
\caption{The first diagram in the \qgrafText output. The numbers associated to the propagators or vertices correspond to the (dual-)field numbers that \qgrafText provides. They are used in the actual \qgrafText output that is given in the listing~\ref{lst:graph1}. Using such information, the Feynman amplitude can be constructed.}
\label{diag:Label1appendix}
\end{figure}

Running \qgrafText then produces the information that there are 244 diagrams. They are written down in the output file, a plain text file: 
\lstset{escapeinside={(*@}{@*)}}
\begin{lstlisting}[caption=excerpt of the \qgraf output file gammaqqbar3Loop.out,numbers=left,frame=lines, breaklines=true,firstnumber=1,label=lst:graph1]{Name}

id sumup = sumup +
 Label(1)*
 (+1/6)* (*@\label{line:symmetryfactor}@*)

 polQ(-1,p1)*
 polQbar(-3,p2)*
 polPhoton(-5,mq)*



 propQ(2,1,-k1)*
 propQ(4,3,-k1+mq)*(*@\label{line:prop1}@*)
 propGluon(6,5,-k1-p1)*
 propGluon(8,7,k1+p1)*
 propGluon(10,9,-k1+k2+k3-p1)*
 propGluon(12,11,-k2)*
 propGluon(14,13,-k3)*

 vert3(1,Qbar(4,k1-mq),Photon(-5,mq),Q(1,-k1))*(*@\label{line:vert2}@*)
 vert3(2,Qbar(2,k1),Gluon(5,-k1-p1),Q(-1,p1))*
 vert3(3,Qbar(-3,p2),Gluon(7,k1+p1),Q(3,-k1+mq))*(*@\label{line:vertex1}@*)
 vert4(4,Gluon(6,k1+p1),Gluon(9,-k1+k2+k3-p1),Gluon(11,-k2),Gluon(13,-k3))*
 vert4(5,Gluon(8,-k1-p1),Gluon(10,k1-k2-k3+p1),Gluon(12,k2),Gluon(14,k3));
.sort

id sumup = sumup +
 Label(2)*
...
\end{lstlisting}
This diagram (the first in \qgraf's output) is shown in fig.~\ref{diag:Label1appendix}. The momentum flow on vertices is always inwards, and the propagator \verb|propQ(4,3,-k1+mq)| has the momentum $-k_1-q$ flowing from the field index 4 to the dual field index 3. Also note the symmetry factor and fermion loop sign in line~\ref{line:symmetryfactor}.

The above code is adapted to be read into a FORM file simply using
\begin{lstlisting}[numbers=left,frame=lines, breaklines=true,firstnumber=1]{Name}
l expr=sumup;
#include gammaqqbar3Loop.out
\end{lstlisting}
The next steps are then:
\begin{itemize}
 \item Provide commuting factors such as $1/p^2$ for quark or gluon propagators, colour and electric charges for fermion vertices, 3- and 4-gluon-vertices. 
 \item Reconstruct external fermion lines: To this end find the vertex that contains the same field index ($-3$) as the antiquark (in our example: vertex 3 in line \ref{line:vertex1}). Then find the quark propagator that connects to the same vertex (\verb|propQ(4,3),-k1+mq)|, line~\ref{line:prop1}), then the vertex that connects to the next field index $4$ (vertex $1$, line \ref{line:vert2}), etc., until the field index of the external quark (-1) is reached.
\item Reconstruct closed fermion loops similarly. 
\item Project onto the form factor; evaluate Dirac traces and colour factors.
\item Collect the integrals into ${\cal T}$'s as defined in eq.~\myref{eq:exampleStandardProps}.
\end{itemize}
Flavour indices on propagators and vertices must be taken into account properly. If after projection onto the form factor in \myref{eq:Formfactorprojection} the two Dirac matrices in the photon vertices $\gamma_\mu$, $\gamma^\mu$  lie in the same Dirac trace, the charge factor is $\sum_q \lr{e_q^2}$, otherwise it is $\lr{\sum_q e_q}^2$.
\pagestyle{empty}
\cleardoublepage
\pagestyle{headings}

\chapter{Equivalent Planar Topologies}\label{app:TopoPermutations}
As described in sec.~\ref{sec:exploitingSymmetries}, quite a number of topologies are usually equivalent, since permutations of (loop or external) momenta change an integral's ID number, but not its value. The ID number defines the topology\footnote{We have seen that several graphs may have the same ID number. The graph given in fig.~\ref{diag:Label1appendix} e.g. has two propagators with the same momentum. This is not reflected in its ID number. More precisely, the ID number only distinguishes  $r=t$, $s=0$ integrals.} according to the auxiliary topology~1 in tab.~\ref{tab:auxtopos}. It is restricted to be smaller than $2^{12}=4096$. Here we list the set of identifications that we obtained. Topologies that are not listed are either vanishing (e.g. all $t\leq3$ integrals) or have $t>9$ and thus are not present in the three-loop quark form factor.
\begin{center}
\begin{tabular}{|l@{  $\to$  }r|}
 \hline
\multicolumn{2}{|c|}{$t=4$} \\ \hline
 178,540,561,2074,2089& 172\\
\hline
\end{tabular}

\vspace{1cm}
\begin{tabular}{|l@{  $\to$  }r|}
 \hline
\multicolumn{2}{|c|}{$t=5$} \\
\hline
 677,2190,2211,2573,2579& 662 \\
 626,1193,1562,2156,2332& 433 \\
 604,1202,1585,2138,2345& 428 \\
 436,620,628,1178,1194,1561,1577,2154,2162,2329,2353& 412 \\
 242,434,625,796,817,1196,1564,2153,2330,3098,3113& 236 \\
 188,569,572,2105,2106& 186 \\
 182,541,565,668,684,2075,2091,2202,2226,2601,2609& 174 \\
 179,542,563,689,690,2078,2093,2217,2220,2586,2588& 173 \\
 557,692,2099,2218,2585& 158\\
\hline
\end{tabular}
\end{center}

\newpage

\begin{center}

\tablefirsthead{
\hline
\multicolumn{2}{|c|}{$t=6$} \\  
\hline
}
\tablehead{
\hline
$t=6$&\small\sl continued \\  
\hline
}
\tabletail{%
\hline
\multicolumn{2}{|r|}{\small\sl continued on next page} \\
\hline
}
\tablelasttail{\hline}
 \begin{supertabular}{|l@{  $\to$  }r|}
 1699,2445,2469,2638,2646& 1683 \\
 1457,1626,1650,2396,2412& 1449 \\
 1642,2420& 1433 \\
 918,2275,2835,3214,3597& 741 \\
 933,1686,1701,2254,2446,2467,2637,2643,2829,3235,3603& 726 \\
 2221,2590& 691 \\
 685,693,694,2206,2219,2222,2227,2587,2589,2605,2611& 670 \\
 686,2203,2230,2603,2613& 669 \\
 679,2191,2215,2575,2583,2701,2702,2707,2710,2723,2725& 663 \\
 884,1258,1434,1436,1641,1644,1817,2417,2418,3178,3353& 500 \\
 881,1260,1820,3177,3354& 498 \\
 882,1257,1818,3180,3356& 497 \\
 860,1266,1452,1458,1628,1649,1841,2394,2409,3162,3369& 492 \\
 876,1242,1450,1460,1625,1652,1833,2393,2410,3186,3377& 476 \\
 636,1210,1593,2170,2361& 444 \\
 634,1209,1594,2172,2364& 441 \\
 629,748,924,1198,1565,2155,2290,2331,2865,3226,3625& 438 \\
 630,1195,1563,1690,1705,2158,2333,2460,2481,2668,2674& 437 \\
 627,754,945,1197,1566,2157,2284,2334,2844,3241,3610& 435 \\
 605,732,940,1206,1589,2139,2266,2347,2857,3250,3633& 430 \\
 606,1203,1587,1713,1714,2142,2349,2473,2476,2650,2652& 429 \\
 621,756,948,1182,1581,2163,2282,2355,2841,3242,3609& 414 \\
 622,1179,1579,1689,1706,2166,2357,2457,2484,2666,2676& 413 \\
 252,442,633,825,828,1212,1596,2169,2362,3129,3130& 250 \\
 246,797,821,1692,1708,2458,2482,2665,2673,3099,3115& 238 \\
 243,753,798,819,946,2281,2842,3102,3117,3244,3612& 237 \\
 813,1716,2474,2649,3123& 222 \\
 573,700,2107,2234,2617& 190 \\
 189,571,574,697,698,2109,2110,2233,2236,2618,2620& 187 \\
 183,543,567,2079,2095,2714,2716,2729,2732,2737,2738& 175 \\
 559,2103,2713,2730,2740& 159\\
\hline
\end{supertabular}
\end{center}\newpage
\begin{center}
\tablefirsthead{
\hline
\multicolumn{2}{|c|}{$t=7$} \\  
\hline
}
\tablehead{
\hline
\multicolumn{2}{|c|}{$t=7$ (\small\sl continued)}\\
\hline
}

\tablelasttail{\hline}
\begin{supertabular}{|l@{  $\to$  }r|}
 2951,3719& 2759 \\
 2711,2727& 2703 \\
 1939,2533,2902,3469,3662& 1763 \\
 1957,2510,2893,3491,3667& 1750 \\
 1955,2509,2894,3493,3670& 1747 \\
 2477,2654& 1715 \\
 1707,2461,2485,2670,2678& 1691 \\
 1703,2447,2471,2639,2647,2766,2774,2957,2981,3731,3747& 1687 \\
 1522,1884,1905,3418,3433& 1516 \\
 1521,1882,1906,3420,3436& 1513 \\
 1500,1514,1524,1881,1897,1900,1908,3417,3434,3441,3442& 1498 \\
 1898,3444& 1497 \\
 1658,2428& 1465 \\
 1459,1630,1651,1778,1969,2398,2413,2540,2908,3497,3674& 1453 \\
 1461,1627,1654,1754,1961,2397,2414,2524,2924,3505,3698& 1451 \\
 1437,1643,1646,1770,1945,2421,2422,2548,2932,3481,3690& 1435 \\
 1438,1645,2419,3306,3865& 1012 \\
 1004,1454,1462,1629,1653,2395,2411,3290,3314,3881,3889& 988 \\
 997,1765,1942,2531,2899,3278,3299,3470,3661,3853,3859& 982 \\
 949,1694,1709,2286,2462,2483,2669,2675,2845,3243,3611& 758 \\
 947,2285,2846,3245,3614& 755 \\
 925,1693,1710,2294,2459,2486,2667,2677,2869,3227,3627& 750 \\
 757,926,950,2283,2291,2843,2867,3230,3246,3613,3629& 749 \\
 919,2279,2787,2789,2839,2963,2966,3215,3599,3725,3726& 743 \\
 941,1717,1718,2270,2475,2478,2651,2653,2861,3251,3635& 734 \\
 942,2267,2859,3254,3637& 733 \\
 935,2255,2765,2771,2831,2958,2979,3239,3607,3734,3749& 727 \\
 702,2235,2238,2619,2621& 701 \\
 2237,2622& 699 \\
 2223,2591,2718,2733,2739& 695 \\
 687,2207,2231,2607,2615,2715,2717,2731,2734,2741,2742& 671 \\
 892,1274,1466,1468,1657,1660,1849,2425,2426,3194,3385& 508 \\
 889,1276,1852,3193,3386& 506 \\
 890,1273,1850,3196,3388& 505 \\
 885,1262,1772,1821,1948,2546,2929,3179,3355,3482,3689& 502 \\
 886,1259,1769,1819,1946,2545,2930,3182,3357,3484,3692& 501 \\
 883,1009,1010,1261,1822,3181,3305,3308,3358,3866,3868& 499 \\
 861,1270,1756,1845,1964,2522,2921,3163,3371,3506,3697& 494 \\
 862,1267,1777,1843,1970,2537,2906,3166,3373,3500,3676& 493 \\
 877,1246,1780,1837,1972,2538,2905,3187,3379,3498,3673& 478 \\
 878,1243,1753,1835,1962,2521,2922,3190,3381,3508,3700& 477 \\
 637,764,956,1214,1597,2171,2298,2363,2873,3258,3641& 446 \\
 638,1211,1595,1721,1722,2174,2365,2489,2492,2682,2684& 445 \\
 635,762,953,1213,1598,2173,2300,2366,2876,3257,3642& 443 \\
 631,1199,1567,2159,2335,2796,2802,2972,2993,3738,3753& 439 \\
 607,1207,1591,2143,2351,2778,2780,2985,2988,3761,3762& 431 \\
 623,1183,1583,2167,2359,2794,2804,2969,2996,3737,3754& 415 \\
 829,1724,2490,2681,3131& 254 \\
 253,761,827,830,954,2297,2874,3133,3134,3260,3644& 251 \\
 247,799,823,2793,2801,2970,2994,3103,3119,3740,3756& 239 \\
 815,2777,2986,3127,3764& 223 \\
 575,2111,2745,2746,2748& 191\\
\end{supertabular}
\end{center}

\begin{center}
\tablefirsthead{
\hline
\multicolumn{2}{|c|}{$t=8$} \\  
\hline
}
\tablehead{
\hline
\multicolumn{2}{|c|}{$t=8$ (\small\sl continued)}\\
\hline
}
 \begin{supertabular}{|l@{  $\to$  }r|}
 3783,3975& 3015 \\
 2967,3727& 2791 \\
 2775,2959,2983,3735,3751& 2767 \\
 2735,2743& 2719 \\
 2021,3534,3555,3917,3923& 2006 \\
 2019,3533,3557,3918,3926& 2003 \\
 1971,2541,2910,3501,3678& 1779 \\
 1949,2550,2933,3483,3691& 1774 \\
 1947,2549,2934,3485,3694& 1771 \\
 1943,2535,2903,3030,3045,3471,3663,3790,3811,3981,3987& 1767 \\
 1782,1965,1973,2526,2542,2909,2925,3499,3507,3675,3699& 1758 \\
 1963,2525,2926,3509,3702& 1755 \\
 1959,2511,2895,3021,3022,3495,3671,3795,3798,4003,4005& 1751 \\
 2493,2686& 1723 \\
 2479,2655,2782,2989,3763& 1719 \\
 1711,2463,2487,2671,2679,2798,2806,2973,2997,3739,3755& 1695 \\
 1532,1913,1916,3449,3450& 1530 \\
 1914,3452& 1529 \\
 1526,1885,1909,2012,2028,3419,3435,3546,3570,3945,3953& 1518 \\
 1523,1886,1907,2033,2034,3422,3437,3561,3564,3930,3932& 1517 \\
 1525,1883,1910,2010,2025,3421,3438,3548,3569,3948,3954& 1515 \\
 1901,2036,3443,3562,3929& 1502 \\
 1501,1899,1902,2009,2026,3445,3446,3545,3572,3946,3956& 1499 \\
 1469,1659,1662,1786,1977,2429,2430,2556,2940,3513,3706& 1467 \\
 1463,1631,1655,2399,2415,3036,3052,3802,3826,4009,4017& 1455 \\
 1647,2423,3060,3818,3993& 1439 \\
 1470,1661,2427,3322,3897& 1020 \\
 1014,1773,1950,2547,2931,3307,3310,3486,3693,3867,3869& 1013 \\
 3309,3870& 1011 \\
 1005,1781,1974,2539,2907,3294,3315,3502,3677,3885,3891& 990 \\
 1006,1757,1966,2523,2923,3291,3318,3510,3701,3883,3893& 989 \\
 999,3027,3043,3279,3303,3789,3813,3855,3863,3982,3990& 983 \\
 957,1725,1726,2302,2491,2494,2683,2685,2877,3259,3643& 766 \\
 958,2299,2875,3262,3645& 765 \\
 955,2301,2878,3261,3646& 763 \\
 951,2287,2797,2803,2847,2974,2995,3247,3615,3742,3757& 759 \\
 927,2295,2795,2805,2871,2971,2998,3231,3631,3741,3758& 751 \\
 943,2271,2779,2781,2863,2987,2990,3255,3639,3765,3766& 735 \\
 2239,2623,2747,2749,2750& 703 \\
 893,1278,1788,1853,1980,2554,2937,3195,3387,3514,3705& 510 \\
 894,1275,1785,1851,1978,2553,2938,3198,3389,3516,3708& 509 \\
 891,1017,1018,1277,1854,3197,3321,3324,3390,3898,3900& 507 \\
 887,1263,1823,3057,3058,3183,3359,3817,3820,3994,3996& 503 \\
 863,1271,1847,3034,3049,3167,3375,3804,3825,4012,4018& 495 \\
 879,1247,1839,3033,3050,3191,3383,3801,3828,4010,4020& 479 \\
 639,1215,1599,2175,2367,2810,2812,3001,3004,3769,3770& 447 \\
 831,2809,3002,3135,3772& 255\\
\end{supertabular}
\end{center}
\begin{center}
\tablefirsthead{
\hline
\multicolumn{2}{|c|}{$t=9$} \\  
\hline
}
\tablehead{
\hline
\multicolumn{2}{|c|}{$t=9$ \small\sl(continued) }\\  
\hline
}
 \begin{supertabular}{|l@{  $\to$  }r|}
 3047,3791,3815,3983,3991& 3031 \\
 3799,4007& 3023 \\
 2807,2975,2999,3743,3759& 2799 \\
 2991,3767& 2783 \\
 3565,3934& 2035 \\
 2029,2037,2038,3550,3563,3566,3571,3931,3933,3949,3955& 2014 \\
 2030,3547,3574,3947,3957& 2013 \\
 2027,3549,3573,3950,3958& 2011 \\
 2023,3535,3559,3919,3927,4045,4046,4051,4054,4067,4069& 2007 \\
 1981,2558,2941,3515,3707& 1790 \\
 1979,2557,2942,3517,3710& 1787 \\
 1975,2543,2911,3038,3053,3503,3679,3806,3827,4013,4019& 1783 \\
 1951,2551,2935,3061,3062,3487,3695,3819,3822,3995,3997& 1775 \\
 1967,2527,2927,3037,3054,3511,3703,3803,3830,4011,4021& 1759 \\
 2495,2687,2814,3005,3771& 1727 \\
 1917,2044,3451,3578,3961& 1534 \\
 1533,1915,1918,2041,2042,3453,3454,3577,3580,3962,3964& 1531 \\
 1527,1887,1911,3423,3439,4058,4060,4073,4076,4081,4082& 1519 \\
 1903,3447,4057,4074,4084& 1503 \\
 1663,2431,3068,3834,4025& 1471 \\
 1022,1789,1982,2555,2939,3323,3326,3518,3709,3899,3901& 1021 \\
 3325,3902& 1019 \\
 3059,3311,3821,3871,3998& 1015 \\
 1007,3035,3051,3295,3319,3805,3829,3887,3895,4014,4022& 991 \\
 959,2303,2811,2813,2879,3003,3006,3263,3647,3773,3774& 767 \\
 895,1279,1855,3065,3066,3199,3391,3833,3836,4026,4028& 511\\
\end{supertabular}
\end{center}

\chapter{Master Integrals}\label{app:ListOfIntegrals}
The irreducible integrals that appear in the reduction of the three-loop quark form factor were given diagrammatically in fig.~\ref{diag:1790} (t=9), fig.~\ref{fig:t7IrreducibleTopos} (t=7), fig.~\ref{fig:t6IrreducibleTopos} (t=6) and fig.~\ref{fig:t5irredtopos} (t=5). Here we shall collect results for these integrals as far as they are known in the literature. They are also listed in~\cite{Gehrmann:2006wg,Heinrich:2007at,Heinrich:2009be}. 

For the integral $A_{9,1}$ (ID=1790, see fig.~\ref{diag:1790} and eq.~\myref{eq:exampleStandardProps}), a result has recently been obtained~\cite{Heinrich:2009be}:
\begin{equation}
\begin{split}
  A_{9,1}^{(n)}\;=\;& i \, S_\Gamma^3 \, \left[ - q^2-i \, \eta
\right]^{-2-3\, \epsilon}  \\ &
\times \Big[-\frac{1}{36
    \epsilon^6}-\frac{\pi^2}{18\epsilon^4}-\frac{14\zeta_3}{9\epsilon^3}-\frac{47\pi^4}{405\epsilon^2}\\ &
  \hspace*{18 pt}
  +\left(-\frac{85}{27}\pi^2\zeta_3-20\zeta_5\right)\frac{1}{\epsilon}
  +\left(-\frac{1160\pi^6}{5103}-\frac{137}{3}\zeta_3^2\right)+{\cal{O}}(\epsilon)\Big]\;.
\end{split}
\end{equation}
$A_{7,3}$ (ID=758) reads
\begin{equation}
   A_{7,3} = i \, S_\Gamma^3 \left[ - q^2-i \, \eta \right]^{-1-3\, \epsilon} \,\left[ \left( -\frac{\pi^2\,\zeta_3}{6}-10
\,\zeta_5\right)\frac{1}{\epsilon} - \frac{119 \,\pi^6}{2160} - \frac{31}{2} \, \zeta_3^2 + {\cal O} (\epsilon)\right] 
\end{equation}
with $\eta\ge 0$ and 
\begin{equation}
 S_{\Gamma} = \frac{1}{\left( 4\pi\right)^{d/2}\,\Gamma(1-\epsilon)}\;.
\end{equation}
$A_{6,1}$ (ID=1433) is integrable loop-by-loop. We may express the result using eqns.~\myref{eq:f} and \myref{eq:g}:
\begin{equation}
 A_{6,1}=\lr{-q^2-i\,\eta}^{-3\epsilon}f_{1,1}^2 g_{1,2\epsilon,1}\;.
\end{equation}
$A_{6,2}$ (ID=444) and $A_{6,3}$ (ID=429) are much more difficult to evaluate. The authors of~\cite{Heinrich:2007at} found
\begin{equation}
\begin{split}
 A_{6,2}=i & S_\Gamma^3\left[-q^2-i\,\eta\right]^{-3\epsilon}\\
&\times
 \biggl[-\frac{2\zeta_3}{\epsilon}-18\zeta_3-\frac{7\pi^4}{180}+\left(-122\zeta_3-\frac{7\pi^4}{20}+\frac{2\pi^2}{3}\zeta_3-10\zeta_5\right)\epsilon\\
&\hspace{3pt}\phantom{\times\biggl[} +\lr{-738\zeta_3-\frac{427\pi^4}{180}+6\pi^2\zeta_3-90\zeta_5+\frac{163\pi^6}{7560}+76\zeta_3^2}\epsilon^2 +{\cal O}\lr{\epsilon^3}\biggr]
\end{split}
\end{equation}
and
\begin{eqnarray*}
 A_{6,3}&=&i  S_\Gamma^3\left[-q^2-i\,\eta\right]^{-3\epsilon}\\
&&\times\biggl[ -\frac{1}{6\epsilon^3}-\frac{3}{2\epsilon^2}-\lr{\frac{55}{6}+\frac{\pi^2}{6}}\frac{1}{\epsilon}-\frac{95}{2}-\frac{3\pi^2}{2}+\frac{17\zeta_3}{3}\\
&&\phantom{\times\biggl[}+\lr{-\frac{1351}{6}-\frac{55\pi^2}{6}-\frac{\pi^4}{90}+51\zeta_3}\epsilon\\
&&\phantom{\times\biggl[}+\lr{-\frac{2023}{2}-\frac{95\pi^2}{2}-\frac{\pi^4}{10}+\frac{935\zeta_3}{3}+\frac{10\pi^2\zeta_3}{3}+65\zeta_5}\epsilon^2\\
&&\phantom{\times\biggl[}+\biggl(-\frac{26335}{6}-\frac{1351\pi^2}{6}-\frac{11\pi^4}{18}+\frac{7\pi^6}{54}\\
&&\phantom{\times\biggl[+\biggl(\:}+1615\zeta_3+30\pi^2\zeta_3-\frac{268\zeta_3^2}{3}+585\zeta_5\biggr)\epsilon^3+{\cal O}\lr{\epsilon^4}\biggr]\qquad.
\end{eqnarray*}
The $t=4$ and $t=5$ topologies in fig.~\ref{fig:t5irredtopos} read
\begin{align*}
 &B_{4\phantom{,1}}=\lr{-q^2-i\,\eta}^{2 - 3\epsilon}f_{1,1}f_{1,\epsilon}f_{1, - 1 + 2\epsilon}& \mathrm{(ID=172)}\phantom{\;,}\\
 &B_{5a\phantom{,}}=\lr{-q^2-i\,\eta}^{1 - 3\epsilon}f_{1,1}^2f_{2\epsilon,1}& \mathrm{(ID=158)}\phantom{\;,}\\
 &A_{5,1}=\lr{-q^2-i\,\eta}^{1 - 3\epsilon} f_{1,1} f_{1,\epsilon} g_{1, - 1 + 2\epsilon,1}&\mathrm{(ID=433)}\phantom{\;,}\\
 &A_{5,2}=\lr{-q^2-i\,\eta}^{1 - 3\epsilon}f_{1,1}^2 g_{\epsilon,\epsilon,1} &\mathrm{(ID=412)}\phantom{\;,}\\
 &C_{2,3}=\lr{-q^2-i\,\eta}^{1 - 3\epsilon} f_{1,1}^2f_{1,\epsilon}& \mathrm{(ID=662)}\;,
\end{align*}
with the $f$'s and $g$'s as defined in~\myref{eq:f} and \myref{eq:g}.
 
\end{appendix}

\end{document}